\documentclass[11pt,a4paper]{article}
\usepackage[utf8]{inputenc}

\usepackage{mathtools}
\usepackage{amsthm}
\usepackage{amssymb}
\usepackage{bbm}
\usepackage{diagbox}
\usepackage{bbold}
\usepackage{dsfont}
\usepackage{units}
\usepackage{titling}
\usepackage{makecell}
\usepackage{graphicx}
\usepackage{caption}
\usepackage{float}
\usepackage{cite}
\usepackage{subfig}
\usepackage{braket}
\usepackage{slashed}
\usepackage{physics}
\usepackage{vmargin}
\usepackage{xcolor}
\usepackage{todonotes}
\usepackage[colorlinks=false]{hyperref}
\usepackage{comment}
\usepackage{cleveref}
\usepackage{authblk}

\numberwithin{equation}{section}
\title{Skyrme crystals, nuclear matter and compact stars} 

\author[1]{Christoph Adam \thanks{adam@igfae.usc.es}}
\author[1]{Alberto Garc\'ia Mart\'in-Caro \thanks{alberto.martin-caro@usc.es}}
\author[1]{Miguel Huidobro \thanks{miguel.huidobro.garcia@usc.es}}
\author[2]{Andrzej Wereszczynski \thanks{andrzej.wereszczynski@uj.edu.pl}}
\affil[1]{Departamento de F\'isica de Part\'iculas, Universidad de Santiago de Compostela, Spain, {\it and} \;
Instituto
Galego de F\'isica de Altas Enerxias (IGFAE) E-15782 Santiago de Compostela, Spain}
\affil[2]{Institute of Physics, Jagiellonian University, Lojasiewicza 11, Krak\'ow, Poland}
\date{\today}

\newcommand{\lag}{\mathcal{L}}
\newcommand{\fpi}{f_{\pi}}
\newcommand{\mpi}{m_{\pi}}
\newcommand{\dmu}{\partial_{\mu}}

\begin{document}
\maketitle
\begin{abstract} 
A general review of the crystalline solutions of the generalized Skyrme model and their application to the study of cold nuclear matter at finite density and the Equation of State (EOS) of neutron stars is presented. For the relevant range of densities, the ground state of the Skyrme model in the three torus is shown to correspond to configurations with different symmetries, with a sequence of phase transitions between such configurations. The effects of nonzero finite isospin asymmetry are taken into account by the canonical quantization of isospin collective coordinates, and some thermodynamical and nuclear observables (such as the symmetry energy) are computed as a function of the density. We also explore the extension of the model to accommodate strange degrees of freedom, and find a first order transition for the condensation of kaons in the Skyrme crystal background in a thermodynamically consistent, non-perturbative way. 
Finally, an approximate EOS of dense matter is constructed by fitting the free parameters of the model to some nuclear observables close to saturation density, which are particularly relevant for the description of nuclear matter. The resulting neutron star mass-radius curves already reasonably satisfy  current astrophysical constraints. 
\\ \\ \\
{\large \bf Dedicated to the memory of our unforgettable friend and colleague Ricardo V\'azquez.}
\\ \\ \\ \\ \\ \\ \\
Invited contribution to  the Special Issue: Symmetries and Ultra Dense Matter of Compact Stars, {\bf Symmetry} 2023, 15(4), 899

\end{abstract}
\newpage

\tableofcontents
\section{Introduction}

With the advent of gravitational wave and multi-messenger astronomy, the available constraints on the equation of state (EOS) of neutron stars, namely, strongly interacting matter at finite density, have improved significantly in the last decade.  Furthermore, additional gravitational wave measurements from neutron star mergers with improved precision from LIGO/Virgo are expected to impose even stricter constraints on the EOS in the near future.

\begin{figure}[hbt]
    \centering
    \includegraphics[scale=0.4]{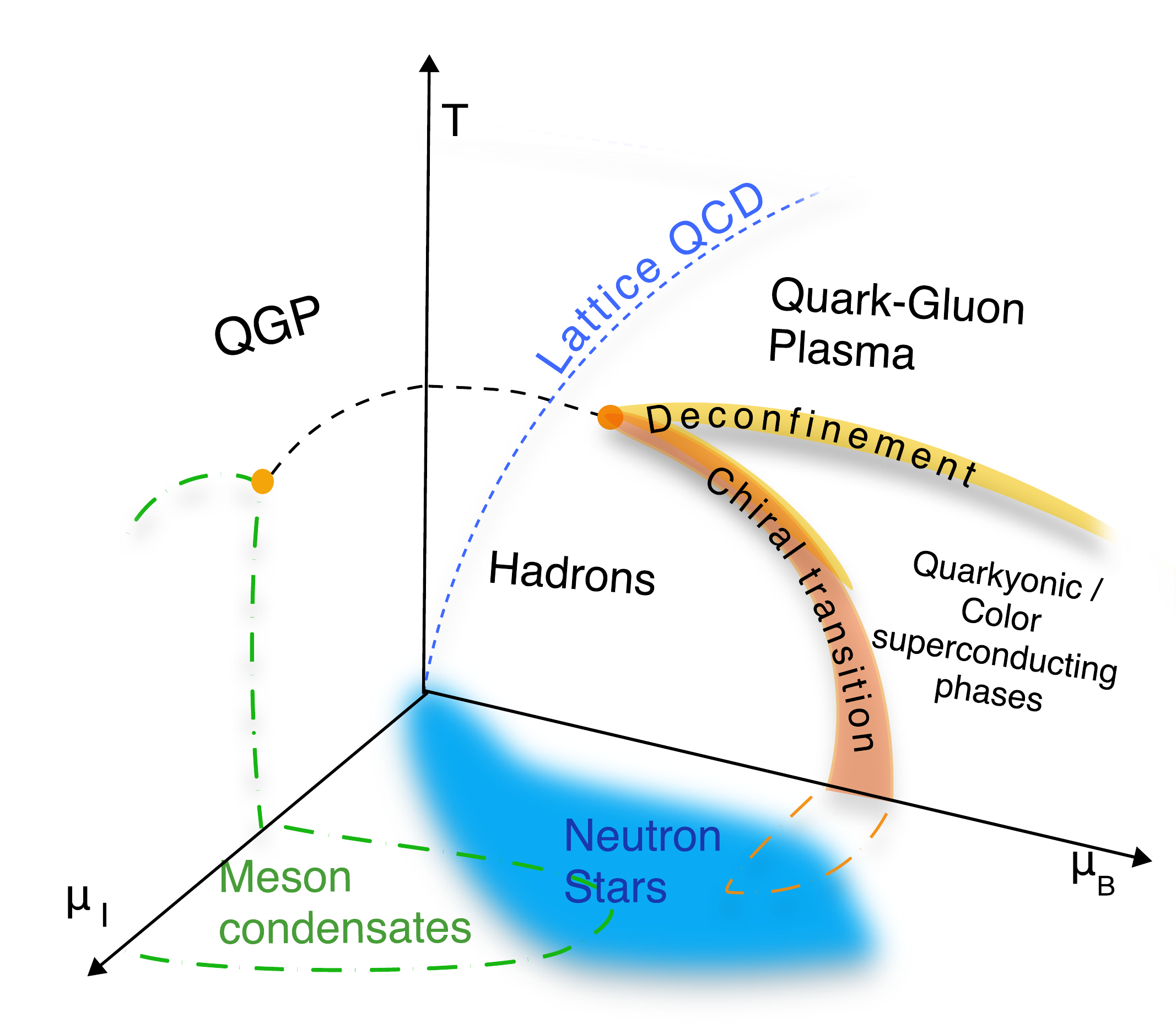}
    \caption{Schematic phase diagram of QCD}
    \label{fig:PhasesQCD}
\end{figure}

The physics of strong interactions is described by a nonabelian gauge theory, Quantum Chromodynamics (QCD), whose nonperturbative nature at small and intermediate energy or density regimes makes the computation of nuclear matter properties from first principles extremely challenging. Indeed, a perturbative treatment is only available at asymptotically high densities, while the main nonperturbative computational tool, lattice QCD, can be used for arbitrarily low temperatures but only small densities, due to the fermion sign problem. Hence, for sufficiently high densities (but not asymptotically high), the phase diagram of QCD is poorly understood, and there is a big theoretical uncertainity on basic observables such as the relevant degrees of freedom or the EOS in this regime, which, on the other hand, is precisely the relevant region for the matter in the interior of neutron stars. Indeed, as depicted in the schematic phase diagram of \cref{fig:PhasesQCD}, matter inside neutron stars is expected to be both at finite baryonic and isospin densities, and almost zero temperature (when compared with the relevant density scale). 

As first principle approaches are currently not suitable for this task, a plethora of different effective theories and phenomenological models have been proposed to either qualitatively or quantitatively study the QCD phase diagram in the intermediate density regime. 

Standard nuclear physics methods like relativistic mean field theory \cite{Walecka:1974qa} or chiral perturbation theory \cite{Weinberg:1990rz,Weinberg:1991um} can be used for this purpose (for a recent review we refer to \cite{Burgio:2021vgk}). 
These theories are effective field theories (EFT) in the sense that they introduce nucleons and mesons as their basic fields, instead of the quarks and gluons of QCD.  Further, they depend on a rather large number of a priori unknown parameters which are usually determined by fitting to nuclear forces and scattering data in the (non-relativistic) low energy regime. Also, nucleons (and baryons in general) are treated as point particles. The extrapolation of these models to baryon densities beyond nuclear saturation might, therefore, be affected by rather large uncertainties \cite{Tews:2018iwm}. 

The Skyrme model represents a slightly different type of EFT. This model, proposed more than 60 years ago by Tony Skyrme \cite{skyrme1961non, skyrme1962unified}, is an EFT of chiral mesons only, whereas nucleons and baryons are realized as topological solitons ("skyrmions") of the mesonic field, and their interactions are described by the same mesonic lagrangian. 
In addition, the topological degree of these solitonic solutions can be identified with the baryon number. The argument for the identification of skyrmions with baryons and nuclei was further strengthened within the \textit{large $N_c$ limit} of QCD \cite{tHooft:1973alw,Witten:1979kh}, for which the theory becomes an effective weakly interacting model of mesons, where baryons satisfy the usual properties of solitons.
The Skyrme model incorporates in a completely natural fashion several important features of strong interaction physics, like chiral symmetry and its breaking, the conservation of baryon number, or the extended character of nucleons. As a consequence of the latter, it also avoids short-distance singularities in nucleon-nucleon interactions. In addition, it naturally incorporates the spin-statistics theorem in that skyrmions quantized with a half-odd integer spin are fermions, whereas for integer spin they are bosons \cite{Finkelstein:1968hy}. For a recent review we refer to \cite{Manton:2022fcb}.

Skyrme´s idea of baryons as topological solitons is shared with a number of other approaches.
First of all, holographic models based on the conjectured gauge/gravity duality have proven to be useful as tools for studying the nonperturbative regimes of QCD-like theories. For a recent review on the application of holographic models to the description of neutron stars, see \cite{Jarvinen:2021jbd}. 
The Skyrme model has, in fact, been shown to appear as a holographic boundary theory in several holographic QCD models such as the Sakai-Sugimoto model. A flat space version of these holographic models was proposed in \cite{Sutcliffe:2010et}, such that the holographic dual of the flat space instantons provides the Skyrme field coupled to an infinite tower of vector mesons. The full holographic dual maintains the conformal symmetry inherited from the instanton, whereas any truncation to a finite number of vector mesons breaks conformal invariance. Secondly, in a different but related approach a class of generalized nuclear effective theories have been developed in the last two decades which maintain the field contents and topological structure of the Skyrme model at low energies while flowing towards an effective theory compatible with the symmetries of QCD at higher energies (for a recent review see \cite{Ma-Yang-2023} in this special issue).
This flow is achieved by a coupling of the Skyrme field to the dilaton \cite{Migdal82} and an infinite tower of vector mesons via hidden local symmetry \cite{Bando1988, Harada2003}, where these fields flow from a spontaneously broken to an unbroken phase. As a consequence, these nuclear effective theories still share some topological structures with the Skyrme model, while their dynamical contents is, in general, different and much harder to calculate. For that reason, instead of the prohibitively difficult full numerical simulations of those field theories, sometimes it is {\em assumed} that they inherit some topological structures from the Skyrme model, and the consequences of these assumptions are worked out using more standard effective field theory methods of nuclear physics. We shall comment on some of these assumptions in our conclusions. A detailed discussion of all the issues mentioned in this last paragraph can be found in \cite{multifaceted}.

\subsection{The Skyrme model}
As said, Skyrme´s original motivation was to reproduce baryons as topological soliton solutions from a purely mesonic field theory.  In its simplest version, the basic fields of the Skyrme model are just the pion fields combined in an SU(2) valued field $U$ as
\begin{equation}
    U = \sigma \mathbbm{I} + i\pi_a \tau_a, \quad U^{\dagger}U = \mathbbm{I} \longrightarrow \sigma^2 + \pi_a\pi_a = 1, \quad a = 1,2,3
    \label{U_exp}
\end{equation}
where $\tau_a$ are the Pauli matrices.
Left ($L$) and right ($R$) chiral transformations act on this field like $U\to LUR$. The two simplest terms of the resulting effective (Skyrme model) lagrangian can be easily guessed from low-energy considerations, namely the so-called nonlinear sigma model (or Dirichlet) term
\begin{equation} 
 \lag_2 = -\frac{\fpi^2}{16}\Tr\{L_{\mu}L^{\mu}\} \; , \quad L_{\mu} = U^{\dagger}\dmu U
 \end{equation}
 providing a kinetic term for the pions, and the pion mass potential
 \begin{equation}
 \lag_0 =  \frac{\mpi^2\fpi^2}{8}\Tr\{U - \mathbbm{I}\}.
 \end{equation}
 Here, $f_\pi $ is the pion decay constant, whose physical value in the conventions used here is $f_\pi = 186 $ MeV. Further, $L_\mu$ are the components of the left-invariant Maurer-Cartan form of the $SU(2)$ group, $m_\pi$ is the pion mass, and  $\mathbbm{I}$ is the $2\times2$ identity matrix. These two terms alone, however, cannot support the existence of static (soliton) solutions, because they are unstable under a spatial rescaling $x \rightarrow \Lambda x$. To stabilize them, Skyrme added the term
 \begin{equation}
 \lag_4 = \frac{1}{32e^2}\Tr\{[L_{\mu}, L_{\nu}]^2\}
 \end{equation}
 where $e$ is a dimensionless parameter. Here the notation $\lag_n$ means that the corresponding term contains $n$ first derivatives. $\lag_4$ is the only Poincare invariant four derivative term that leads to a positive hamiltonian which is quadratic in momenta. 

In the original version \cite{skyrme1961non, skyrme1962unified}, Skyrme only considered the model $\lag_{24} = \lag_2 +\lag_4$. The resulting  static solutions of the Skyrme field are maps from real space $\mathbbm{R}^3$ to the $SU(2)$ group manifold, which is $S^3$. Besides, in order to have finite energy solutions, the field must tend to the vacuum of the theory at spatial infinity. We take $U(x\rightarrow \infty) = \mathbbm{I}$, such that the infinity of $\mathbbm{R}^3$ is compactified into a point. Then the base manifold has the topology of the $S^3$, so the Skyrme field maps the $S^3$ onto itself and we may conclude from homotopy theory that the Skyrme model allows for the existence of topologically nontrivial configurations \cite{manton2004topological}. These solitonic solutions can be classified according to their different topologies via the topological number $B$, and the idea of Skyrme was to identify this integer number with the baryon number, and these topological solutions, which are called \textit{skyrmions}, with baryons \cite{ma2016lecture,zahed1986skyrme}. An explicit expression for $B$ in terms of the Skyrme field can be found from the topological current
\begin{equation}
    B^{\mu} = \frac{\epsilon^{\mu\nu\rho\sigma}}{24\pi^2}\Tr\{L_{\nu}L_{\rho}L_{\sigma}\}.
\end{equation}
It is straightforward to see that the divergence of the topological current vanishes identically, ($\dmu B^{\mu} = 0$), which implies the conservation of the topological number,
\begin{equation}
    B = \int d^3x B^0,
\end{equation}
which is an integer.
 Furthermore, the choice of a vacuum value for the Skyrme field at spatial infinite represents the \textit{spontaneous chiral symmetry breaking} of QCD. These analogies between the Skyrme model and QCD at low energies support the idea that the fundamental fields of the Skyrme model, $\pi_a$, may be identified with the physical pions.


In our approach  we will also add a term which is of sixth order in first derivatives,
\begin{equation}
\lag_6 = \lambda^2\pi^4B_{\mu}B^{\mu} \label{sextic}
\end{equation}
where $\lambda$ is a parameter. This term is, again, singled out as being the only Poincare invariant term of sixth order which leads to a standard hamiltonian quadratic in momenta. We will argue that this term is, in a certain sense, the most important one for our purposes. 
The generalized Skyrme model that we will consider is, therefore,
\begin{equation}
     \lag = \lag_2 + \lag_4 +\lag_6 + \lag_0 
        \label{Lagrangian}
\end{equation}
and we will refer to the individual terms as the \textit{quadratic}, \textit{quartic}, \textit{sextic} and \textit{potential} terms, respectively.

For the simplest model $\lag_{24}$, the collective coordinate quantization of the spin and isospin degrees of freedom of the $B=1$ skyrmion allowed to describe the proton, the neutron and some higher excitations (e.g., the delta resonance) and calculate some of their observables with a precision   of about 30\% \cite{adkins1983static}, as could be expected naively from the large $N_c$ arguments with an expansion parameter $N_c^{-1}$ for $N_c=3$.  A better precision, therefore, requires the inclusion of further terms and further (meson) fields into the effective model. The simplest Skyrme model $\lag_{24}$ also has some problems in the description of higher $B$ skyrmions which should correspond to atomic nuclei with weight number $A=B$. While the model is partially successful in describing some nuclear spectra in terms of spin and isospin excitations, one major problem is that the binding energies of  skyrmions of baryon charge $B$ 
against their decomposition into $B$ nucleons are up to ten times higher than the binding energies of the corresponding physical nuclei. Also, the resulting skyrmions for large $B$ are rather hollow structures, at variance with the quite constant baryon densities inside physical nuclei.  

The inclusion of the mass term for the pions, apart from reproducing the explicit chiral symmetry breaking, already improves some of these shortcomings. Indeed, this term induces, \textit{e.g.}, the $\alpha$-clustering for larger values of $B$ \cite{Battye:2006na}, which is a known property of some nuclei. As a consequence, for some nuclei which are known to have $\alpha$ particle subclusters, the Skyrme model with pion mass term already provides an excellent description of nuclear spectra
\cite{Lau:2014baa}, particularly when vibrational degrees of freedom are taken into account in addition to spin and isospin \cite{Halcrow:2016spb}. However, the binding energy problem remains. 

Finally, the sextic term was first considered in \cite{Jackson:1985yz}. In \cite{Adam:2010fg} it was shown that combined with a potential term it leads to a BPS model which reproduces many important features of physical nuclei, like small binding energies and the spherical and compact shapes \cite{Adam:2013wya}. This potential should be chosen different from the pion mass term (e.g. $\lag_{0,k} \simeq \Tr\{U - \mathbbm{I}\}^k$, $k\in \mathbbm{N}$), because otherwise the BPS submodel would correspond to the unphysical limit of infinite pion mass.
Based on additional BPS bounds for generalized Skyrme models discovered in \cite{Harland:2013rxa,Adam:2013tga}, it was later found that these additional potentials $\lag_{0,k}$ serve to reduce binding energies already on their own, both without \cite{Gillard:2015eia} and with \cite{Gudnason:2016tiz,Gudnason:2018jia} the sextic term.

A further improvement was achieved by the inclusion of the rho mesons \cite{Naya:2018kyi}, based on the instanton-inspired approach to vector meson coupling in the Skyrme model of Ref. \cite{Sutcliffe:2010et}. The cases $B\le 8$ were investigated numerically, and it was found that realistic cluster structures emerged for all skyrmions, even for substructures different from $\alpha$ particles. In addition, the binding energies were reduced significantly.
\\
The upshot of all this is that \; 
{\em i)} there has been significant progress in the last years in the Skyrme model as a model for nuclei and nuclear matter, where several shortcomings of the model have been improved significantly, and clear strategies for their resolution have been found; and
\;
{\em ii)} progress is still slower than one would like, not because of a fundamental problem of the model, but because calculations are hard. Already the starting point for any investigation, i.e., the skyrmion solution for a given $B$, is difficult to find, especially for the extended versions of the model where more terms and/or more fields are included.   A parameter scan of the extended models with the aim of fitting the parameters and coupling constants of a model to physical observables in order to identify promising regions in parameter space is a challenging numerical problem with current techniques.

We shall, therefore, use a simpler version of the Skyrme model and fit only those observables which are most relevant for nuclear matter at sufficiently high densities because, as we will see,  the most widely used configuration for Skyrme matter (the Skyrme crystal) allows to model nuclear matter only above nuclear saturation density\footnote{In section 2.4 we will discuss non-crystalline solutions consisting of regularly arranged clusters of skyrmions surrounded by empty space. These solutions have a much better behavior at low density, but we will not attempt to model nuclear matter below saturation with these configurations in the present paper.}. At such high densities potential terms like $\lag_{0,k}$ are irrelevant, as follows from simple scaling arguments, and may be ignored. For reasons of simplicity, we will also ignore vector mesons, although the adequacy of this approximation is more difficult to gauge\footnote{We shall, however, take into account the effect of kaon condensation in section 4, because a kaon condensate can be studied in the background of an unmodified Skyrme crystal in leading approximation.}. On the other hand, the sextic term \eqref{sextic} will provide the leading contribution to the Skyrme crystal energy at high density, as follows from scaling arguments, again.

\subsection{Skyrme matter}
The inclusion of the sextic term is, in fact, mandatory for any realistic modeling of nuclear matter by Skyrme matter at sufficiently high densities, and its inclusion should provide already a rather reasonable description there. First of all, this term precisely describes the repulsive nuclear force in the high density regime \cite{Adam:2015lra}. Secondly, the Skyrme model without the sextic term leads to a maximum neutron star mass of about $1.5 M_\odot$ 
\cite{Naya:2019rlm}, whereas NS of up to 2 solar masses are firmly established, and there are clear indications that the maximum possible NS mass may, in fact, be as high as $2.3-2.5 M_\odot$. Thirdly, a related fact is that the Skyrme model without the sextic term always leads to an EOS with a speed of sound which is below the conformal bound $c_s^2 < 1/3$ for all pressures. But EOS with such a restricted speed of sound are strongly disfavored according to a recent analysis \cite{Altiparmak:2022bke}.   

In principle, the ultimate goal of a sufficiently general Skyrme model would be to describe nuclear matter covering a large range of densities, from individual baryons to isolated nuclei where $n_B \sim n_0 = 0.16$ fm$^{-3}$, to the cores of neutron stars $n_B \sim 10 n_0$. 
We will find, however, that the currently available Skyrme matter solutions only allow to model nuclear matter above nuclear saturation density $n_0$. We will, therefore, restrict to the model \eqref{Lagrangian} which contains four parameters, the pion decay constant $\fpi$, the Skyrme parameter $e$, the pion mass $\mpi$, and $\lambda^2$. From these parameters, we fix the pion mass to its physical value, $\mpi = 140$ MeV, whereas we will use the other three to fit our solutions to reproduce the nuclear matter properties most relevant for our considerations. In particular, as is common practice in the Skyrme model, we will not fix the pion decay constant $f_\pi$ to its physical value. The idea is that in the still rather restricted version \eqref{Lagrangian} of the Skyrme model this modified value of $f_\pi$ partly takes into account the effect of neglected terms, whereas in more complete, general versions of the model the optimum value of $f_\pi$ should flow to its physical value.

We will consider static solutions of the Skyrme field and we adopt the usual Skyrme units of energy and length to work with more manageable expressions,
\begin{equation}
    E_s = \frac{3\pi^2 \fpi}{e} , \quad x_s = \frac{1}{\fpi e}.
\end{equation}
Then the static energy functional in these units becomes
\begin{align}
    \notag E = \frac{1}{24\pi^2}\int d^3x &\left[ -\frac{1}{2}\Tr\left\{L_i^2\right\} - \frac{1}{4}\Tr\left\{\left[L_i,L_j\right]^2\right\} + 8\lambda^2 \pi^4 f^2_{\pi}e^4 (\mathcal{B}^0)^2 + \left(\frac{\mpi}{\fpi e}\right)^2\Tr (\mathbbm{I} - U) \right] = \\[2mm]
    \frac{1}{24\pi^2}\int d^3x &\left[ (\partial_i n_A)^2 + \left( \partial_i n_A \partial_j n_B - \partial_i n_B\partial_j n_A \right)^2 + c_6\left( \epsilon_{ABCD}n_A\partial_1 n_B\partial_2 n_C \partial_3 n_D \right)^2 + c_0(1-\sigma) \right],
    \label{Energy_integral}
\end{align}
where we have defined $c_6 = 2\lambda^2\fpi^2 e^4$ and $c_0 = 2\mpi^2/(\fpi e)^2$. In the last expression, we have introduced the field configuration \cref{U_exp} and adopted the vectorial notation $n_A = (\sigma, \pi_a)$. Recall that $U \in SU(2)$ implies that $n_A$ is a unit vector, $n_An_A = 1$, with $A = 0,1,2,3$. It is also possible to find from \eqref{Energy_integral} a lower bound for the energy per baryon number, which is known as the BPS (Bogomol'nyi-Prasad-Sommerfield) bound \cite{Harland:2013rxa,Adam:2013tga}. For this choice of units the standard Skyrme model ($\lambda^2 = \mpi = 0$) becomes independent of the value of the parameters $\fpi$ and $e$, and its BPS bound is equal to 1.

Unfortunately, trying to numerically calculate the skyrmion solution for $B>>1$, corresponding to a macroscopic amount of nuclear matter, is not feasible. Some simplifying assumptions, therefore, must be made for a description of nuclear matter within the Skyrme model. The simplest and most widely used assumption is that skyrmionic matter forms a cubic lattice \cite{Klebanov:1985qi,Goldhaber:1987pb,KUGLER1988491,kugler1989skyrmion,Castillejo1989DENSESS,Baskerville:1996he,Perapechka:2017yyc,Adam:2021gbm}. That is to say, there exists a cubic unit cell with a baryon content $B_{\rm cell}$ and side length $l_{\rm cell}$ (i.e., volume $V_{\rm cell} = l_{\rm cell}^3$), such that the Skyrme field is periodic in all three Cartesian directions under the translation $x^i \to x^i + l_{\rm cell}$. It is then sufficient to minimize the energy functional within one unit cell, and the extension to an arbitrary amount of Skyrme matter is trivial. In addition to periodicity, usually some further discrete symmetries are assumed, leading to different cubic crystal types like, e.g., simple cubic (SC), face-centered cubic (FCC), or body-centered cubic (BCC), and the resulting solutions are known as Skyrme crystals.

It must be emphasized, however, that the mere existence of solutions for any of these crystal types by itself does not imply their physical relevance. Indeed, the minimization of the Skyrme model energy functional leads to an elliptic PDE, and elliptic systems typically always have a solution for sufficiently regular boundary conditions. The reason that some crystals are considered relevant is that the corresponding solutions are of rather low energy. More precisely, the crystal energy per unit cell as a function of the lattice length, 
$E_{\rm cell}(l_{\rm cell})$ is a convex function with a minimum for a certain $l_{0, {\rm cell}}$. For some crystals, the resulting energy at the minimum $l_{0, {\rm cell}}$ is only slightly above the topological energy bound. For the FCC crystal for the original Skyrme model, e.g.,  $E_{\rm cell}(l_{0,{\rm cell}})$ is only about 
$3.7 \%$ above the bound, which is very close, because it is well-known that the bound cannot be saturated.

For $l_{\rm cell} \not= l_{0, {\rm cell}}$, the energy grows rather quickly. Here, the region 
$l_{\rm cell} > l_{0, {\rm cell}}$ defines a thermodynamically unstable regime (formally, the pressure is negative). In this region, it is expected that equilibrium configurations of nuclear matter are, instead, given by larger clusters of nuclear matter (large nuclei, or a “nuclear pasta” phase in an intermediate region close to $l_{0, {\rm cell}}$) surrounded by almost empty space.
 We shall find some indications for the formation of larger nuclei in our investigations. This more inhomogeneous phase ameliorates the thermodynamical instability without resolving it completely. That is to say, the energy for $l_{ {\rm cell}}>l_{0, {\rm cell}}$ still is slightly above $E_{\rm cell}(l_{0,{\rm cell}})$, but the difference between $E_{\rm cell}(l_{0,{\rm cell}})$ and, say, $\lim_{l_{{\rm cell}}\to \infty}E_{\rm cell}(l_{{\rm cell}})$ can be made smaller than 5\%. 

The nuclear pasta phase between $l_{0, {\rm cell}}$ and the low density phase where large nuclei appear could, in principle, lead to a thermodynamically stable description for all densities.
The formation of nuclear pasta, however, is expected to result from a subtle interplay between the strong and the Coulomb forces. While the coupling of the Skyrme model to the electromagnetic field is known \cite{Callan:1983nx}, and promising results concerning the computation of Coulomb effect in the case of $\alpha$-particle skyrmions have been reported \cite{Ma:2019fvk}, a viable treatment of macroscopic amounts of skyrmionic matter coupled to electromagnetism which could give rise to the strong inhomogeneities implied by nuclear pasta is currently not available. As a consequence, all currently existing descriptions of nuclear matter based on Skyrme crystals approach zero pressure already at a finite baryon density $n_B=B_{\rm cell}/V_{\rm cell}$, corresponding to the nuclear saturation density $n_0=B_{\rm cell}/V_{{0,\rm cell}}$.
This implies that compact stars based on Skyrme crystals alone have no outer core and crust. On the other hand, the region $n_B \le n_0$ is well described by standard methods of nuclear physics and  can be joined with skyrmionic matter at higher densities. We shall make use of this possibility in several occasions. 

The region $l_{\rm cell} < l_{0, {\rm cell}}$ of the Skyrme crystal is thermodynamically stable, but this does not necessarily imply that it provides the true minimum energy configuration for all baryon densities $n_B>n_0$. However, other good candidates for these minimizers have not been found within the Skyrme model, therefore we shall simply {\em assume} that they are given by Skyrme crystals and work out the consequences of this assumption as far as possible.  
\\ \\
The paper is organized as follows:
In Section 2, we introduce different types of classical Skyrme crystal solutions. We discuss their properties and possible phase transitions between them in several versions of Skyrme models. We also present evidence for a high density crystal-fluid transition and a low density transition from a crystal to a non-homogeneous phase. This section is partly based on our previous work \cite{Adam:2021gbm}, but most of the results have been newly computed. Section 3 is devoted to the semiclassical quantization of the Skyrme crystals, which allows to pass from symmetric to asymmetric nuclear matter. Here we show how to compute the symmetry energy and particle (proton, neutron and lepton) fractions within the Skyrme theory. This part reviews the material recently published in \cite{Adam:2022aes}. Next, in Section 4, we take into account the strangeness degrees of freedom by extending the Skyrme field to a $SU(3)$-valued matrix field. In particular, we compute the kaon condensation in the semiclassical Skyrme crystal, briefly reviewing the very recent results of \cite{Adam:2022cbs}. In Section 5 we apply all the previously investigated crystal solutions to the description of neutron stars. Concretely, in section 5.2 we briefly review the construction of an EOS motivated by the generalized Skyrme model \cite{Adam:2020yfv}, which already leads to a very successful description of NS. In Section 5.3 we use the full semiclassical Skyrme crystal of the generalized model to describe both nuclear matter and NS, and we scan the model parameter space to find particularly promising parameter values. In Section 5.4, we include the effects of the kaon condensate on NS properties. Finally, Section 6 contains our conclusions and an outlook to further research. 


\section{Skyrme crystals}
Skyrmions have been extensively studied, and solutions for finite values of $B$ were found both in the standard Skyrme and the BPS submodels with different shapes and properties. The usual procedure to find a minimal energy solution considers the different possible symmetries for the skyrmion and then the solution is the one with the lowest energy. However, it becomes more difficult to find the minimal energy solutions for increasing $B$ since the number of possible configurations grows quickly \cite{Gudnason:2022jkn}. A simple estimation shows that the number of nucleons inside a NS must be of the order of $M_{\odot}/m_p \sim 10^{57}$, then obtaining a solution via the usual procedure becomes an impossible task.

Skyrme crystals are solutions obtained imposing periodic boundary conditions, then they are infinitely spatially extended solutions so they formally have infinite baryon number. For this reason, skyrmion crystals are good candidates to describe infinite nuclear matter and to reproduce the conditions inside NS. To construct these periodic solutions, we split the crystal in finite unit cells where we construct the skyrmion configuration, then the main difference to obtain the crystal with respect to the isolated skyrmions lies in the boundary conditions. Now Skyrme crystals compactify the real space into $\mathbbm{T}^3$, however since the $\mathbbm{T}^3$ is still an oriented and compact manifold, the Hopf's degree theorem ensures the existence of topological solitons labelled by an integer number.

From all the possible unit cells in three dimensions that we may use to construct a Skyrme crystal, we will consider cubic unit cells throughout this work, but we will allow for different symmetries within them. Additionally, since the crystal is infinitely extended it has infinite energy and baryon number, however the unit cell is finite in size, hence it carries a finite amount of energy and baryon number. Then the energy per unit cell as well as the energy per baryon number of the crystal are completely well defined and finite,
\begin{equation}
    \frac{E_{\rm crystal}}{B_{\rm crystal}} = \frac{N_{\rm{cell}s}E_{\rm{cell}}}{N_{\rm{cell}s}B_{\rm{cell}}} = \frac{E_{\rm{cell}}}{B_{\rm{cell}}}.
\end{equation}

The first Skyrme crystal was proposed in 1985 for the standard Skyrme model by Klebanov \cite{Klebanov:1985qi}, motivated by the phenomenological application of crystals to the interior of NS. He considered the simplest possible crystal with a simple cubic (SC) unit cell, in which eight $B = 1$ skyrmions were located in the corners of the cube in the maximal attractive channel with respect to their nearest neighbours. Then, he computed the minimal energy field configuration respecting these conditions for different values of the unit cell side length and found that the lowest value of the energy was just an $8\%$ above the BPS bound. In the following, we will explain how we construct Skyrme crystals via the procedure given in \cite{KUGLER1988491} with the different symmetries that have been proposed, and we will compare them within the generalized Skyrme model.

\subsection{Crystal ansatz}
The starting point in the construction of the Skyrme crystal proposed in \cite{KUGLER1988491} is the expansion of the fields in Fourier series,
\begin{equation}
    \begin{aligned}
    &\sigma = \sum^{\infty}_{a,b,c} \beta_{abc}\cos\left( \frac{a\pi x}{L} \right) \cos\left( \frac{b\pi y}{L} \right) \cos\left( \frac{c\pi z}{L} \right) \\[2mm]
    &\pi_1 = \sum^{\infty}_{h,k,l} \alpha_{hkl} \sin\left( \frac{h\pi x}{L} \right) \cos\left( \frac{k\pi y}{L} \right) \cos\left( \frac{l\pi z}{L} \right) \\[2mm]
    &\pi_2 = \sum^{\infty}_{h,k,l} \alpha_{hkl} \cos\left( \frac{l\pi x}{L} \right) \sin\left( \frac{h\pi y}{L} \right) \cos\left( \frac{k\pi z}{L} \right) \\[2mm]
    &\pi_3 = \sum^{\infty}_{h,k,l} \alpha_{hkl} \cos\left( \frac{k\pi x}{L} \right) \cos\left( \frac{l\pi y}{L} \right) \sin\left( \frac{h\pi z}{L} \right).
    \label{Field_ansatz}
    \end{aligned}
\end{equation}
Here, the unit cell extends from $-L$ to $L$ in all three Cartesian directions, so
the side length of the unit cell is $l=2L$. Then the symmetries of a given crystal impose some conditions on the Skyrme field which, in the end, are translated into some constraints on the coefficients of the expansions $\beta_{abc}$ and $\alpha_{hkl}$. Finally, the constrained coefficients are varied in order to obtain the lowest energy configuration. Although the expansion series of the fields are infinite, even the truncation to the first coefficients provides a good approximation to the solution, then the addition of higher modes produces corrections to the energy which become smaller for higher orders. This conclusion is also seen numerically, while the first coefficients are of order $\sim 1$, we have calculated that the next orders decay to the $\sim 4\%$, $\sim 0.3\%$ and $\sim 0.06 \%$ of the first order results. Hence we may safely truncate the series to a finite number of coefficients, we take around 30 coefficients to obtain the solution for each crystal. Finally, these expansions of the fields break the normalization condition of the vector $n_A$, then we have to renormalize it
\begin{equation}
    n_A \longrightarrow \frac{n_A}{\sqrt{n_An_A}}.
\end{equation}

The crystal considered by Klebanov has the simplest unit cell. It is invariant under cubic symmetry transformations:
\begin{align}
    \notag\text{A}_1&: (x,y,z) \rightarrow (-x,y,z), \\[2mm] &(\sigma,\pi_1,\pi_2,\pi_3) \rightarrow (\sigma,-\pi_1,\pi_2,\pi_3), \label{A1}\\[2mm]
    \notag\text{A}_2&: (x,y,z) \rightarrow (y,z,x), \\[2mm] &(\sigma,\pi_1,\pi_2,\pi_3) \rightarrow (\sigma,\pi_2,\pi_3,\pi_1),
    \label{A2}
\end{align}
and it has an additional periodicity symmetry on the side length of the unit cell,
\begin{align}
    \notag \text{A}_3&: (x,y,z) \rightarrow (x+L,y,z), \\[2mm] &(\sigma,\pi_1,\pi_2,\pi_3) \rightarrow (\sigma,-\pi_1,\pi_2,-\pi_3).
\end{align}
Indeed all the symmetries shown in this work are based on the cubic symmetry, hence they will all satisfy symmetries $A_1$ and $A_2$. The last symmetry, $A_3$, repeats the location of a skyrmion with period $L$ and performs the mutual isorotation between nearest neighbours.
Under these symmetries, the energy is periodic in $L$ but the fields are periodic in $2L$, then we take the ranges of the unit cell to be [$-L$, $L$] and perform the integrals of the energy and baryon numbers within these limits. 

We proceed now to show how to construct other symmetries of interest and show the numerical results.

\subsubsection{Face centered cubic crystal of skyrmions}
This symmetry was proposed in \cite{kugler1989skyrmion} in order to have a new solution with lower energy for very large values of $L$. It shares symmetries $A_1$ and $A_2$ and also two additional symmetries,
\begin{align}
    \notag \text{C}_3&: (x,y,z) \rightarrow (x,z,-y), \\[2mm] &(\sigma,\pi_1,\pi_2,\pi_3) \rightarrow (\sigma,-\pi_1,\pi_3,-\pi_2), \\[2mm]
    \notag\text{C}_4&: (x,y,z) \rightarrow (x+L,y+L,z), \\[2mm] &(\sigma,\pi_1,\pi_2,\pi_3) \rightarrow (\sigma,-\pi_1,-\pi_2,\pi_3).
\end{align}
In this case, the energy and baryon number are periodic in $2L$, and the unit cell has the shape on an FCC lattice of skyrmions. We have eight $B = 1$ skyrmions in the corners of the cube, symmetry $C_4$ locates other six skyrmions in centre of the faces and it also isorotates them by $\pi$ with respect to their nearest neighbours. Hence, this lattice differs from the first in that each skyrmion is surrounded by 12 nearest neighbours all of them in the maximal attractive channel. Since we have the eight skyrmions in the corners and other six in the faces of the cube, the total baryon number in this unit cell is $B_{\rm{cell}} = 4$.

\begin{figure}[h!]
    \centering
    \includegraphics[scale=0.3]{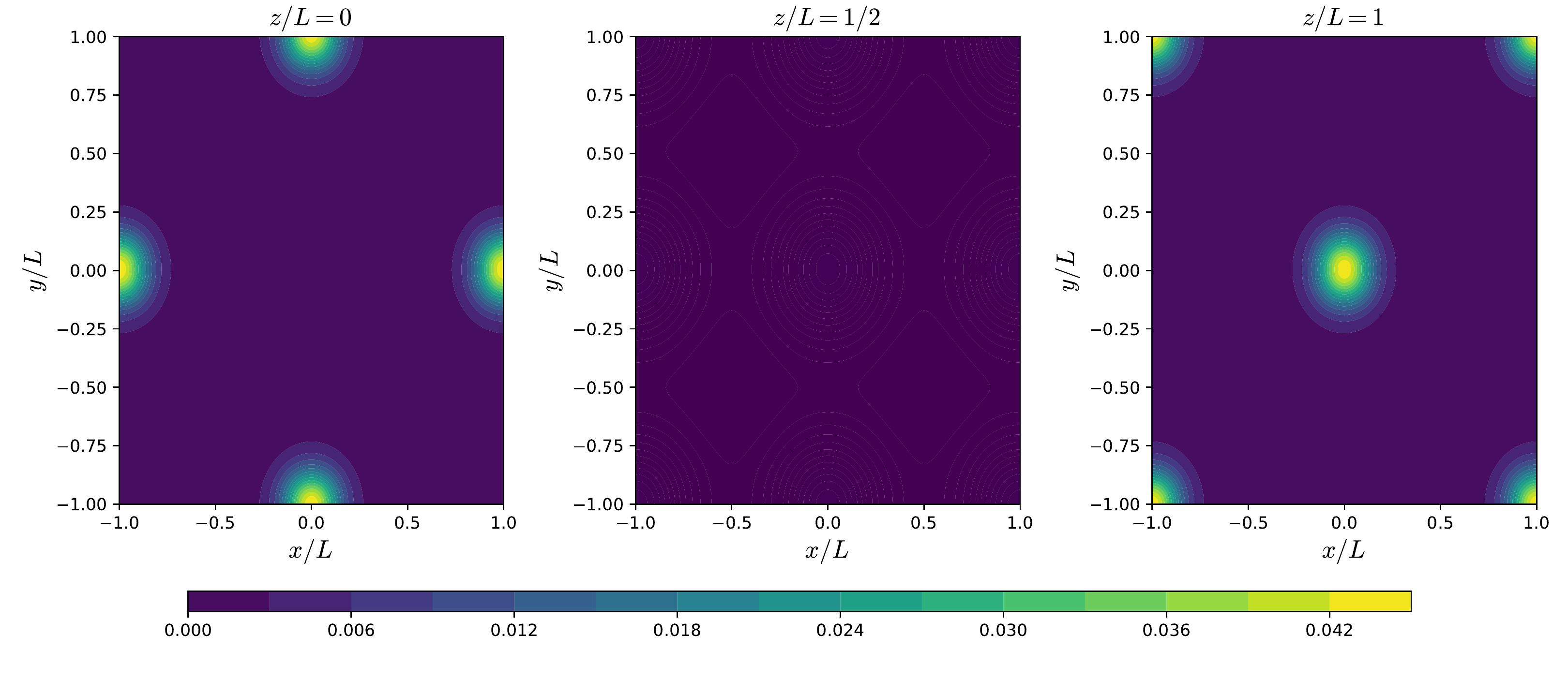}
    \caption{Energy density contour plots of the face-centered cubic (FCC) unit cell of skyrmions for a large value of $L = 10$. Each plot shows different heights of the unit cell, which are $z=0$, $L/2$ and $L$ because in this case also the energy density has periodicity in $2L$.}
    \label{Figure.C}
\end{figure}

\begin{figure}[h!]
    \centering
    \includegraphics[scale=0.4]{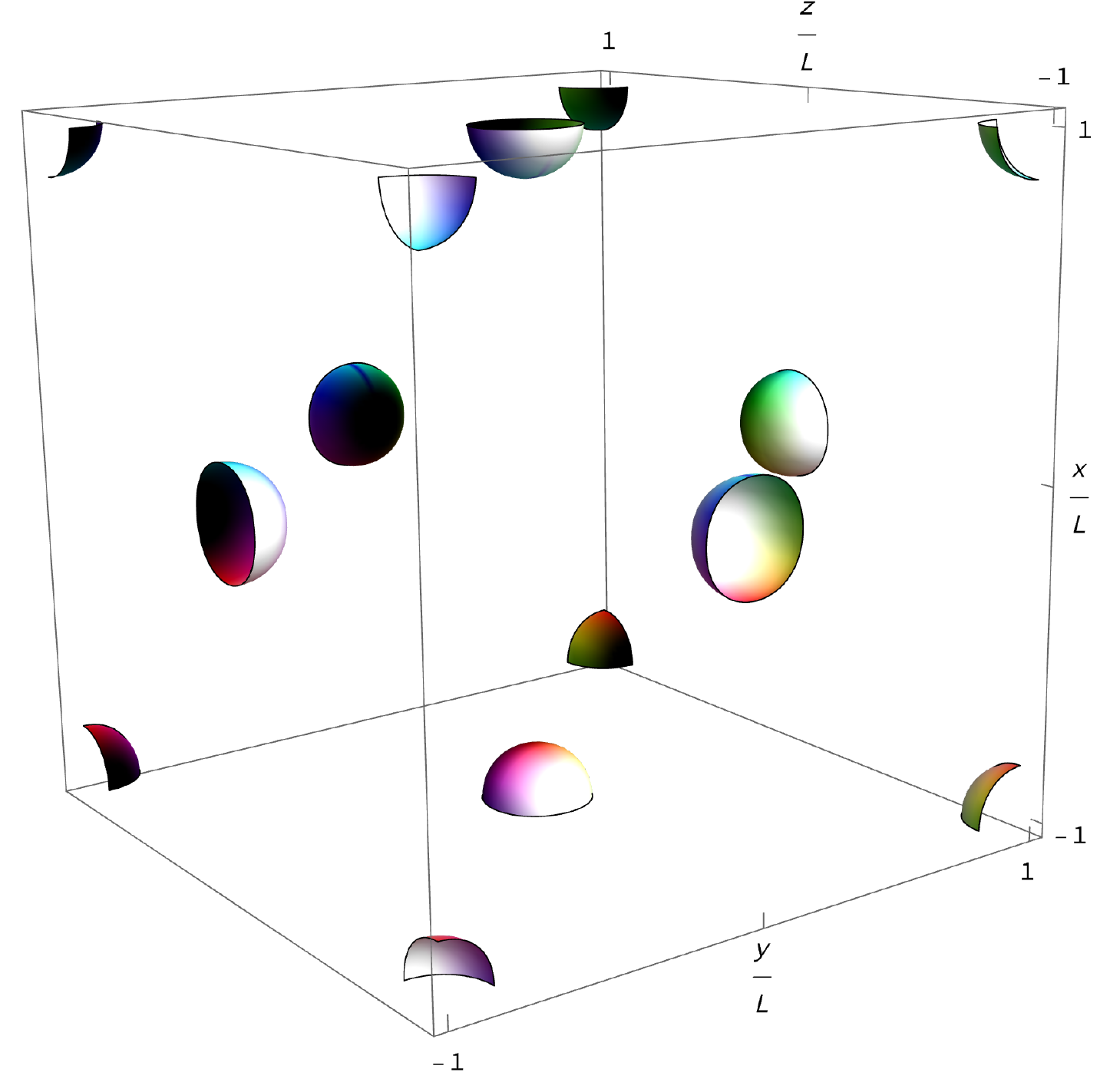}
    \caption{The energy density of the FCC unit cell is shown in 3D. The FCC symmetry is clearly visible as well as in the contours \cref{Figure.C} plot. We adopt the Runge colouring convention \cite{Feist:2012ps} in this figure to represent the pion fields.}
    \label{Figure.FC3D}
\end{figure}

As we mentioned before, these symmetries impose some constraints on the Fourier coefficients. They can be easily obtained imposing the symmetries on the field ansätze \cref{Field_ansatz}. In this case the non-vanishing coefficients are obtained from the combination of the following restrictions,
\begin{itemize}
    \item $h$ is odd, $k$ and $l$ are even \hspace{3mm} or \hspace{3mm} $h$ is even, $k$ and $l$ are odd,    
    \item $a$, $b$, $c$ are all odd \hspace{3mm} or \hspace{3mm} $a$, $b$, $c$ are all even.
\end{itemize}
We show the resulting energy density contours in \cref{Figure.C}, and the three-dimensional energy density plot in \cref{Figure.FC3D}. The crystal symmetry is clearly visible in the plots. 

\subsubsection{Body centered cubic crystal of half-skyrmions}
This unit cell was proposed in \cite{Goldhaber:1987pb} to be the one with the lowest energy for small values of $L$. It introduces an additional symmetry,
\begin{align}
    \notag\text{B}_4&: (x,y,z) \rightarrow (L/2-z,L/2-y,L/2-x), \\[2mm] &(\sigma,\pi_1,\pi_2,\pi_3) \rightarrow (-\sigma,\pi_2,\pi_1,\pi_3).
\end{align}
to those of the Klebanov crystal ($A_1, A_2$ and $A_3$). The motivation of this new crystal comes from the appearance of an additional symmetry when two $B = 1$ skyrmions are brought together and form the lower energy $B = 2$ field configuration, in which the $B = 1$ skyrmions have lost their individual identity. This new symmetry produces a unit cell which may be seen as a BCC of half-skyrmions, which are solutions for which $\sigma = -1$ at the centre of the cube of side length $L$ until some radius $r_0$ for which $\sigma = 0$. This solution carries a half of baryon charge and it is undefined outside $r_0$. However, a new half-skyrmion solution can be defined via the transformation $(\sigma, \pi_a) \rightarrow (-\sigma, -\pi_a)$, these new solutions are located in the corners of the cube, connected to the $\sigma = 0$ value at $r = r_0$ of the central half-skyrmion, forming a cube of side length $L$. As a result, the mean value of the $\sigma$ field in this cube is exactly 0, so the energy coming from the potential term $\lag_0 \sim (\sigma - 1)$ will scale exactly as $8c_0 L^3$. Further, the 8 half-skyrmions in the corners contribute a total baryon number of $1/2$, so the cube of side length $L$ contains a baryon charge of $B=1$.

\begin{figure}[h!]
    \centering
    \includegraphics[scale=0.3]{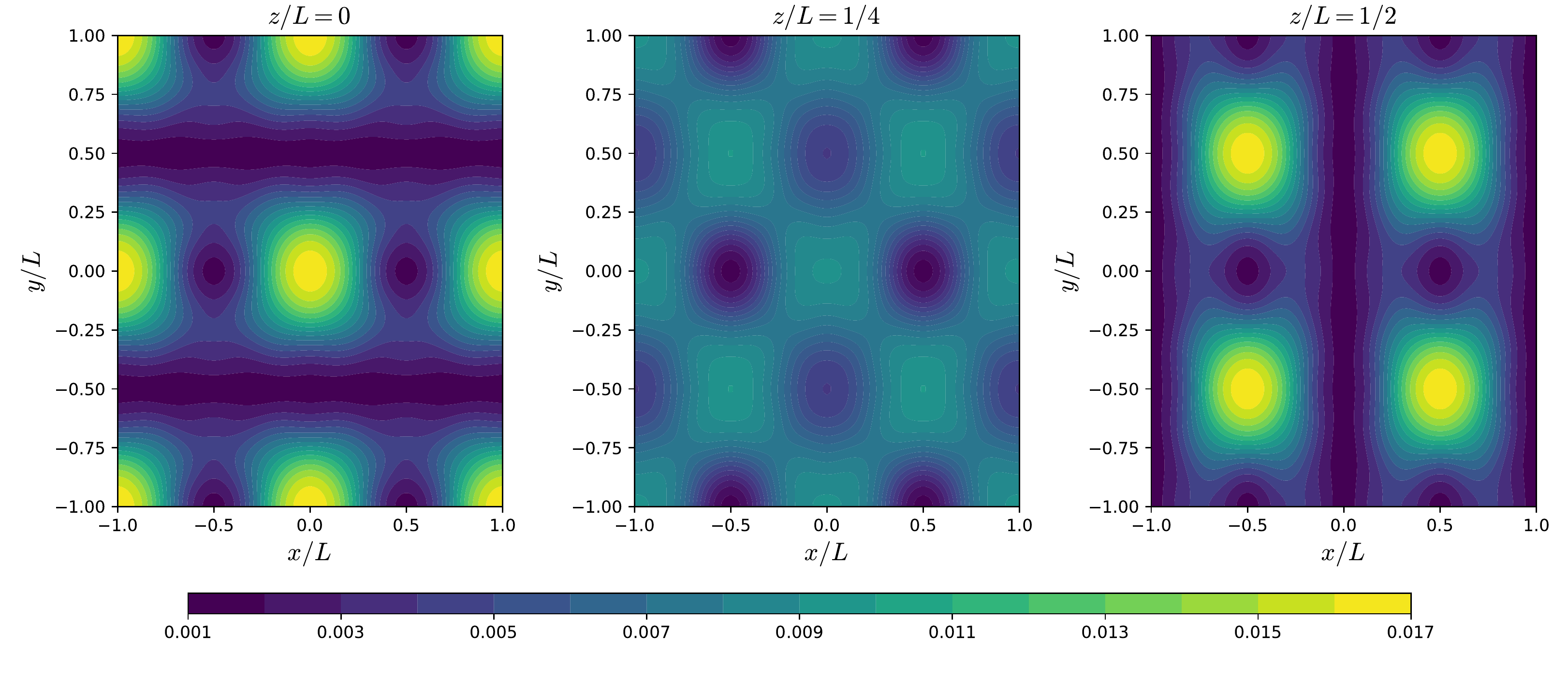}
    \caption{Energy density contour plots for the unit cell of the body-centered-cubic (BCC) crystal at the minimum of energy. In this case, the energy density has period $L$ so we show the cuts at $z = 0$, $z = L/4$ and $z = L/2$.}
    \label{Figure.B}
\end{figure}

\begin{figure}[h!]
    \centering
    \includegraphics[scale=0.4]{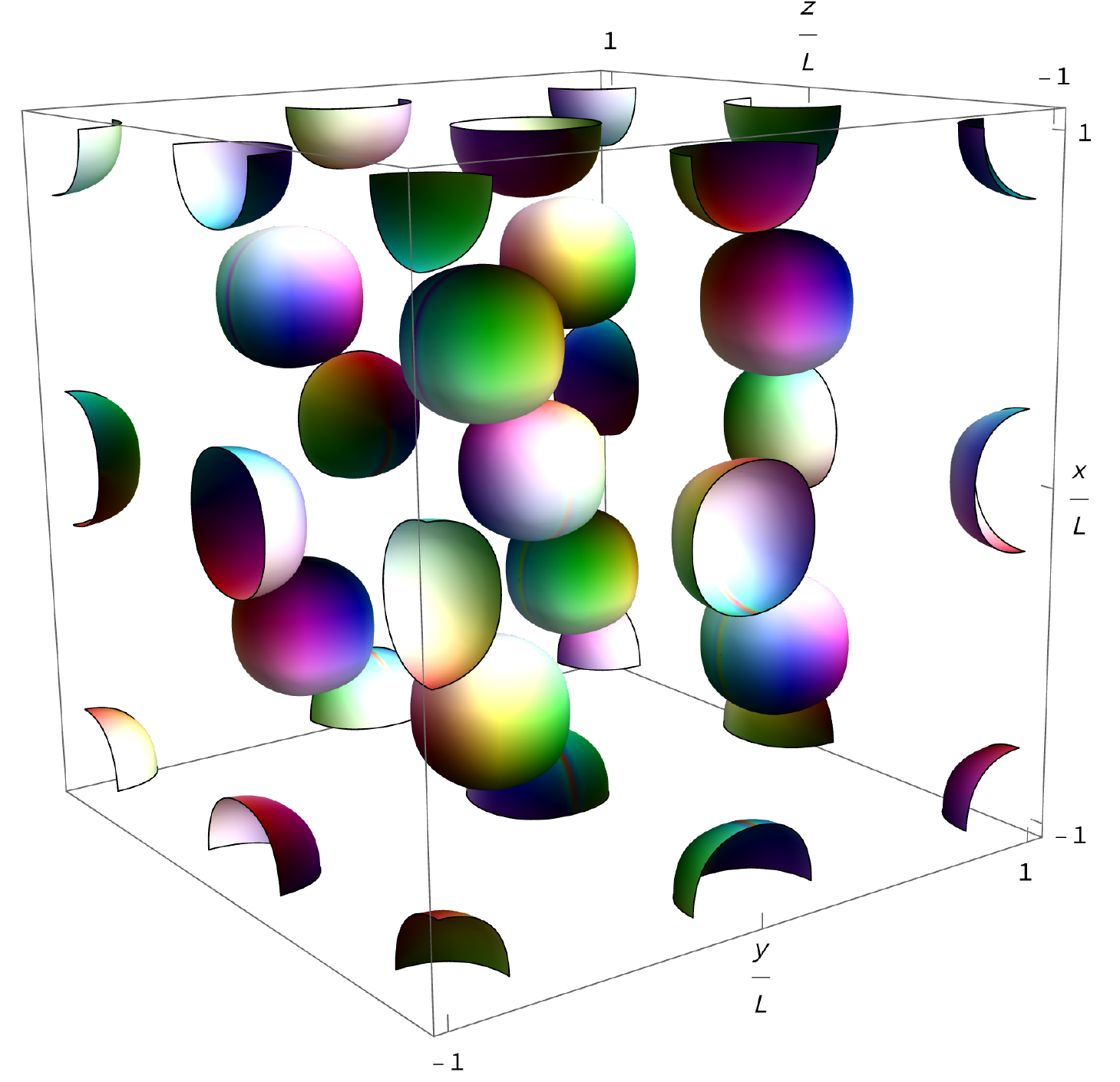}
    \caption{The tridimensional energy density plot of the BCC unit cell is shown. Again we represent the pion fields using the Runge colouring convention.}
    \label{Figure.B3D}
\end{figure}

The energy and baryons densities are also periodic in $L$ but, again, the fields have a $2L$ periodicity, then we have $B_{\rm{cell}} = 8$ within our unit cell of side length $2L$. The restrictions imposed on the Fourier coefficients by the last symmetries are
\begin{itemize}
    \item $h$, $k$ are odd, $l$ is even.
    \item $a$, $b$ and $c$ are even.
    \item $\beta_{abc} = \beta_{bca} = \beta_{cab}$.
    \item $\alpha_{hkl} = -(-1)^{\frac{h+k+l}{2}}\alpha_{khl}$.
    \item $\beta_{abc} = -(-1)^{\frac{a+b+c}{2}}\beta_{bac}$.
\end{itemize}
We show the resulting energy density plots in \cref{Figure.B} and \cref{Figure.B3D}. 

\subsubsection{Enhanced face centered cubic crystal of skyrmions}
This new crystal configuration was almost simultaneously found in two different publications \cite{KUGLER1988491, Castillejo1989DENSESS} to be the one with the lowest energy in the standard Skyrme model. It may be seen as the half-skyrmion version of the FCC crystal explained before. Indeed it shares the symmetries $A_1, A_2, C_3$ and an additional symmetry,
\begin{align}
    \notag\text{D}_4&: (x,y,z) \rightarrow (x+L,y,z), \\[2mm] &(\sigma,\pi_1,\pi_2,\pi_3) \rightarrow (-\sigma,-\pi_1,\pi_2,\pi_3),
\end{align}
then some of the FCC crystal Fourier coefficients are set to 0 in this crystal.

Concretely, this new unit cell only allows the Fourier coefficients which satisfy the conditions
\begin{itemize}
    \item $h$ is odd, $k$ and $l$ are even,    
    \item $a$, $b$, $c$ are all odd.
\end{itemize}

\begin{figure}[h!]
    \centering
    \includegraphics[scale=0.3]{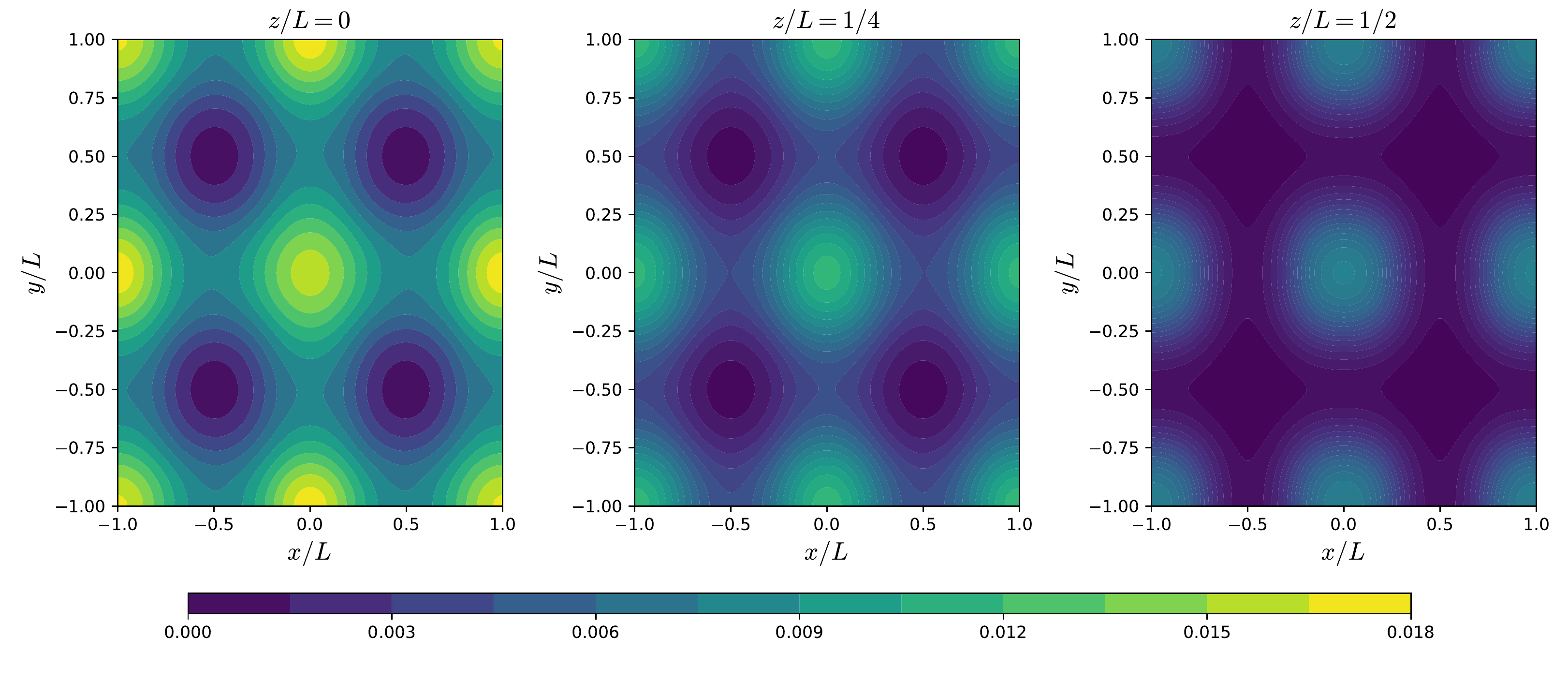}
    \caption{Energy density contour plots for the unit cell of the enhanced face-centered cubic crystal (FCC$_{+}$) at the minimum of energy. It can be appreciated in the figure that this crystal can also be viewed as a simple cubic crystal of half-skyrmions. The energy density has the same period as the BCC, so the cuts are the same.}
    \label{Figure.F}
\end{figure}

\begin{figure}[h!]
    \centering
    \includegraphics[scale=0.4]{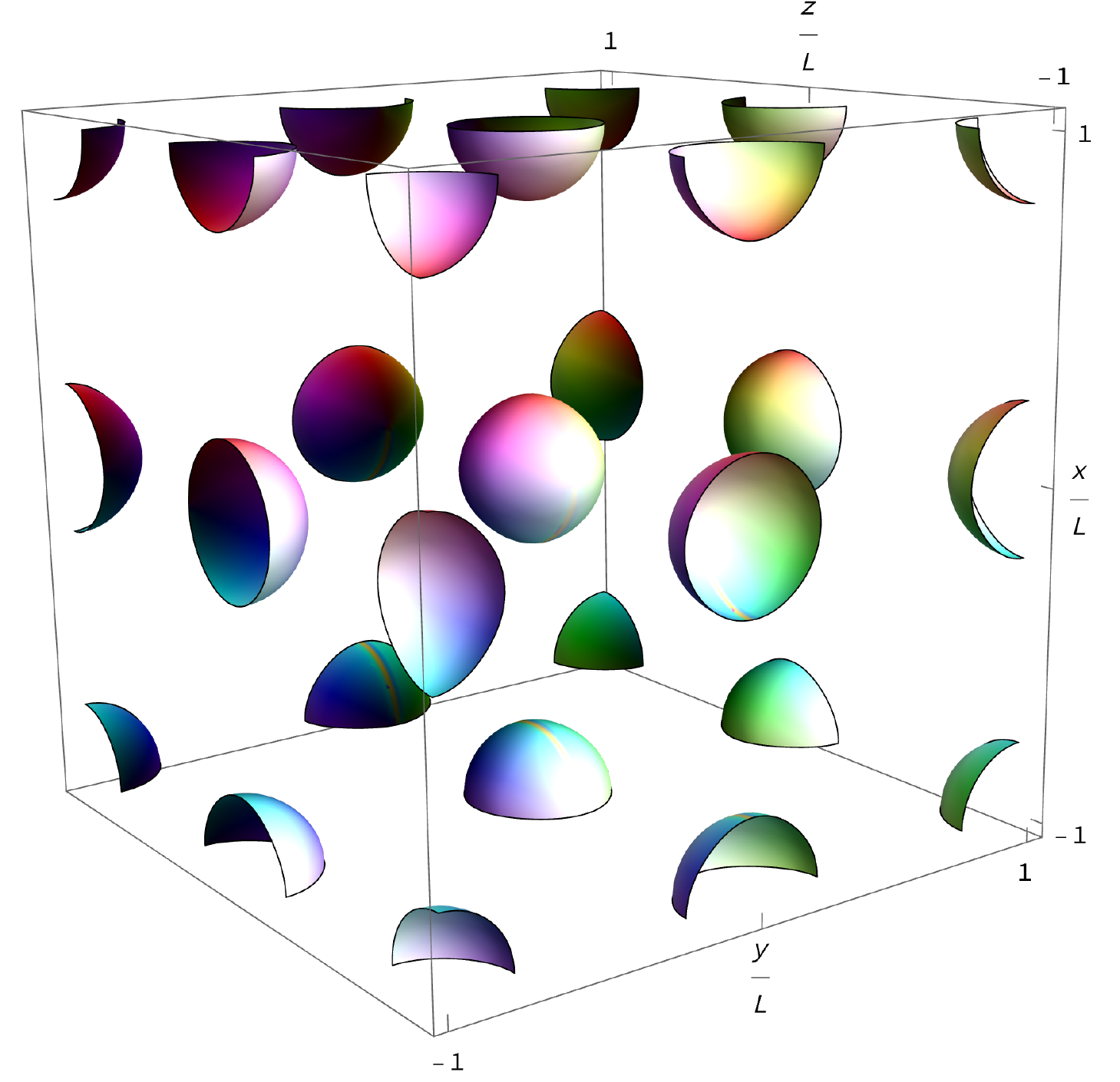}
    \caption{The energy density in three dimensions for the FCC$_+$ crystal. Again the Runge colouring convention is used as in the last two cases.}
    \label{Figure.F3D}
\end{figure}

Since this crystal has less free Fourier coefficients than the FCC crystal, it is a particular case of the last one which we shall call FCC$_+$. As a result, the FCC$_{+}$ crystal will always have equal or larger energy than the FCC crystal. This may lead to phase transitions between the crystals at some length of the unit cell, as we will see later.
As in the FCC crystal, the half-skyrmion solutions with $\sigma = -1$ in their centre are located at the corners and faces of the unit cell. Further, the opposite half-skyrmions with 
$\sigma = 1$ occupy the body center and the link centers of the unit cube.
As a consequence, the mean value of the $\sigma$ field is 0, again, as in the BCC crystal.
The energy and baryon densities are periodic in $L$ and they have the appearance of a simple cubic unit cell of half-skyrmions. However, since the fields are periodic in $2L$ we take that to be the side length of the unit cell, hence the unit cell still has the shape of an FCC crystal with the alternating half-skyrmion solutions. Then, the baryon content within our unit cell is again $B_{\rm{cell}} = 4$.
We show the resulting energy density plots in \cref{Figure.F} and in \cref{Figure.F3D}.

\subsection{Energy curves}
Each crystal configuration explained before is constructed imposing the corresponding symmetry transformations, then we fix the value of $L$ and use a Nelder-Mead algorithm \cite{10.1093/comjnl/7.4.308} from the GSL C++ library to find the optimal values of the remaining free Fourier coefficients that minimize the energy functional. For different values of $L$ we may construct the curve $E(L)$ for all the different crystals and combinations of terms in the lagrangian \cref{Lagrangian}. For all cases we always find a convex curve with a minimum located at some value of $L$ which is different in each case.

It will be important for the next sections to obtain an analytical expression of the energy curve. Then we may try to guess a specific fitting curve just by studying the scaling of each term in the energy functional. We find that $E_i \sim L^{3-i}$, where $i$ represents the number of derivatives in the $i-$term, and the three comes from the integration over $3$D space. Hence we use the following fit for the energy curves in each case,
\begin{equation}
    E(L) = k + k_2 L + \frac{k_4}{L} + c_6\frac{k_6}{L^3} + c_0 k_0 L^3.
    \label{Fit}
\end{equation}

In order to show the main properties of Skyrme crystals and the impact that the different terms have in the $E(L)$ curve we fix the value of the physical constants that appear in the generalized Skyrme model to some standard values,
\begin{equation}
    \fpi = 129\: \text{MeV}, \quad e = 5.45, \quad \lambda^2 = 5\: \text{MeV fm}^3, \quad \mpi = 140\: \rm{MeV}.
    \label{Param_vals}
\end{equation}

\begin{figure}[h!]
    \centering
    \includegraphics[scale=0.55]{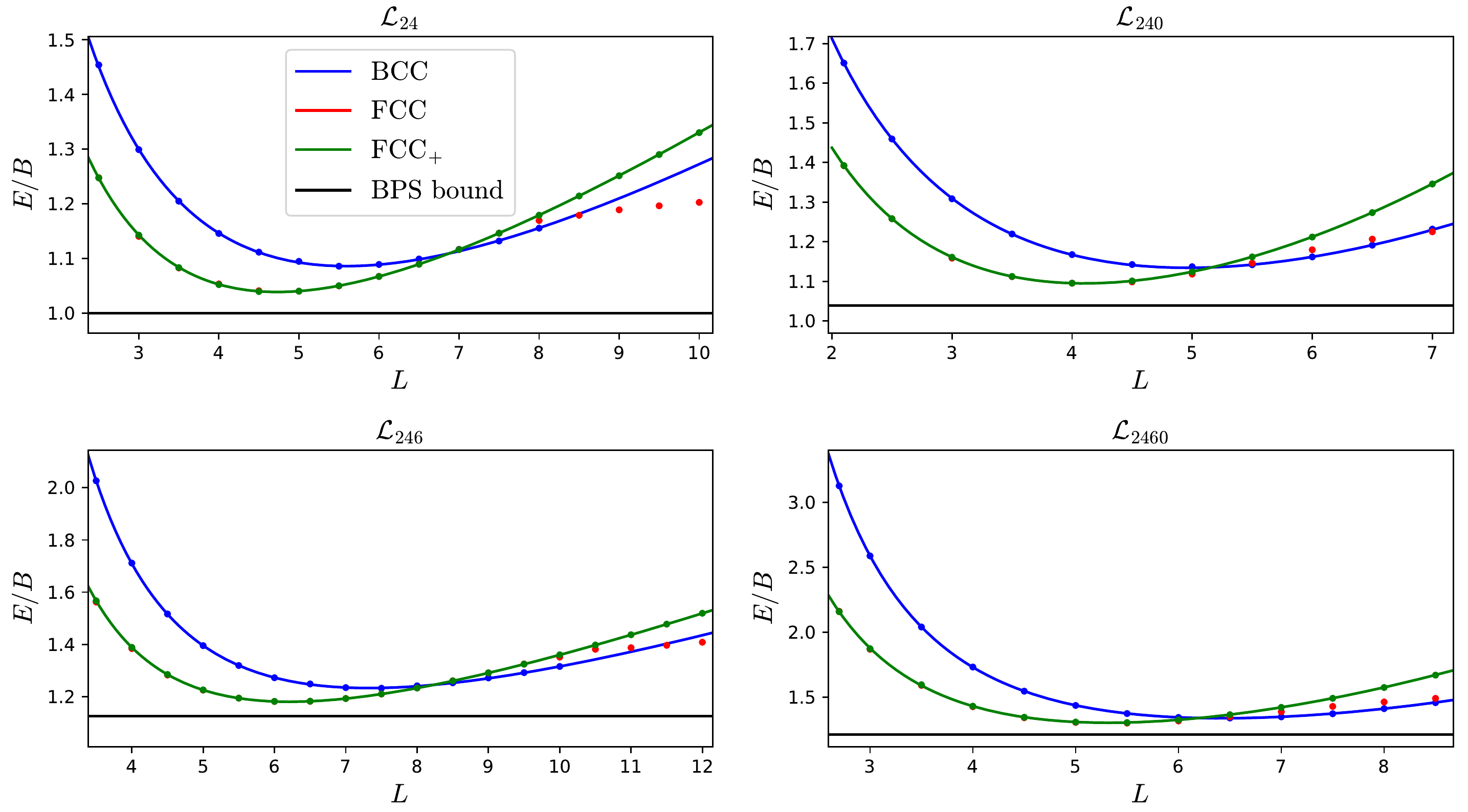}
    \caption{ Energy curves of the different Skyrme crystals considering the different terms in the generalized Skyrme lagrangian. We also show the BPS bound in each case to see how close it is to the minima of the crystals.}
    \label{Figure.E_L}
\end{figure}

We may see from \cref{Figure.E_L} how the $E(L)$ curves change turning on and off the different terms. The pion-mass potential is an attractive term, so we would expect that more compact solutions are preferable. On the other hand the sextic term is repulsive so the opposite effect is expected. These behaviors are visible in \cref{Figure.E_L} where the length parameter value at which the minimal energy configuration is achieved, denoted by $L_0$, is shifted to smaller values when the potential term is introduced, whereas it increases when the sextic term is present. Larger values of $\mpi$ or $\lambda^2$ will increase these effects, and values of the parameters different from \cref{Param_vals} lead to completely different results. However the Skyrme units, in which we are computing the curves, produce a universal $E(L)$ curve in the $\lag_{24}$ case in the sense that it does not depend on the value of the parameters. We also note that the expression \cref{Fit} that we are using to fit the energy of the crystals is quite accurate for the FCC$_{+}$ and BCC cases, however the FCC crystal has a non-trivial behaviour for large values of $L$, invalidating the expression of our fit in this region.

For comparison, we also include the topological lower energy bounds in each case in \cref{Figure.E_L}. For the $\lag_{24}$ case, this bound coincides with the Skyrme-Faddeev bound originally given already in \cite{skyrme1962unified}, whereas more stringent bounds can be derived once more terms are included in the lagrangian \cite{Harland:2013rxa,Adam:2013tga}. Here, we always use the most stringent bound. 

As in previous work \cite{Adam:2021gbm,Perapechka:2017yyc}, we conclude that the FCC$_{+}$ crystal reaches the lowest energy in the standard Skyrme model between the crystals considered here. In that case, the minimum is only a $3.7\%$ above the BPS bound, which also places the Skyrme crystal as the skyrmion solution with the lowest energy ever achieved in the standard Skyrme model. Obviously, when the other terms are included the energy increases, but also the BPS bound of the model which is shown in each plot. We also conclude that for the parameters that we consider here, the lowest energy configuration is also achieved for the FCC$_+$ crystal in the generalized model. The location and the value of the energy in the minimum, $L_0$ and $E_0$ respectively, for this case, as well as the value that it is above the BPS bound in percentage, are shown in  tables 1 and 2 below,
\begin{table}[!htb]
    \begin{minipage}{.5\linewidth}
      \centering
        \begin{tabular}{|c|c|c|c|}
			\hline
			Model & $L_0$ & $E_0/B$ & Minimum ($\%$) \\ \hline
			$\lag_{24}$ & 4.7 & 1.04 & 4 \\ \hline
			$\lag_{240}$ & 4.1 & 1.09 & 5 \\ \hline
			$\lag_{246}$ & 6.2 & 1.18 & 5 \\ \hline
			$\lag_{2460}$ & 5.3 & 1.30 & 7 \\ \hline
		\end{tabular}
        \caption{FCC$_{+}$ crystal.}
    \end{minipage}%
    \begin{minipage}{.5\linewidth}
      \centering
        \begin{tabular}{|c|c|c|c|}
			\hline
			Model & $L_0$ & $E_0/B$ & Minimum ($\%$) \\ \hline
			$\lag_{24}$ & 5.5 & 1.08 & 8 \\ \hline
			$\lag_{240}$ & 4.9 & 1.13 & 9 \\ \hline
			$\lag_{246}$ & 7.3 & 1.23 & 9 \\ \hline
			$\lag_{2460}$ & 6.4 & 1.34 & 10 \\ \hline
		\end{tabular}
        \caption{BCC crystal}
    \end{minipage}
    \caption*{ The side length and the value at which the $E(L)$ finds the minimum. In the last column we show $[(E_{\rm min} - E_{\rm bound})/E_{\rm bound}] \times 100$, i.e., the percentage deviation of the minimum crystal energy from the bound for the lattice.}
\end{table}

The values of the coefficients in \cref{Fit} are obtained in each case using the Python optimization library GEKKO.
\begin{table}[!htb]
      \centering
        \begin{tabular}{|c|c|c|c|c|c|}
			\hline
			Model & $k$ & $k_2$ & $k_4$ & $k_6$ & $k_0$ \\ \hline
			$\lag_{24}$ & 0.047 & 0.105 & 2.344 & 0 & 0 \\ \hline
			$\lag_{240}$ & 0.022 & 0.109 & 2.384 & 0 & 0.008 \\ \hline
			$\lag_{246}$ & 0.006 & 0.106 & 2.750 & 0.905 & 0 \\ \hline
			$\lag_{2460}$ & 0.334 & 0.074 & 1.747 & 1.062 & 0.010 \\ \hline
		\end{tabular}
        \caption{Fitting constants for the numerically obtained $E(L)$ curves for the FCC$_{+}$ crystal.}
  \end{table}
  \begin{table}
  \centering
        \begin{tabular}{|c|c|c|c|c|c|}
			\hline
			Model & $k$ & $k_2$ & $k_4$ & $k_6$ & $k_0$ \\ \hline
			$\lag_{24}$ & 0.017 & 0.096 & 2.988 & 0 & 0 \\ \hline
			$\lag_{240}$ & -0.022 & 0.101 & 3.061 & 0 & 0.004 \\ \hline
			$\lag_{246}$ & 0.040 & 0.093 & 3.150 & 1.710 & 0 \\ \hline
			$\lag_{2460}$ & 0.172 & 0.078 & 2.798 & 1.751 & 0.005 \\ \hline
		\end{tabular}
        \caption{Fitting constants for the numerically obtained $E(L)$ curves for the BCC crystal.}
\end{table}

We remark that the fitting constants shown in tables 3 and 4 will change if we use different values of $c_6$ and $c_0$. However it will be shown in the next section that a value for the constants of the energy curve fit independent of the parameters may be obtained for the FCC$_+$ crystal solution.

\subsection{Phase transitions}
We may anticipate from \cref{Figure.E_L} that even though the FCC$_{+}$ crystal reaches the lowest energy at the minimum, it may not be the crystal with the lowest energy for all values of $L$. This is clear in the region with large values of $L$, for which the FCC crystal has lower energy than the FCC$_{+}$. We will also see that there is a phase transition from the FCC to the BCC crystal at small values of $L$, however since they do not have the same baryon content within the unit cell, a more careful comparison is necessary. We will study in the following the possible phase transitions that we may have since it may lead to an interesting phenomenology of the Skyrme crystals. For simplicity, since $L$ is a measure of the size of unit cell it is also a measure of the baryon density, then we will also refer to the region of small values of $L$ as the high density regime and for large values of $L$ the low density regime.

\subsubsection{Low density phase transition}
As we noted in the construction of the crystals, the FCC$_{+}$ crystal will always have larger or equal energy than the FCC. In \cref{Figure.E_L} we see that the energies of both crystals are indistinguishable at high densities, but at some value of the length the curves split and the FCC crystal becomes the ground state. This behaviour in the $E(L)$ curves suggests a phase transition between the crystals, but the presence of the pion-mass potential term is crucial in the understanding of this possible transition. Concretely, this potential term explicitly breaks chiral invariance, then it is not compatible with the FCC$_+$ symmetry $\sigma \rightarrow -\sigma$. However, the relevance of the potential term in the energy decreases at high densities, so both crystals tend to the same energy in the chiral limit. Hence when this term is not included in the lagrangian both crystals are allowed and we find a FCC to FCC$_+$ second order phase transition, but when the potential term is present the FCC crystal is always the ground state and the energy curves approach asymptotically.

This phase transition has been extensively studied in \cite{Lee:2003aq}, where the $\sigma$ field was proposed to be the order parameter of the transition.
\begin{figure}[h!]
    \centering
    \includegraphics[scale=0.5]{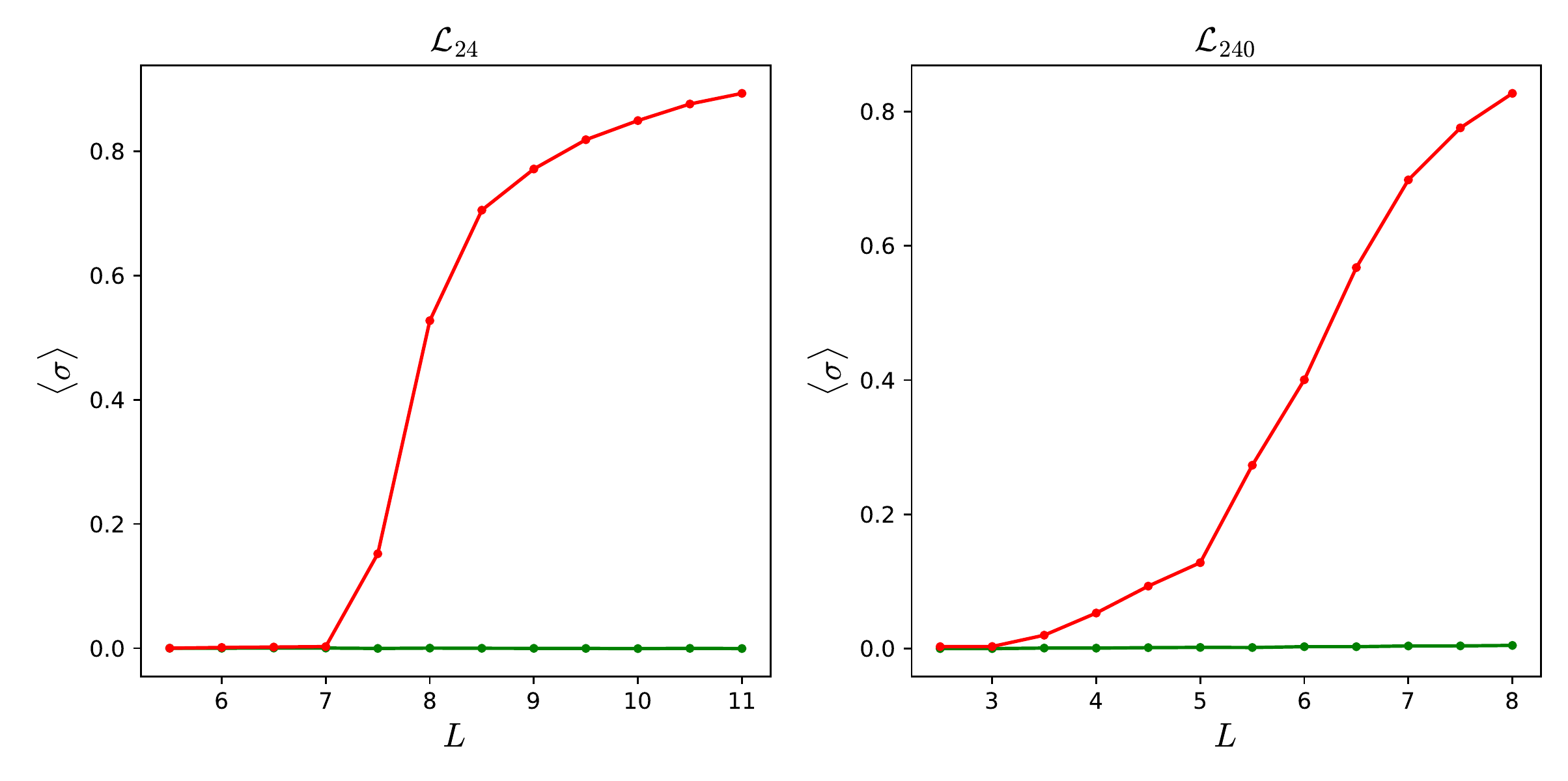}
    \caption{ The mean value of the $\sigma$ field within the unit cell shows the second order transition from the FCC to the FCC$_+$ crystal when the pion-mass term is not included, and the asymptotic approach when it is included.}
    \label{Figure.SigmaMV}
\end{figure}

We show in \cref{Figure.SigmaMV} the mean value of the $\sigma$ field in the unit cell for the cases $\lag_{24}$ and $\lag_{240}$ since they represent the cases without and with pion-mass potential term respectively. The addition of the sextic term does not qualitatively change the curves.

Although the FCC$_+$ is not the ground state crystal it is a good approximation to the FCC crystal at large densities. Indeed, for the values of the parameters that we have considered, the transition point (in the case without potential term) always occurs at densities smaller than the minimum of energy, and even with potential term the FCC$_+$ crystal is already a good approximation to the FCC crystal.

\subsubsection{High density phase transition}
Now we want to compare the energy curves between the BCC and the FCC crystals. An important point here is that whilst the FCC unit cell contains 4 baryons, the BCC unit cell has 8 baryon units. If we want to compare both crystals we need to do it at the same baryon density, which may be easily defined,
\begin{equation}
    n_B = \frac{B_{\rm{cell}}}{V_{\rm{cell}}}.
    \label{density}
\end{equation}
Hence, if we want to compare the energies we may calculate the density of both crystals and find the point at which the BCC crystal is more energetically favourable than the FCC crystal.

\begin{figure}[h!]
    \centering
    \includegraphics[scale=0.45]{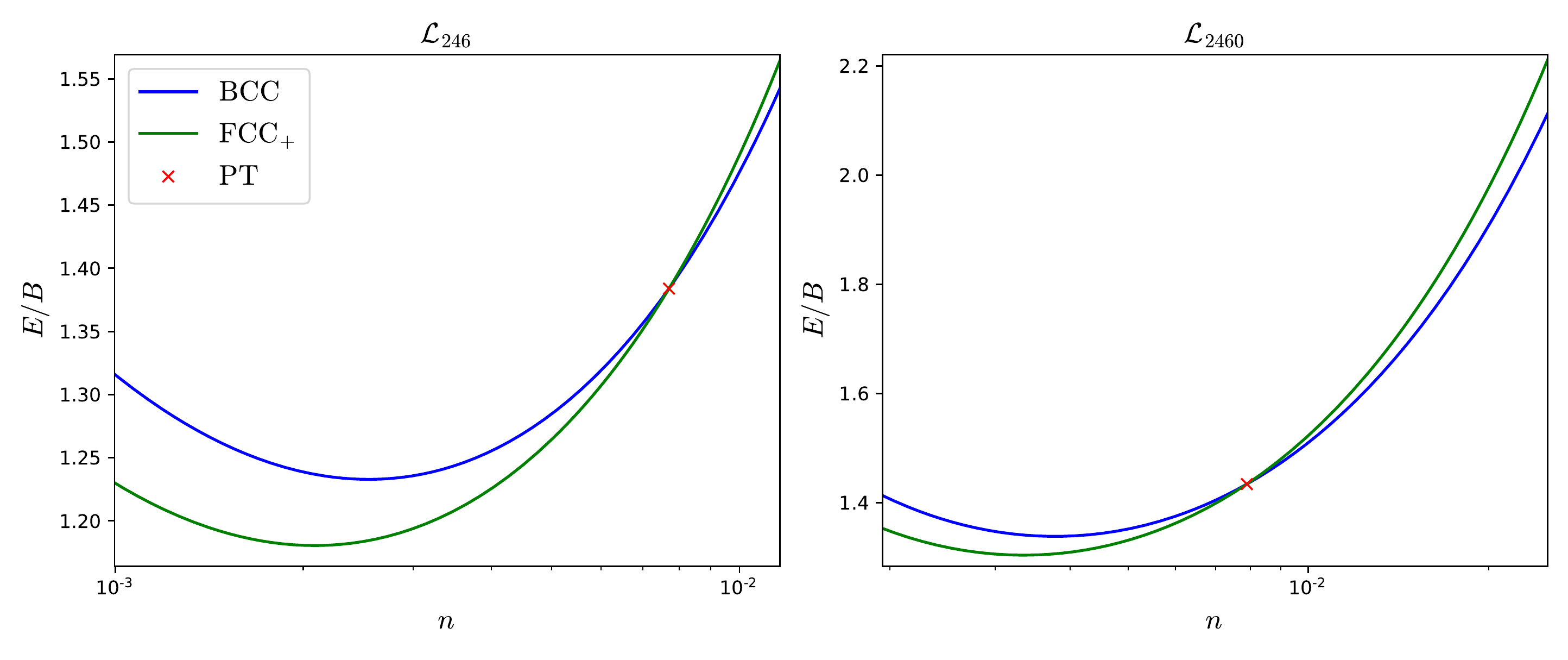}
    \caption{ The comparison between the energies of the BCC and FCC crystals at the same baryon density shows that the first one becomes more favourable at some point, denoted by the red cross in the plots. The sextic term is crucial to have this transition at realistic densities. Both $n$ and $E$ are shown in Skyrme units.} 
    \label{Figure.Evsn}
\end{figure}

We find that the different terms that we consider in the lagrangian have an important impact on the transition point. Concretely, the sextic term locates the transition at physically reasonable densities, \textit{i.e.} the same order of magnitude as the density at the energy minimum. Without the sextic term we find the transition point at very high densities, and the addition of the pion-mass term shifts the transition density to even higher values, therefore we only plot the cases in which we have the sextic term, see \cref{Figure.Evsn}.

\begin{table}[h!]
	\centering
		\begin{tabular}{|c|c|}
            \hline
            Model & $n_{\rm PT}/n_0$ \\ \hline
            $\lag_{246}$ & 3.7 \\ \hline
            $\lag_{2460}$ & 2.4 \\ \hline
        \end{tabular}
        \caption{Ratio between the transition density and the density at which the minimum of the energy is achieved for the FCC$_{+}$ crystal.}
\end{table}

Since the energy curves have different slopes at the transition point there is a discontinuity in the derivative of the energy. This implies that the FCC-to-BCC is a first order phase transition and we must perform a Maxwell construction (MC) in order to avoid unphysical regions.

The pressure of a system acquires great relevance when phase transitions are present since it must remain finite and continuous in order to have a physical transition. The pressure, as well as the energy density of the crystal, can be obtained from their thermodynamical definition,
\begin{align}
    p &= -\frac{\partial E}{\partial V} = -\frac{1}{24L^2}\frac{\partial E_{\rm cell}}{\partial L},\\[2mm]
    \rho &= \frac{E}{V} = \frac{E_{\rm cell}}{8L^3}.
    \label{press_enden}
\end{align}
From these expressions we may conclude that there is a discontinuity in the pressure of the crystal and this is contradiction with the Gibbs conditions that must be preserved in every phase transition,
\begin{equation}
    p_1 = p_2, \quad \mu_B^{1} = \mu_B^2.
\end{equation}
For this analysis we will identify the FCC crystal as phase 1 and the BCC crystal as phase 2. Besides, in our system the baryon charge is conserved so we must find the mixed phase which has the associated chemical potential ($\mu_B$) common to both phases.

The MC introduces a mixed phase which preserves the Gibbs conditions in this case. The main idea of this construction is to find one point in each of the energy curves which have the same pressure, we denote it by $p_{PT}$, and join them with the curve which has the same value of $\mu_B$ for the two phases. Mathematically this means that we have to find the points $(V_1, V_2)$ of each phase that have the same slope in the $E(V)$ diagram and are both tangent to the straight line with the same slope (which is $p_{PT}$).

We must be careful for this calculation since we are dealing now with the volumes of the unit cells. This means that same volumes have different baryon content in each crystal, so we need to rescale them in order to have the same baryon number,
\begin{equation}
    n_{\rm FCC} = n_{\rm BCC} \longrightarrow V_{\rm FCC} = \frac{V_{\rm BCC}}{2}.
\end{equation}

The final energy curve with a physical phase transition starts at low densities in the FCC crystal until we find the mixed phase which joins to the BCC crystal,
\begin{equation}
    E(V) = \left\{
        \begin{array}{ll}
          E_{\rm FCC}(V), & \quad V\geq V_1 \\
          E_{\rm FCC}(V_1) - p_{PT}(V - V_1), \quad &V_1 \geq V \geq V_2 \\
          E_{\rm BCC}(V), & \quad V \leq V_2
    \end{array}
    \right.
\end{equation}
\begin{figure}[h!]
    \centering
    \includegraphics[scale=0.4]{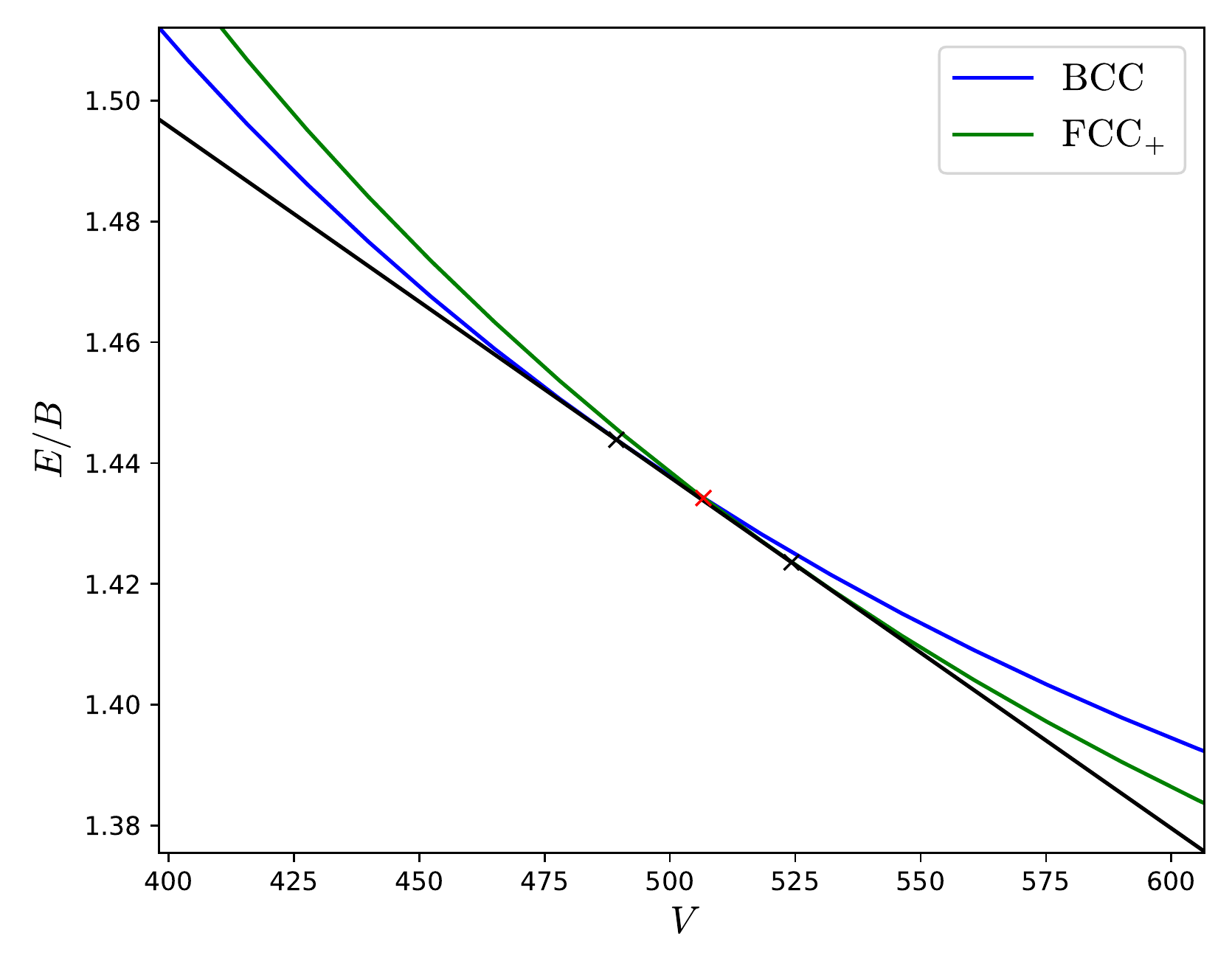}
    \caption{Maxwell construction in the FCC-to-BCC phase transition. The red cross denotes again the transition point, but now the black crosses denote where the mixed phase starts. At low densities we are in the FCC phase (green line) until we reach the mixed phase (black line), which joins to the BCC phase (blue line).}
    \label{Figure.Maxwell_Cons}
\end{figure}
see \cref{Figure.Maxwell_Cons}.
We may use the FCC$_+$ energy curve for these calculations since the density at which the transition occurs the FCC and FCC$_+$ crystals are the same in the $\lag_{246}$ case and the difference between them in the $\lag_{2460}$ case is negligible.

\subsubsection{Fluid-like transition}
From the previous calculations we know how to calculate the most important thermodynamical magnitudes (pressure, energy and baryon densities) for the Skyrme crystal. We also know that decreasing the size of the unit cell increases the densities as well as the pressure, hence we may try to calculate how homogeneous the crystal becomes for increasing density and if a transition to a fluid is possible.

Indeed, it is known that the BPS Skyrme submodel ($\lag_{60}$) describes a perfect fluid due to the properties of the sextic term. Then it seems reasonable that, since the sextic term is the most important one at small values of $L$, the crystal becomes more homogeneous within the unit cell. To study the degree of homogeneity we will compare the energy density profiles obtained from the numerical minimization with a constant density profile with the value of the mean energy density of the unit cell, $\rho_{\rm mean} = E_{\rm cell}/V_{\rm cell}$.

\begin{figure}[h!]
    \centering
    \includegraphics[scale=0.4]{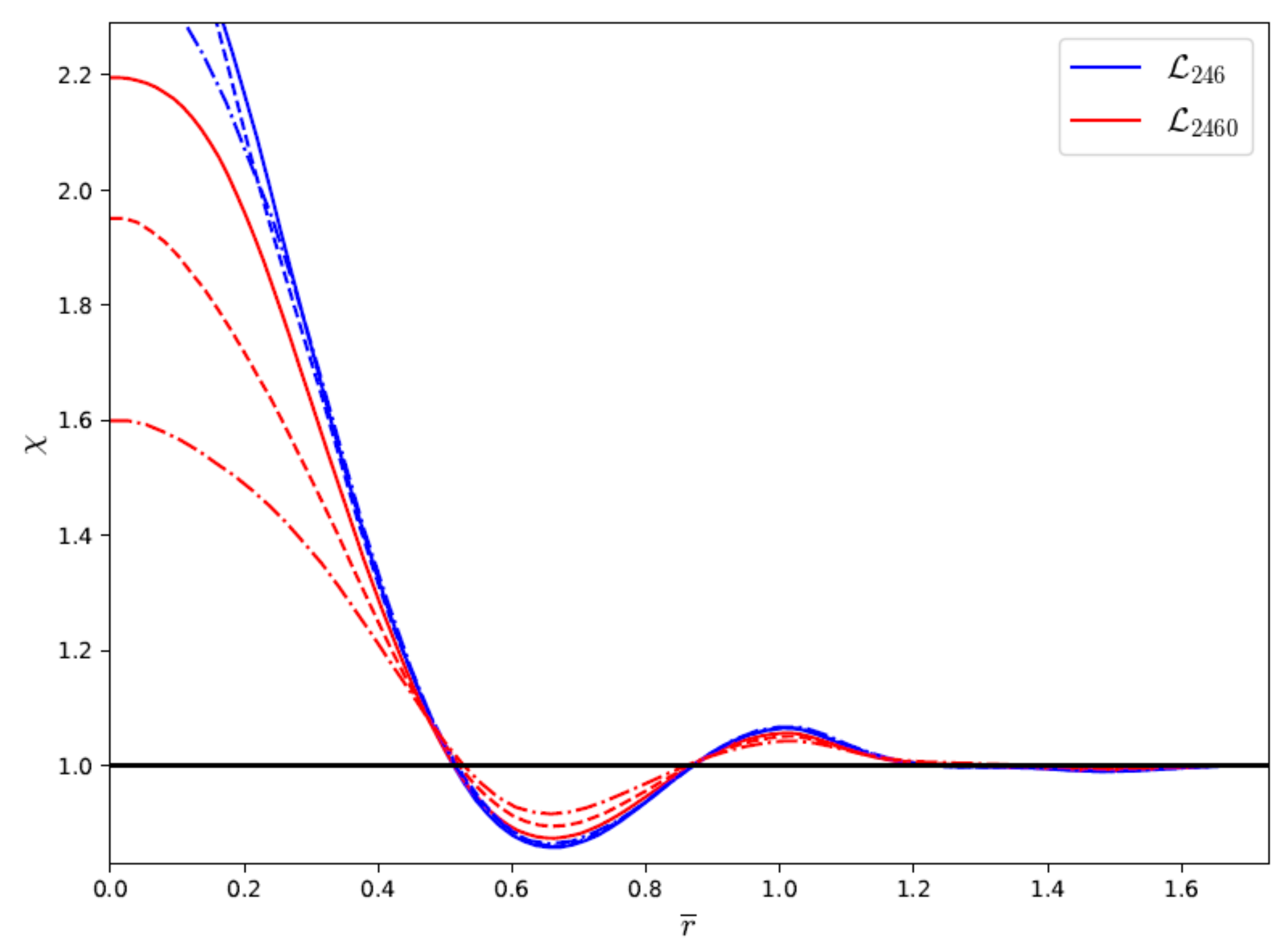}
    \caption{REP for the $\lag_{240}$ (blue) and $\lag_{2460}$ (red) cases, for $n_0$ (continuous lines), $3n_0$ (dashed lines), and $7n_0$ (dashed-dotted lines). The degree of homogeneity almost does not change for $\lag_{240}$, whereas the energy density becomes much more homogeneous with increasing baryon density for $\lag_{2460}$.}
    \label{Figure.Fluid_Plot}
\end{figure}

At the minimum of the energy we expect to have a highly inhomogeneous crystal, where the skyrmions are surrounded by regions of vacuum, and reducing the size of the unit cell will decrease the inhomogeneity. However, we also expect to increase this effect with the addition of the sextic term, so we will compare the $\lag_{240}$ and $\lag_{2460}$ cases, since we want to consider the more realistic cases in which pions have mass. To compare the energy densities within the unit cell we define the radial energy profile (REP) enclosed within a sphere of radius $r$,
\begin{equation}
    E(r) = \int_0^r d^3x\: \rho,
\end{equation}
where $\rho$ is the $\rho_{\rm mean}$ in the case of the constant energy density unit cell and the integrand of \cref{Energy_integral} in the real case. We calculate both REP for each case at the baryon density of the minimal energy ($n_0$), and at the higher densities $3n_0$ and $7n_0$. We show in \cref{Figure.Fluid_Plot} the ratio $\chi =E(r)/E_{\rm mean} (r)$ of the two REP for each case.

\subsection{New lattice solutions}
For the moment, we have mostly focused on the behaviour of Skyrme crystals in the region to the left of the minimum of the $E(L)$ curve. The reason is that, from \cref{press_enden} we may observe that the minimum corresponds to the point $p = 0$, and the region $L \geq L_0$ has negative pressure, hence it is unstable. It is the aim of this section to show that there is a new branch of solutions which have different energies in the low density regime and tend to the FCC$_+$ crystal at high energies.

The fact that the Skyrme crystal has a minimum is not a bad behaviour, since this is expected to occur in symmetric nuclear matter. However, the energy of the crystals seems to diverge with $L$, but this is due to the Fourier expansion that we use to construct the Skyrme crystal, in which we impose the skyrmions to be in fixed positions and we do not allow them to move freely within the unit cell to find the lowest energy configuration. This is a correct procedure for small values of $L$, however if we increase the size of the unit cell the skyrmion can only spread instead of clustering to form a compact configuration surrounded by vacuum.

This motivated us to find new lower energy configurations with a new numerical minimization method which lets the skyrmions move freely within the unit cell. We use a gradient flow method to find the field configurations with minimal energy, locating a $B = 4$ skyrmion in the centre of the unit cell. The motivation for this new lattice starts with the similarities between the isolated $B = 4$ skyrmion, which has cubic symmetry, and the FCC$_{+}$ symmetry. Indeed the $B = 4$ skyrmion is quite similar to the crystal in the sense that it is composed by eight half-skyrmions located in the corners of a cube.

Besides, the study of the $B = 4$ skyrmion in periodic boundary conditions under different deformations showed the phase transitions it may experiment \cite{SilvaLobo:2010acs}. Concretely, the phase transition between the FCC$_{+}$ and the $B = 4$ skyrmion lattice was found at a certain value of $L$, then the new lattice becomes a more energetically favourable crystal. Since the isolated $B = 4$ skyrmion aims to describe an alpha particle, we will refer to this configuration as the $\alpha$-lattice. 

We calculated the energy for the $\alpha$-lattice at different values of $L$ for the $\lag_{240}$ and $\lag_{2460}$ models. This new lattice has lower energy than the crystal at a certain value of $L$ and, furthermore, it tends to a constant value at $L \rightarrow \infty$. Indeed it tends to the value of the isolated $B = 4$ skyrmion, so we may construct other cubic lattices with larger values of $B$, since it is know that
the energy per baryon of skyrmions decreases for increasing values of the baryon charge. We show the energy of the next simplest cubic lattice, which is a multiple of the $\alpha-$lattice, the $B = 32$ lattice.
\begin{figure}[h!]
    \centering
    \includegraphics[scale=0.5]{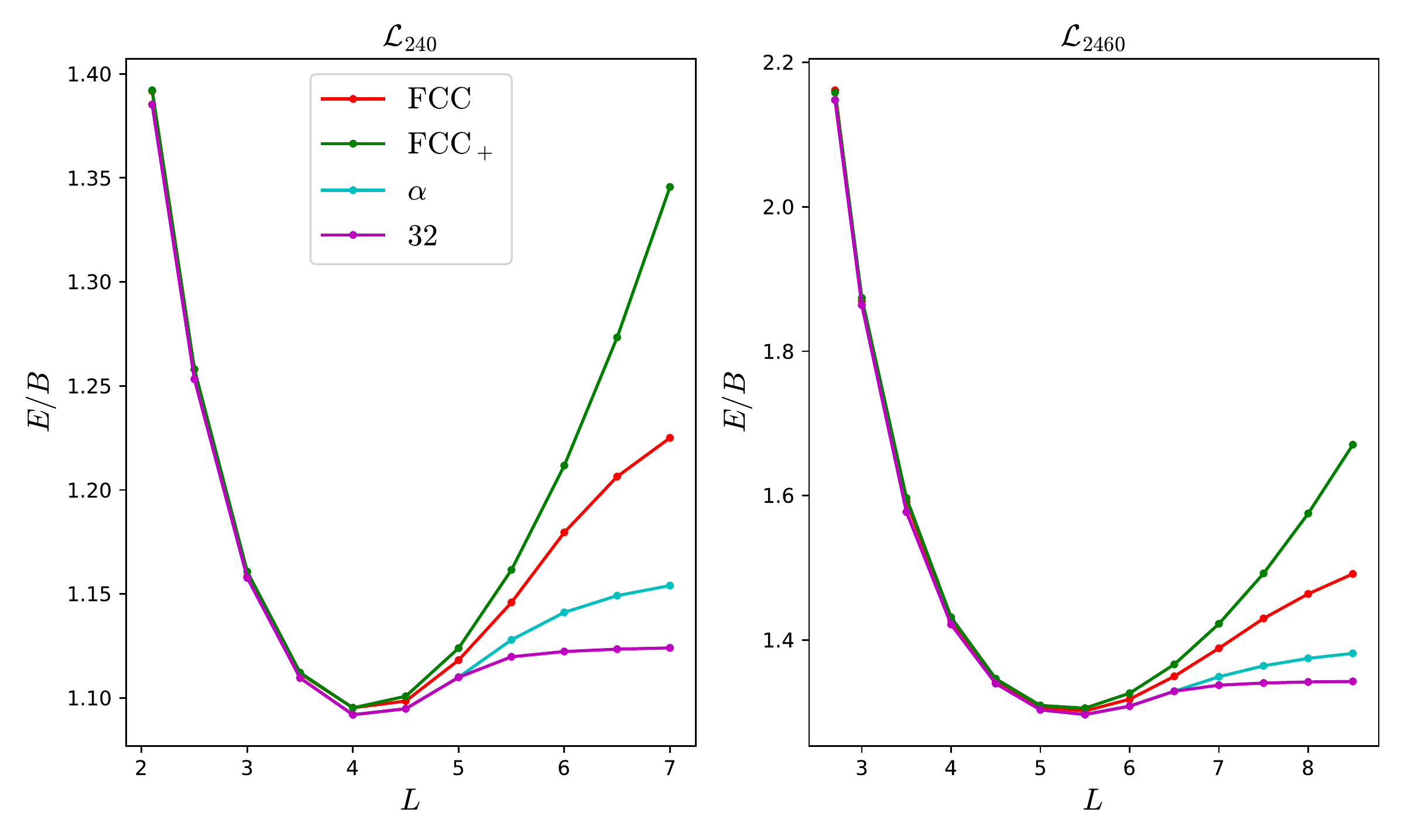}
    \caption{Energy per baryon number for the different lattices that we consider with and without sextic term. These $\alpha$ and $B=32$ lattices are important in the low density regime where they have much lower energies than the standard crystal. However, already for densities which are slightly smaller than at the minimum of the energy all these lattices tend to the same FCC crystal.}
    \label{Figure.Lattice}
\end{figure}

\begin{figure}[!htb]
   \begin{minipage}{0.48\textwidth}
     \centering
     \includegraphics[scale=0.33]{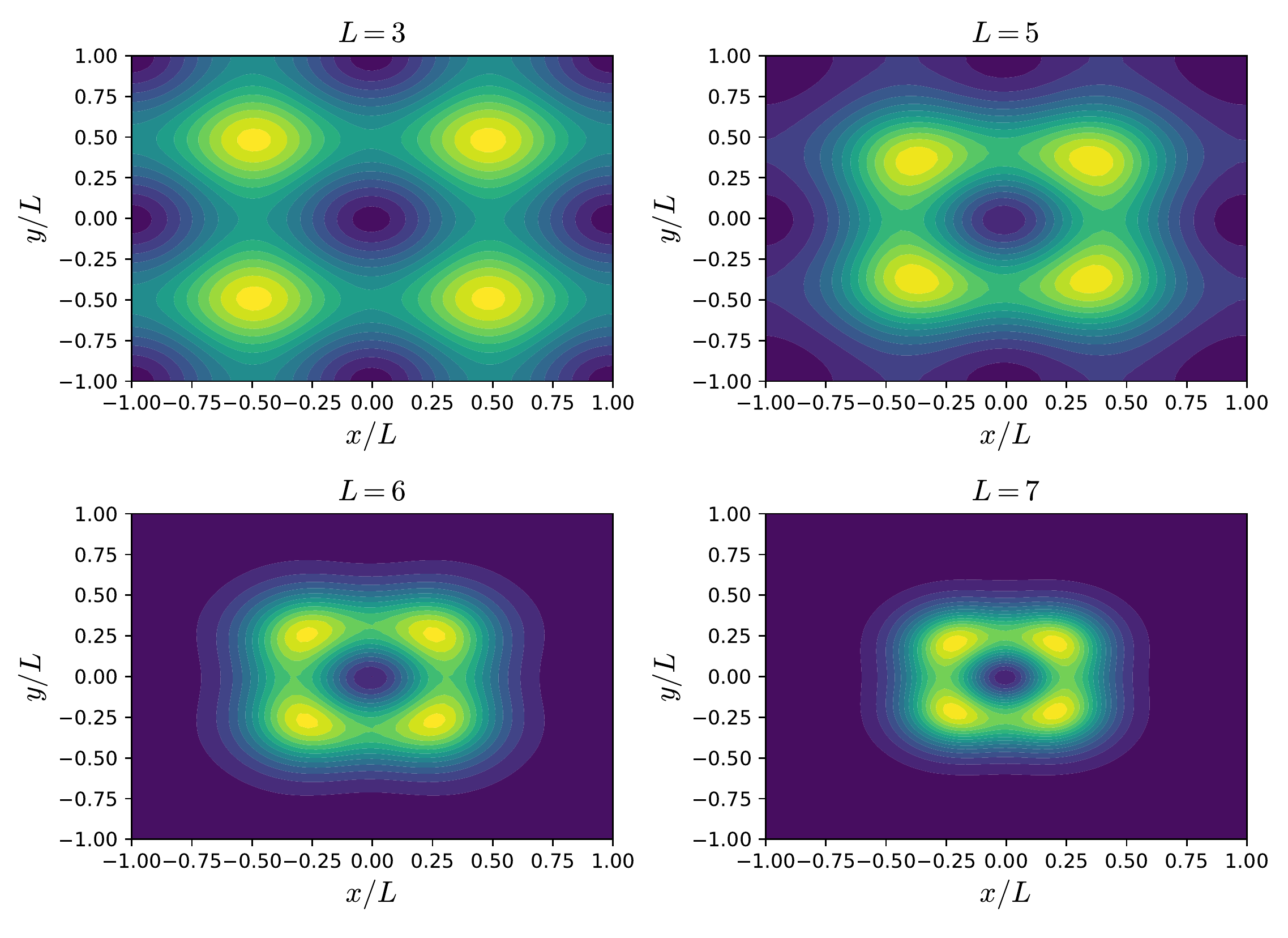}
   \end{minipage}\hfill
   \begin{minipage}{0.48\textwidth}
     \centering
     \includegraphics[scale=0.33]{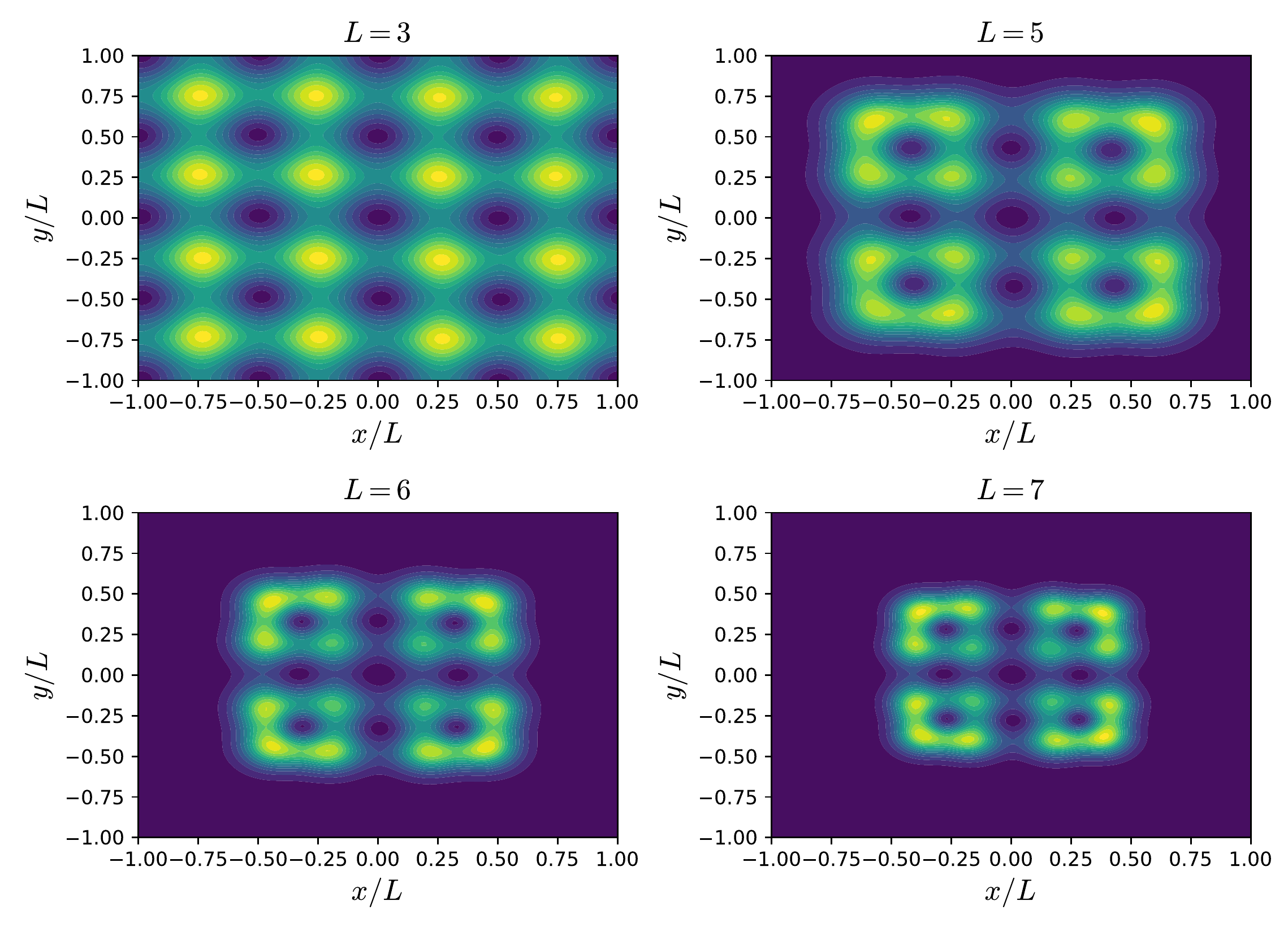}
   \end{minipage}
   \caption{Energy density contour of the $\alpha$ and $B=32$ lattices at different sizes of the unit cell. The shape of the isolated skyrmion is visible at very large values of $L$, whereas the FCC$_+$ crystal is recovered at high densities. The legend of the colours is as in \cref{Figure.F}}
   \label{edens-lattice}
\end{figure}

As expected, both the $\alpha$ and $B=32$ lattices have less energy than the FCC crystal at low densities, since they achieve a more compact configuration surrounded by vacuum. Decreasing the size of the unit cell forces the skyrmion within the unit cell to recover the FCC$_+$ crystal configuration. The transition for both lattices may be seen in \cref{Figure.Lattice}. In \cref{edens-lattice} the corresponding energy density contours are shown for the $\lag_{240}$ model.

\section{Isospin quantization and symmetry energy}
The classical cristalline solutions presented in the previous sections can be understood as models for infinite, isospin symmetric nuclear matter, that is, with the same number of protons and neutrons. However, it is well known that nuclear matter in the interior of neutron stars cannot be completely isospin symmetric, with all but a small fraction of the total baryonic degrees of freedom being protons at a given density. The magnitude that determines the fraction of protons over the total baryon number is the so-called \emph{symmetry energy}, namely, the change in the binding energy of the system as the neutron-to-proton ratio is changed at a fixed value of the total baryon number, and its knowledge is essential to determine the composition of nuclear matter at high densities. 

For an finite nuclear system with total baryon number $A=N+Z$, with $N\, (Z)$ the number of neutrons (protons), the isospin asymmetry parameter is defined as $\delta=(N-Z)/A=(1-2\gamma)$, with $\gamma=Z/A$ the proton fraction. For an infinite system, the above quantities can be defined with densities instead of total numbers. The binding energy of infinite nuclear matter is thus parametrized as a function of both the baryon density and the asymmetry parameter,
\begin{equation}
    \frac{E}{A}(n_B,\delta)=E_N(n_B)+S_N(n_B)\delta^2+\order{\delta^3},
\end{equation}
with $E_N(n_B)$ the binding energy of isospin-symmetric matter, and $S_N(n_B)$ the symmetry energy.
Although its dependence on the density has proven difficult to measure experimentally, it is usually parametrized as an expansion in powers of the baryon density around nuclear saturation $n_0$,
\begin{equation}
    S_N(n_B)=S_0 +\frac{1}{3}L \epsilon +\frac{1}{18}K_{\rm sym}\epsilon^2 +\cdots
\end{equation}
with $\epsilon=(n-n_0)/n_0$, and 
\begin{equation}
    L=3 n_0\pdv{S_N}{n}\eval_{n=n_0}, \quad  K_{\rm sym}=9n_0^2\pdv[2]{S_N}{n}\eval_{n=n_0}
    \label{eq_symetobs}
\end{equation}
the slope and curvature of the symmetry energy at saturation, respectively.The symmetry energy at saturation is well constrained ($S_0\sim 30$ MeV) by nuclear experiments  \cite{FiorellaFantina:2018dga}, but the values of the slope
 and higher order coefficients are still very uncertain. However, recent efforts on the analysis of up to date combined astrophysical and nuclear observations have allowed to constrain the value of these quantities with reasonable uncertainty above nuclear saturation \cite{Landry:2021ezp,Tang:2021snt,deTovar:2021sjo,Gil:2021ols,Li:2021thg}.

In the Skyrme model, due to the $SU(2)$ isospin symmetry of the Lagrangian, isospin degrees of freedom represent zero-modes, which are quantized using standard canonical quantization in terms of some collective coordinate parametrization (see, e.g., \cite{adkins1983static,adkins1984skyrme,battye2009light,Lau:2014baa}). 

Following this approach, we consider a (time-dependent) isospin transformation of a static Skyrme field configuration,
\begin{equation}
    U(\vec{x})\rightarrow \tilde U(\vec{x},t)\equiv g(t) U(\vec{x}) g^\dagger(t).
    \label{IsorotF1}
\end{equation}
Then, the time component of the left invariant form $L_\mu$ becomes $ \tilde U^\dagger\partial_0 \tilde U=g T_ag^\dagger \omega_a$, where $ T_a$ is the $\mathfrak{su}(2)$-valued current,
\begin{equation}
    T_a=\frac{i}{2}U^\dagger[\tau_a,U]=i(\pi_a\pi_b-\pi_c\pi_c\delta_{ab}+\sigma\pi_c\epsilon_{abc})\tau_b\equiv iT_b^a\tau_b,
    \label{UdcommtU1}
\end{equation}
and  $\vec{\omega}\doteq g^\dagger\dot{g}=\tfrac{i}{2}\omega_a\tau_a$ is the isospin angular velocity.
The time dependence of the new Skyrme field induces a kinetic term in the energy functional, given by \footnote{Remember that we are using the mostly minus convention for the metric signature.}
\begin{align}
    \notag T=\frac{1}{24\pi^2}\int\{&a\Tr\{L_0L_0\}-2b\Tr\{[L_0,L_k][L_0,L_k]\} -c\,\,\mathcal{B}^i\mathcal{B}_i\}d^3x=\frac{1}{2}\omega_i\Lambda_{ij}\omega_j
\end{align}
where $\Lambda_{ij}$ is the isospin inertia tensor, given by
\begin{equation}
    \Lambda_{ij}=\frac{1}{24\pi^2}\int d^3x\left\{2a\Tr\{T_iT_j\}-4b\Tr\{[T_i,L_k][T_j,L_k]\}\right.-\left.\frac{c}{32\pi^4}\varepsilon^{abc}\Tr\{T_iL_bL_c\}\varepsilon_{ars}\Tr\{T_jL_rL_s\}\right\}\,,
    \label{Inertia_Tensor1}
\end{equation}
and the values of $a,b,c,d$ are easily obtained from \eqref{Lagrangian},
\begin{equation}
    a=-\frac{1}{2}, \quad b=\frac{1}{4},\quad c=-8\lambda^2\pi^4\frac{f_\pi^2e^4}{\hbar^3}, \quad 
    d = \frac{m^2_{\pi}}{f^2_{\pi}e^2}.
    \label{adimpars}
\end{equation}

Due to the symmetries of the cristalline phases, the complete isospin inertia tensor for the unit cell of a cubic crystal turns out to be proportional to the identity, and the associated eigenvalue (the isospin moment of inertia) can be written
\begin{equation}
    \Lambda =\frac{1}{24\pi^2} \left[2a \Lambda^{(2)}-4b\Lambda^{(4)}-\frac{c}{32\pi^4}\Lambda^{(6)} \right],
\label{iso_moment}
\end{equation}
where the contribution from each term in the lagrangian, denoted by $\Lambda^{(n)}$, can be found in \cite{Adam:2022aes}.
This fact enormously simplifies the kinetic term in the Lagrangian of an isospinning cubic crystal with a number $N_{\text{cells}} \equiv N$ of unit cells, which is reduced to
\begin{equation}
    T=\frac{1}{2}\omega_i\Lambda_{ij}\omega_j=N\frac{1}{2}\Lambda\omega_a\omega^a,
\end{equation}
and, by defining the corresponding canonical momentum $J_a=\partial L/\partial \omega^a=N\Lambda\omega_a$, we may write it in Hamiltonian form,
\begin{equation}
    H=\frac{1}{2N\Lambda}J_aJ^a.
\end{equation}
Now, following the standard canonical quantization procedure, we promote the isospin angular momentum variables to operators, so that we may diagonalise the Hamiltonian in a basis of eigenstates with a definite value of the total isospin angular momentum,
\begin{equation}
    H=\frac{\hbar^2}{2N\Lambda}J^{\rm{tot}}(J^{\rm{tot}}+1)
    \label{Hamiltonian}
\end{equation}
 The total isospin angular momentum of the full crystal will be given by the product of the total number of unit cells times the total isospin of each unit cell, which can be obtained by composing the isospin of each of the cells. In the charge neutral case, all cells will have the highest possible value of isospin angular momentum, so that in each unit cell with baryon number  $B_{\text{cell}}$, the total isospin will be $\frac{1}{2}B_{\text{cell}}$, and hence the total isospin of the full crystal will be $J^{\rm{tot}}=\frac{1}{2}N B_{\text{cell}}$.

Therefore, the quantum correction to the energy (per unit cell) in the charge neutral case (completely asymmetric matter) due to the isospin degrees of freedom will be given by (assuming $N\rightarrow \infty$)
\begin{equation}
    E_{\rm{iso}}=\frac{\hbar^2}{8\Lambda}B_{\text{cell}}^2.
    \label{Eiso1}
\end{equation}
Such correction could also have been obtained directly by introducing an 'external' isospin chemical potential $\mu_I$ , and promoting the regular derivatives in the Skyrme Lagrangian to covariant derivatives of the form \cite{Son:2000xc}
\begin{equation}
    \partial_\mu U\rightarrow D_\mu U=\partial_\mu U-\frac{i\mu_I}{2}\delta_{\mu 0}[\tau^3,U],
\label{Covreplacement}
\end{equation}
so that, if $U$ is a static configuration, the time component of the Maurer-Cartan form becomes
\begin{equation}
    L_0=-\frac{i}{2}\mu_IU^\dagger[\tau^3,U]=-\mu_IT_3.
\end{equation}
This expression is equivalent to that of an iso-rotating field with angular velocity $\omega_a=-\mu_I\delta_{3a}$. Thus, it is straightforward to obtain the isospin chemical potential for the Skyrmion crystal using its thermodynamical definition
$
    \mu_I=-\pdv{E}{n_I}
$, 
where $n_I$ is the (third component of) the isospin number density. Given that $(J^{\rm{tot}})^2=J_1^2+J_2^2+J_3^2$ and $n_I=J_3/N$, we may write the isospin energy per unit cell as
\begin{equation}
    E_{\rm{iso}}=\frac{\hbar^2 }{2\Lambda}\qty(n_I^2+\frac{J_2^2}{N^2}+\frac{J_1^2}{N^2})
\end{equation}
and then
\begin{equation}
    \mu_I=-\pdv{E_{\rm{iso}}}{n_I}=-\frac{\hbar^2 }{\Lambda}n_I.
\end{equation}

For $B<\infty$ Skyrmions, isospin quantization (together with spin) induces a set of constraints (Finkelstein-Rubinstein constraints \cite{Finkelstein:1968hy,Krusch:2002by}) which characterize the allowed states with a given value of total spin and isospin angular momentum, and thus the ground states and lowest energy spin and isospin excitations have been studied for many Skyrmions with $B\le 12$  \cite{battye2009light,Lau:2014baa}. 
However, in the case of a crystal, to compute the contribution of the quantization of the global isospin zero modes to the total energy we would need to know, in principle, the quantum isospin state of the whole crystal. This is of course impossible in the thermodynamic limit, and some additional approximation becomes necessary.

Following \cite{Adam:2022aes}, since the total third component of isospin is a good quantum number for the total quantum state of the crystal, we perform a mean field approximation and consider that the isospin density in an arbitrary skyrmion crystal quantum state is approximately uniform so that
\begin{equation}
    \ev{I^0_3}=\frac{\ev{I_3}}{\int d^3x}=\frac{\bra{\Psi}\int I^0_3 d^3x\ket{\Psi}}{NV_{\rm cell }}\doteq \frac{n_I}{V_{\rm cell}}
\end{equation}
where $n_I$ is the effective isospin charge  per unit cell in this arbitrary quantum state.  The (mean field) effective proton fraction corresponding to such an isospin charge per unit cell with baryon number $B_{\text{cell}}$ is given by
    \begin{equation}
       n_I=-\frac{1}{2}(1-2\gamma)B_{\text{cell}}=-\frac{B_{\text{cell}}}{2}\delta .
    \end{equation}
Hence, the isospin energy per unit cell of the Skyrmion crystal can be written in terms of the asymmetry parameter
\begin{equation}
    E_{\rm iso}=\frac{\hbar^2 B_{\text{cell}}^2}{8\Lambda}\delta^2,
    \label{Eiso}
\end{equation}
from where we can readily read off the symmetry energy for Skyrme crystals
\begin{equation}
    S_N(n_B)=\frac{\hbar^2 L^3}{\Lambda}n_B.
\end{equation}
Therefore, the symmetry energy of a Skyrmion crystal is uniquely determined by the (isospin) moment of inertia of each unit cell, which has an intrinsic dependence on the density. On the other hand, the isospin energy correction depends, in the mean-field approximation, both on $\Lambda$ (hence the density) and the asymmetry parameter (or equivalently, the proton fraction). 

However, a nonzero proton fraction  would rapidly lead to a divergence in the Coulomb energy of the total crystal in the infinite crystal limit. Therefore, 
 a neutralizing background of negatively charged leptons (electrons and possibly muons) must be included in the system, such that the Coulomb forces are screened. The total system is thus characterized at equilibrium by two conditions, namely the \emph{charge neutrality condition}
\begin{equation}
    n_p=\frac{Z}{V}=n_e+n_\mu ,
\label{chargecond}
\end{equation}
 and the \emph{$\beta$-equilibrium condition}
\begin{equation}
    \mu_n=\mu_p+\mu_l\implies\mu_I=\mu_l,\qquad l=e,\mu ,
\label{betacond}
\end{equation}
i.e., the isospin chemical potential must equal that of charged leptons, which in turn means that the direct and inverse beta decay processes such as
$    n\leftrightarrow p+l+\bar{\nu}_l$
take place at the same rate.
Leptons inside a neutron star can be described as a non-interacting, highly degenerate fermi gas, so that the chemical potential for each type of leptons can be written
\begin{equation}
    \mu_l=\sqrt{(\hbar k_{F})^2+m_l^2}
\end{equation}
where $k_{F}=(3\pi^2n_l)^{1/3}$ is the corresponding Fermi momentum, and $m_l$ is the mass of the corresponding lepton. For sufficiently high densities, then, the electron chemical potential becomes larger than the mass of the muon, $\mu_e\geq m_\mu$, and the production of muons is preferred by the system. We can estimate the total proton fraction by enforcing both beta equilibrium and
charge neutrality. Neglecting the contribution of muons in a first step, to the charge density, from  \eqref{chargecond} we can relate the electron density to the proton fraction parameter, $n_e=\gamma B_{\text{cell}} /(2L)^3$. The $\beta$ equilibrium condition then provides an equation which defines $\gamma$ implicitly as a function of the lattice length parameter,
\begin{equation}
    \frac{\hbar L}{\Lambda}(1-2\gamma)=\qty(\frac{3\pi^2}{B_{\text{cell}}^2})^{1/3}\gamma^{1/3}
\end{equation}
where we also assumed the ultrarelativistic electron approximation, i.e. $m_l/k_F\simeq 0$. 


Including the muon contribution to the charge density yields a slightly more complicated expression for the $\beta$-equilibrium condition, given by
\begin{equation}
    \frac{\hbar B_{\text{cell}}}{2\Lambda}(1-2\gamma)=\qty[3\pi^2\qty(\frac{\gamma B_{\text{cell}}}{8L^3}-n_\mu)]^{\tfrac{1}{3}},
\end{equation}
where 
\begin{equation}
    n_\mu=\frac{1}{3\pi^2}\qty[\qty(\frac{\hbar B_{\text{cell}}(1-2\gamma)}{2\Lambda})^2-\left(\frac{m_\mu}{\hbar}\right)^2]^{\tfrac{3}{2} }.
\end{equation}
 The proton fraction inside beta-equilibrated matter determines, in addition, whether a proto-neutron star will go through a cooling phase via the emission of neutrinos through the direct Urca (DU) process $n\rightarrow p +e+\bar \nu_e$. This process is expected to occur if the proton fraction reaches a critical value, $\gamma_p> x_{DU}$, the so-called DU-threshold \cite{PhysRevLett.66.2701,Klahn:2006ir}. The DU process allows for an enhanced cooling rate of NS. Whether it takes place or not in the hot core of proto-neutron stars or during the merger of binary NS systems \cite{universe7110399}, therefore, determines the proton fraction (and the symmetry energy) of matter at ultra-high densities. It is, however, not clear whether this enhanced cooling actually occurs, although there is recent evidence that supports it \cite{Brown:2017gxd}.

 In $npe\mu$ matter, the DU threshold is given by \cite{Klahn:2006ir}
 
 \begin{equation}
     x_{DU}=\frac{1}{1+(1+(\frac{n_e}{n_e+n_\mu})^{1/3})^3}.
 \end{equation}

\begin{figure*}[htb!]
    \centering
    \includegraphics[scale=0.5]{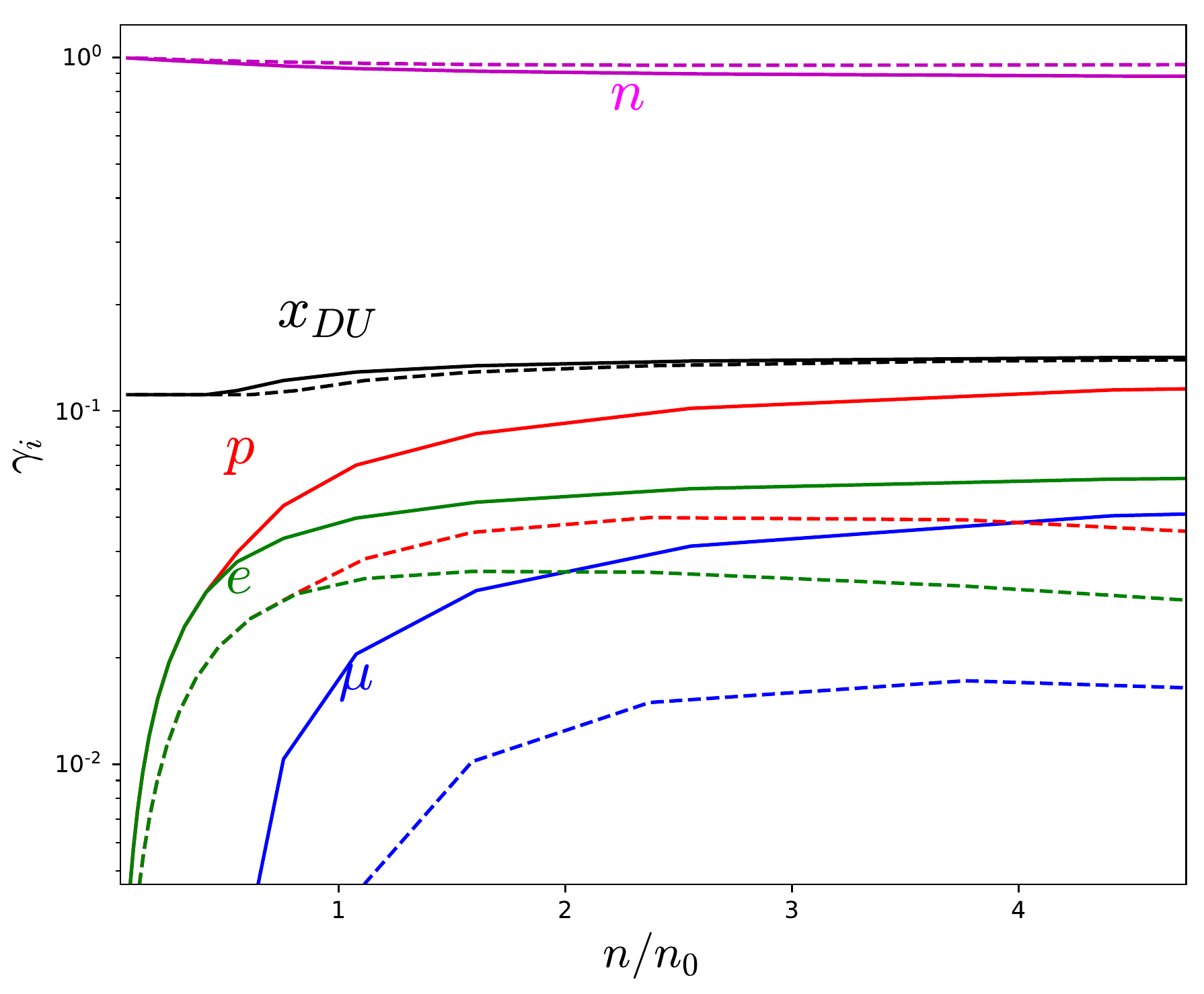}
    \caption{Fraction density $\gamma_i$ for each particle as a function of the baryon density for $\lambda^2=0$ (solid) and $\lambda^2=1.5$ (dashed). The corresponding DU threshold is also shown in black. This figure was originally published in \cite{Adam:2022aes}.}
    \label{fig:Protonfractionse}
\end{figure*}
We show the particle populations $\gamma_i$ in beta-equilibrated skyrmion matter in \cref{fig:Protonfractionse} for the cases $\lambda^2=0,1.5$ MeV fm$^3$. 

In both cases, a persistent population of protons and leptons at higher nucleon density is expected,  although we can see that in the case with the sextic term the fraction of charged particles is smaller. This is explained by the lower symmetry energy of the sextic term, which makes it much easier to convert protons into neutrons. Finally, for all values of the parameters ($f_{\pi}, e, \lambda^2$) we considered, the DU-threshold is not reached. One should, however, not consider this fact as a prediction of the Skyrme model, because it strongly depends on the parameter values. Besides, it is also generally assumed that around $2-3$ times the nuclear saturation density, additional degrees of freedom (strange baryons) appear and become important for the description of nuclear matter which, in particular, may affect the proton fraction at these densities.

\section{Kaon condensate in skyrmion crystals}
Up to this point, we have shown how to describe $\beta$-equilibrated $npe\mu$-matter at finite density in terms of a semi-classical skyrmion crystal and a leptonic sea of electrons and muons, and we have determined the different particle fractions as a function of density in the mean field approximation. However, for densities about twice nuclear saturation and above, additional degrees of freedom are expected to become relevant for the physics of dense nuclear matter. In particular, strange mesons (kaons) and hyperons are believed to  modify the EOS at sufficiently high density, which motivates the question of how to describe an additional quark flavor using the Skyrme model approach. 

 Below we shall briefly review the crucial steps to determine whether kaon fields may condense inside a Skyrme crystal for a sufficiently high density, and, if so, whether this critical density value is relevant for the description of matter inside compact stars. A more detailed discussion can be found in \cite{Adam:2022cbs}.


\subsection{ The kaon condensate effective potential}
Following the bound-state approach first proposed in \cite{callan1985bound} we may extend the skyrmion field to a $SU(3)$-valued field $U$ and consider kaon fluctuations on top of a $SU(2)$ skyrmion-like background $u$. With the only requirement that unitarity must be preserved, different ansätze have been proposed in the literature for the total $SU(3)$ field describing both pions and kaons. In this work, we choose the ansatz proposed by Blom et al in \cite{Blom1989HyperonsAB}:
\begin{equation}
    U = \sqrt{U_K}U_{\pi}\sqrt{U_K}.
    \label{BDR ansatz}
\end{equation}
In this ansatz $U_{\pi}$ represents the $SU(3)$ embedding of the purely pionic part $u$, and the field $U_K$ are the fluctuations in the strange directions. It can be shown that this ansatz is equivalent to the one first proposed by Callan and Klebanov in \cite{callan1985bound} when computing static properties of hyperons, although both may differ in other predictions of the model \cite{nyman1990low}.

In the simplest $SU(3)$ embedding, the $SU(2)$ field $u$ is extended to $U_\pi$ by filling the rest of entries with ones in the diagonal and zeros outside. On the other hand, the kaon ansatz is modelled by a $\mathfrak{su}$(3) -valued matrix $\mathcal{D}$ which is non trivial in the off-diagonal elements:
\begin{equation}
    U_{\pi} = \begin{pmatrix}
                u & 0\\
                0 & 1
              \end{pmatrix}, \hspace{3mm}
    U_K = e^{ i \frac{2\sqrt{2}}{f_{\pi}}
                   \mathcal{D} },\quad
u=\sigma+i\pi_a\tau_a,\quad \mathcal{D}=\mqty( 0 & K\\ K^{\dagger} & 0)
\label{Kparam}
\end{equation}
where $K$ consists of a scalar doublet of complex fields representing  charged and neutral kaons:
\begin{equation}
    K=\mqty(K^+\\K^0),\quad K^\dagger = (K^-, \bar{K}^0).
\end{equation}
We now extend the Generalized Skyrme Lagrangian \eqref{Lagrangian} to include strange degrees of freedom. First, we need to replace the potential term $\lag_0$ in order to correctly take into account the mass of the kaon fields. The new potential term has the form \cite{nyman1990low}:
\begin{align}
    \notag \lag_0^{\rm new} = \frac{f^2_{\pi}}{48}\left( m^2_{\pi} + 2 m^2_K \right) \Tr{U + U^{\dagger} - 2} +\frac{\sqrt{3}}{24}f^2_{\pi}\left( m^2_{\pi} - m^2_K \right)\Tr{\lambda_8\left(U + U^{\dagger}\right)},
\end{align}
where $\lambda_8$ is the eighth Gell-Mann matrix and $m_K$ is the vacuum kaon mass. In addition, the effects of the chiral anomaly must be taken into account in the effective theory by means of the Wess-Zumino-Witten (WZW) term, which can be expressed in terms of a 5-dimensional action:
\begin{equation}
    S_{WZ} = -i\frac{N_c}{240\pi^2}\int d^5x \: \epsilon^{\mu\nu\alpha\beta\gamma}\Tr{L_{\mu}L_{\nu}L_{\alpha}L_{\beta}L_{\gamma}}.
\end{equation}

The onset of kaon condensation in the Skyrme model takes place at a critical density $n_{\rm cond}$ at which $\mu_{e}$ becomes greater than the energy of the kaon zero-momentum mode (s-wave condensate). Thus, for baryon densities $n\geq n_{\rm cond}$, the presence of kaons will be more energetically favorable than electrons at the fermi surface, so kaons will be produced at expense of electrons, and the macroscopic contribution of the kaon condensate to the energy must be taken into account when obtaining the EOS. To do so, we write the field condensates (i.e. the non-zero vacuum expectation values (vev), $\ev{K^\pm}$ ) as a homogeneous field whose time dependence is given by:
\begin{equation}
    \ev{K^\mp}=\phi e^{\mp i\mu_K t}
\end{equation}
The real constant $\phi$ corresponds to the zero-momentum component of the fields, which acquires a nonvanishing, macroscopic value after the condensation. Its exact value is determined from the minimization of the corresponding effective potential, which is determined from the Skyrme lagrangian. On the other hand, the phase $\mu_K$ is nothing but the corresponding kaon chemical potential. First, we will need an explicit form of the $SU(3)$ Skyrme field in the kaon condensed phase. Assuming the charged kaons will be the first mesons to condense, we drop the neutral kaon contribution and define the following matrix
\begin{equation}
    \tilde{\mathcal{D}}=\mqty(0&0&\phi e^{i\mu_K t}\\
    0&0&0\\
    \phi e^{-i\mu_k t}&0&0)
\end{equation}
which results from substituting the kaon fields in $\mathcal{D}$ as defined in \eqref{Kparam} by their corresponding vev in the kaon condensed phase. Also, taking advantage of the property $\mathcal{D}^3=\phi^2\mathcal{D}$, we may write the $SU(3)$ element generated by $\tilde{\mathcal{D}}$ explicitly in matrix form:
\begin{equation}
    \Sigma=e^{i \tfrac{\sqrt{2}}{f_\pi}\tilde{\mathcal{D}}}=\mqty(\cos\tilde{\phi}&0&ie^{i\mu_K t}\sin \tilde\phi\\
    0&1&0\\
    ie^{-i\mu_K t}\sin \tilde\phi&0&\cos\tilde\phi)
    \label{Sigmaexpl}
\end{equation}
where $\tilde\phi=\tfrac{\sqrt{2}}{f_\pi}\phi$ is the dimensionless condensate amplitude.

Furthermore, assuming the backreaction from the kaon condensate to the skyrmion crystal is negligible, and thus the classically obtained crystal configuration will be the physically correct background even in the kaon condensed phase, we may write the $SU(3)$ field in this phase as $U=\Sigma U_\pi \Sigma$, where $U_\pi $ is the $SU(3)$ embedding of the $SU(2)$ skyrmion background as in \eqref{Kparam}. Introducing this $U$ in the total action yields the standard Skyrme action for the $SU(2)$ field plus an effective potential term for the kaon condensate:
\begin{equation}
    S_{Sk}(U)+S_{\rm WZW}(U)=S_{Sk}(U_\pi)-\int d t V_K(\tilde{\phi}),
\end{equation}
where
\begin{equation}
    V_K = \frac{1}{24\pi^2}\int d^3x\Big[ V^{(2)}_K + V^{(4)}_K + V^{(6)}_K + V^{(0)}_K\Big] + V^{(WZW)}_K,
\end{equation}
with
\begin{align}
    &V_K^{(0)}= 2\frac{m^2_K}{f^2_{\pi}e^2}(1+\sigma)\sin^2\tilde\phi\\[2mm]
    &V_K^{(2)}=\mu_K^2\sin^2\tilde{\phi}[(1+\sigma^2+\pi^2_3)\sin^2\tilde\phi-2(1+\sigma\cos^2\tilde\phi)].\\[2mm]
    &V_K^{(4)}= -2\mu_K^{2}\sin^{2}{\tilde\phi} \big\{(1+\sigma)\partial_i n^2\cos^{2}{\tilde\phi} +  2[ \partial_ i \sigma^{2} (1-\pi_{3}^{2}) +  \partial_i \pi_3^{2} (1-\sigma^{2}) + 2\sigma \pi_{3} \partial_ i \sigma \partial_i \pi_3]\sin^2\tilde{\phi} \big\} \\[2mm]
    &V_K^{(6)}= - \lambda^2 f^2_{\pi} e^4 \mu_K^2 \sin^4(\tilde\phi) (\partial_i\pi_3\partial_j\sigma-\partial_i\sigma\partial_j\pi_3)^2,\\[2mm]
    &V_K^{\rm (WZW)}= -\mu_K N_c B_{\rm cell} \sin^{2}{\tilde \phi }.
\end{align}

\vspace*{0.5cm}
\subsection{ Quantum corrections to the effective potential}
In the above calculations, we have taken separately the contributions of a kaon condensate and an isospin angular momentum of the skyrmion crystal. However, since kaons possess an isospin quantum number, the condensate interacts with the skyrmion isospin both indirectly via the charge neutrality and $\beta$ equilibrium conditions, which relate their corresponding chemical potentials, and also directly, due to the appearance of additional contributions to the total energy when we consider a (time-dependent) isospin transformation of the full Skyrme field + kaon condensate configuration $U=\Sigma U_\pi \Sigma$:
\begin{equation}
    U\rightarrow \tilde U\equiv A(t) U A^\dagger(t),
    \label{IsorotF}
\end{equation}
where $A$ is an element of $SU(3)$ modelling an isospin rotation,
\begin{equation}
    A=\mqty(a&0\\0&1), \quad a\in SU(2).
\end{equation}
in analogy with the $SU(2)$ case, we define the isospin angular velocity $\vec{\omega}$ as $A^\dagger\dot{A}=\tfrac{i}{2}\omega_a\lambda_a$ ($a=1,2,3$),
with $\lambda_\textsc{a}$ the Gell-Mann matrices generating $SU(3)$ for $\textsc{a}=1,\cdots 8$. Notice that $\vec{\omega}$ is a three-vector, since $A^\dagger\dot{A}$ belongs to the isospin $\mathfrak{su}(2)$ subalgebra of $\mathfrak{su}(3)$.
Then, we may write the time component of the Maurer-Cartan current as
$ \tilde U^\dagger\partial_0 \tilde U=AL_0A^\dagger +A T_a A^\dagger \omega_a$, where $ T_a$ is the $\mathfrak{su}(3)$-valued current:
\begin{equation}
    T_a=\frac{i}{2}U^\dagger[\lambda_a,U]\equiv iT_a^\textsc{a}\lambda_\textsc{a},
    \label{UdcommtU}
\end{equation}
where we have made use of the parametrization \eqref{Kparam}. 

Now the kinetic isorotational energy in this case takes the form
\begin{equation}
    T=\frac{1}{2}\omega_a\Lambda_{ab}\omega_b+\Delta_a\omega_a - V_K
\label{kinetic}
\end{equation}
where $\Lambda_{ab}$ is the isospin inertia tensor and $\Delta_a$ is the kaon condensate isospin current, given by

\begin{align}
    \Lambda_{ab}&=\int\left\{2a\Tr\{T_aT_b\}-4b\Tr\{[T_a,L_k][T_b,L_k]\}-\frac{c}{32\pi^4}\varepsilon^{lmn}\Tr\{T_aL_mL_n\}\varepsilon_{lrs}\Tr\{T_jL_rL_s\}\right\}\,d^3x,
    \label{Inertia_Tensor}\\
    \Delta_a&=\int\left\{2a\Tr\{L_0T_a\}-4b\Tr\{[T_a,L_k][L_0,L_k]\}-\frac{c}{32\pi^4}\varepsilon^{lmn}\Tr\{L_0L_mL_n\}\varepsilon_{lrs}\Tr\{T_aL_rL_s\}\right\}\,d^3x,
\end{align}
where $a,b$ and $c$ are those in \cref{adimpars}. As argued previously, the symmetries of the crystalline configuration imply that the isospin inertia tensor becomes proportional to the identity, i.e. $\Lambda^{\rm{crystal}}_{ab}=\Lambda\delta_{ab}$. However, the presence of a kaon condensate breaks this symmetry to a $U(1)$ subgroup, so that $\Lambda_{ab}$ presents two different eigenvalues in the condensate phase, $\Lambda_{\rm cond}=\text{diag}(\Lambda,\Lambda,\Lambda_3)$. Similarly, $\Delta_a=0$ in the purely barionic phase, and its third component acquires a non-zero value in the condensate phase, $\Delta_{\rm cond}=(0,0,\Delta)$. The explicit expressions for $\Lambda_3$ and $\Delta$ in the condensed phase can be found in \cite{Adam:2022cbs}. One can check that in the non-condensed phase, $\phi=0$ and the results of the previous section are recovered, namely, $\Lambda_3=\Lambda$, $\Delta=0$.

The quantization procedure now goes along the same lines as in the previous section. However, the isospin breaking due to the kaon condensate implies that the canonical momentum associated to the third component of the isospin angular velocity will now be different, and given by
$ I_3=\Lambda_3\omega_3+\Delta$.

Thus, after a Legendre transformation to rewrite \eqref{kinetic} in Hamiltonian form, and making the $N\rightarrow\infty $ approximation, one can write the quantum energy correction per unit cell of the crystal in the kaon condensed phase as
\begin{equation}
    E_{\rm{quant}}=\frac{1}{2\Lambda_3}(I_3^2-\Delta^2).
\label{quant_corr_kcond}
\end{equation}
The first term on the rhs is just the isospin correction, while now there is an additional second term due to the isospin of the kaons. Indeed, since the kaon field enters also in the expression of the isospin moment of inertia $\Lambda_3$, both terms will depend nontrivially on the kaon vev field.

When the kaon field develops a nonzero vev, apart from the neutron decay and lepton capture processes, additional processes involving kaons may occur:

\begin{equation}
    n\leftrightarrow p + K^-, \qquad l\leftrightarrow K^-+\nu_l
\end{equation}
such that the chemical equilibrium conditions 
$ \mu_n=\mu_p+\mu_K ,  \quad  \mu_ l = \mu_ K$
are satisfied. These are the extension of \cref{betacond} to the condensate phase.

The total energy within the unit cell may be obtained as the sum of the baryon, lepton and kaon contributions:
\begin{equation}
    E = E_{\rm class} + E_{\rm iso}(\gamma,\tilde{\phi}) + E_K(\mu_e, \tilde{\phi})+ E_{e}(\mu_e)+\Theta(\mu_e^2-m_\mu^2)E_{\mu}(\mu_e)
    \label{eq:Etot}
\end{equation}
The kaon contribution is the effective potential energy
\begin{equation}
    E_K(\mu_e, \tilde{\phi}) = V_K-\frac{\Delta^2}{2\Lambda_3},
\end{equation}
which depends on the condensate $\tilde{\phi}$ and on the lepton chemical potential through the explicit dependence on $\mu_K$ of both $V_K$ and $\Delta$, and $\mu_K=\mu_e$ due to the equilibrium conditions. Therefore, the energy of the full system depends on the proton fraction, the kaon vev field and the electron chemical potential. Their respective values can be obtained, for fixed $n_B$ (or equivalently, fixed $L$) by minimizing the free energy
\begin{equation}
    \Omega = E -\mu_e (N_e +\Theta(\mu_e^2-m_\mu^2)N_\mu-\gamma B)
\end{equation}
 with respect to $\gamma$, $\tilde{\phi}$ and $\mu_e$, i.e.
\begin{equation}
    \pdv{\Omega}{\gamma}\eval_{n_B}\!\!\!(\gamma,\tilde{\phi}, \mu_e)=\pdv{\Omega}{\tilde{\phi}}\eval_{n_B}\!\!\!(\gamma,\tilde{\phi}, \mu_e) = \pdv{\Omega}{\mu_e}\eval_{n_B}\!\!\!(\gamma,\tilde{\phi}, \mu_e)=0.
    \label{minimcond}
\end{equation}
The first equation imposes the expected condition $\mu_e = \mu_I = 2\hbar^2(1-2\gamma)/\Lambda_3$. Then, after substituting into the other two conditions we get:
\begin{align}
    \gamma n_B - &\frac{(\mu^2_I - m^2_{e})^{3/2} + (\mu^2_I - m^2_{\mu})^{3/2}}{3\pi^2\hbar^3} + \frac{n_B}{4}\pdv{E_K}{\mu_e}\eval_{\mu_e = \mu_I} = 0, \label{cond1} \\
    &\pdv{V_K}{\tilde{\phi}} - \frac{\Delta}{\Lambda_3}\pdv{\Delta}{\tilde{\phi}} + \pdv{\Lambda_3}{\tilde{\phi}}\left(\frac{\Delta^2}{2\Lambda_3^2} - \frac{\mu_I^2}{2\hbar^2}\right) = 0,
    \label{cond2}
\end{align}
which are precisely the charge neutrality condition, and the minimization of the grand canonical potential with respect to the kaon field. We note here that we drop the ultrarrelativistic consideration for electrons since the appearance of kaons may decrease hugely the electron fraction.
By solving the system of equations \ref{cond1} and\eqref{cond2} for $\gamma$ and $\tilde{\phi}$ we obtain all the needed information for the new kaon condensed phase. Then we may compare the particle fractions and energies between both phases, which we will call $npe\mu$ and $npe\mu\overline{K}$.

Before solving the full system for different values of the lattice length $L$, we may try to obtain the value of the length at which kaons condense, $L_{\rm cond}$. This value is indeed important since it will determine whether or not a condensate of kaons will appear at some point in the interior of NS. 
This is accomplished with the same system of \cref{cond1,cond2} by factoring the $\sin\tilde{\phi}$ from the second equation and setting $\tilde{\phi} = 0$. Then we may see the system as a pair of equations to obtain the values of $\gamma_{\text{cond}}$ and $L_{\text{cond}}$, the values of the proton fraction and the length parameter for which the kaons condense.

We show in the table below the density at which kaons condense for different values of the parameters as well as the values of some nuclear observables they yield. All the values are given in units of MeV or fm, respectively.
\begin{table*}[h!]
    \centering
    \begin{tabular}{|c|c|c|c|c|c|c|c|c|}
    \hline
         label & $f_{\pi}$ & $e$ & $\lambda^2$ & $E_0$ & $n_0$ & $S_0$ & $L_{\text{sym}}$ & $n_{\text{cond}}/n_0$  \\
         \hline
         set 1 & 133.71 & 5.72 & 5 & 920 & 0.165 & 23.5 & 29.1 & 2.3\\ \hline
         set 2 & 138.11 & 6.34 & 5.78 & 915 & 0.175 & 24.5 & 28.3 & 2.2\\ \hline
         set 3 & 120.96 & 5.64 & 2.68 & 783 & 0.175 & 28.7 & 38.7 & 1.6\\ \hline
         set 4 & 139.26 & 5.61 & 2.74 & 912 & 0.22 & 28.6 & 38.9 & 1.6\\ \hline
        \end{tabular}
    \caption{Sets of parameter values and observables at nuclear saturation}
    \label{tab:Condensation}
\end{table*}

Parameter sets 1 and 2 are chosen so that the energy per baryon and baryon density at saturation are fitted to experimental values, whereas the sets 3 and 4 correctly fit the symmetry energy and slope at saturation. In section 5.4 we shall always use set 1 of parameter values.


\section{Neutron Stars}

\subsection{TOV system of equations}
In order to calculate the mass and radius for a non-rotating NS we have to solve the standard TOV (Tolman-Oppenheimer-Volkoff) system of ODEs. It is obtained inserting a spherically symmetric ansatz of the spacetime metric,
\begin{equation}
    ds^2 = -A(r)dt^2 + B(r)dr^2 + r^2(d\theta^2 + \sin^2\theta d\varphi^2),
\end{equation}
in the Einstein equations,
\begin{equation}
    R_{\mu\nu} - \frac{1}{2}g_{\mu\nu}R = 8\pi G T_{\mu\nu}.
\end{equation}
To describe matter inside the star, in the right-hand side of the equation, it is standard to use the stress-energy tensor of a perfect fluid,
\begin{equation}
    T_{\mu\nu} = (\rho + p)u_{\mu}u_{\nu} + p g_{\mu\nu},
\end{equation}
where the pressure $p$ and the energy density $\rho$ are not independent but related by the EOS. Hence the EOS describes the nuclear interactions inside the NS and different EOS lead to different observables.

The resulting TOV system involves 3 differential equations for $A, B$ and $p$, which must be solved for a given value of the pressure in the centre of the NS ($p(r = 0) = p_0$) until the condition $p(r = R) = 0$ is achieved.
\begin{align}
    \frac{dA}{dr} &= Ar\left(8\pi G Bp - \frac{1-B}{r^2}\right), \\[2mm]
    \frac{dB}{dr} &= Br\left(8\pi G B\rho + \frac{1-B}{r^2}\right), \\[2mm]
    \frac{dp}{dr} &= -\frac{p + \rho}{2A}\frac{dA}{dr}.
\end{align}

We use a 4$^{\rm th}$ order Runge-Kutta method of step $\Delta r = 1$ m to solve the system in order to obtain the main observables from the solutions. The radial point at which the pressure vanishes defines the radius of the NS, and the mass $M$ is obtained from the Schwarzschild metric definition outside the star,
\begin{equation}
    B(r = R) = \frac{1}{(1 - \frac{2GM}{R})}.
\end{equation}

For different values of the pressure in the centre we may obtain different couples of values for the mass and radius, then we can represent the Mass-Radius (MR) curve which is the most important result in this static problem. Different EOS may lead to very different MR curves, then the measurement of these observables can constrain the curve and so the EOS. Indeed, the region of low mass values in the MR curve has been tightly constrained from nuclear experiments at low densities.

The TOV system can be generalized via the Hartle-Thorne perturbative formalism to consider rotating NS in a slow rotation approximation. In that case more equations are added to the TOV system hence a new set of observables like the moment of inertia, rotational Love number and quadrupolar moment may be extracted. These observables are typically obtained from isolated pulsars, hence the importance in the extension of the TOV system to the Hartle-Thorne formalism lies in the appearance of a new source of information for NS to restrict the nuclear EOS. Furthermore it is possible to extract from this formalism an additional equation which describes how a compact object is deformed by an external tidal force. The associated observable is the tidal deformability which can be calculated from the GW spectrum emitted in NS collisions.

In this work we will only consider the MR curves and the tidal deformability to compare with recent observations.

\subsection{A generalized Skyrme EOS}
In this section we want to briefly review the construction of a Skyrme-model based EOS first discussed in \cite{Adam:2020yfv}, based on the following two observations. Firstly, neutron stars based on the standard $\lag_{24}$ model lead to too small maximum NS masses \cite{Naya:2019rlm}, whereas the EOS based on the $\lag_{60}$ submodel imply rather large maximum masses \cite{Adam:2014dqa}. Secondly, the sextic term provides the leading contribution to the EOS at large densities \cite{Adam:2015lra}, whereas it must be subleading at lower densities because of its scaling properties. This
motivates to consider a generalized Skyrme model EOS in the study of NS which interpolates between the standard $\lag_{24}$ crystal EOS at intermediate densities and the $\lag_{60}$ submodel EOS at large densities, as was done in \cite{Adam:2020yfv}. Then we will compare these results with the full numerical generalized Skyrme crystal solutions.

Even though we have not compared the energy of the crystals with other skyrmion solutions, it is known that the crystalline configurations are, so far, the ones with the lowest energy in the standard ($\lag_{24}$) Skyrme model. In one of the works where the FCC$_+$ crystal was found \cite{Castillejo1989DENSESS} a parametrization of the energy in terms of the unit cell side length, similar to \cref{Fit}, is shown
\begin{equation}
    E_{24}(l) = E_0\left[a\left(\frac{l}{l_0} + \frac{l_0}{l}\right)+b\right],
    \label{Castillejo_En}
\end{equation}
where $a = 0.474$ and $b = 0.0515$ are dimensionless parameters, and $E_0$, $l_0$ are, respectively, the energy and length scales of the minimum in the $E(L)$ curve. The relation between the lattice length used in \cite{Castillejo1989DENSESS} and the one used in this work is $l = 2L$. Since the standard Skyrme model is relevant at lower energies, we want to identify the minimum of the crystal with the saturation point of infinite nuclear matter, $E_0 = 923.3$ MeV, $l_0^{-3} = n_0 = 0.16$ fm$^{-3}$. The change in the values with respect to the ones originally used may be seen as a redefinition for $\fpi$ and $e$. From \cref{Castillejo_En} we may obtain the pressure and the energy density using \cref{press_enden},
\begin{equation}
    p_{24}(l) = a\frac{E_0}{3l^2}\left(\frac{l_0}{l^2} - \frac{1}{l_0}\right),
    \label{pressure_Sk}
\end{equation}
which vanishes at the minimum $l_0$.

On the other hand, for high densities the sextic term is the most relevant one since the EOS yielded by this term is maximally stiff ($\rho_{6} = p$), with a speed of sound ($c^2_s = \partial p/\partial \rho$) equal to 1. The BPS Skyrme model, even with a general potential term (with no derivatives of $U$), $\lag_0 = -\mu^2\mathcal{U}$, retains this property, besides its stress-energy tensor is of the perfect fluid form from which we may easily obtain the EOS.
\begin{align}
    T^{\mu\nu}_{\rm BPS} = (p + \rho)u^{\mu}u^{\nu} + pg^{\mu\nu}, \quad u^{\mu} = \frac{B^{\mu}}{\sqrt{g_{\alpha\beta}B^{\alpha}B^{\beta}}},\\[2mm]
    p = \lambda^2\pi^4g_{\alpha\beta}B^{\alpha}B^{\beta} - \mu^2\mathcal{U}, \quad \rho = p + 2\mu^2\mathcal{U}.
\end{align}
Additionally, since the topological current $B^{0}$ may be identified with the baryon density $n_B$ in the Skyrme model, a full thermodynamical relation between $p$, $\rho$ and $n_B$ can be obtained
\begin{equation}
    p + \rho = 2\lambda^2\pi^4 n_B^2.
    \label{p_rho_n_BPS}
\end{equation}
The dependence on the Skyrme field of the potential $\mathcal{U}$ may lead to a non-barotropic EOS, but the idea to finally join both submodels is to take a constant potential which will take into account the residual contributions when we are at high densities. Then, we define the EOS at high densities as,
\begin{equation}
    \rho = \rho_6 + \rho_0 = p + \rm{const.}
\end{equation}

Besides the different properties that each submodel has, some attempts to describe NS also motivate a generalized model from the combination of these two submodels to describe nuclear matter from the saturation point $n_0$ to very high densities ($n \sim 10n_0$). Skyrmion crystals described by \cref{Castillejo_En} with the original parameters were used to calculate NS observables \cite{PhysRevD.85.123004}, however the masses turn out to be too small ($M_{\rm max} \approx 1.5M_{\odot}$) compared with the most recent measurements of NS ($M_{\rm max} \geq 2M_{\odot}$).
Later, the BPS Skyrme model was also coupled to gravity \cite{Adam:2014dqa}, using different potentials, and the NS masses were too large ($M_{\rm max} \approx 3 - 4M_{\odot}$) compared to observations. These results are themselves an evident motivation to combine the submodels and study the NS properties, since we intuitively may think that a generalized model would lead to intermedium NS masses, which are precisely in the region of observed masses ($M \sim 2 - 2.5M_{\odot}$).

In \cite{Adam:2010fg} the parameters of the BPS submodel ($\lambda^2$, $\mu^2$) were fitted to reproduce the nuclear saturation point. However from the previous discussions we use the parameters of the standard Skyrme model to fit the minimum of the crystal to these values and construct a generalized Skyrme EOS from the asymptotic values,
\begin{equation}
    \rho_{\rm Gen}(p \ll p_{PT}) = \rho_{\rm Sk}, \quad \rho_{\rm Gen}(p \gg p_{PT}) = p + \text{const.}
\end{equation}
The EOS $\rho_{\rm Sk}(p)$ may be obtained from $E_{24}(l)/l^3$ and the relation between $l$ and $p_{24}$ in \cref{pressure_Sk}. A simple construction, continuous and with a smooth transition between the two descriptions is an interpolation of the form,
\begin{align}
    \rho_{\rm Gen}(p) =& (1 - \alpha(p,p_{PT},\beta))\rho_{\rm Sk} + \alpha(p,p_{PT},\beta)(p + \rho_{\rm Sk}(p_{PT})), \label{generalized} \\[2mm]
    \notag &\alpha(p, p_{PT}, \beta) = \frac{(p/p_{PT})^{\beta}}{1 + (p/p_{PT})^{\beta}}.
\end{align}
For this interpolation, the value of $\beta$ measures how fast the transition is. Here we take the value $\beta = 0.9$ since depending on the value of $p_{PT}$ a larger value of $\beta$ would lead to a speed of sound larger than 1 in some regions inside the star. We will constrain the range of values for $p_{PT}$ comparing the resulting maximum masses with the observations. 

\begin{figure}[h!]
    \centering
    \includegraphics[scale=0.45]{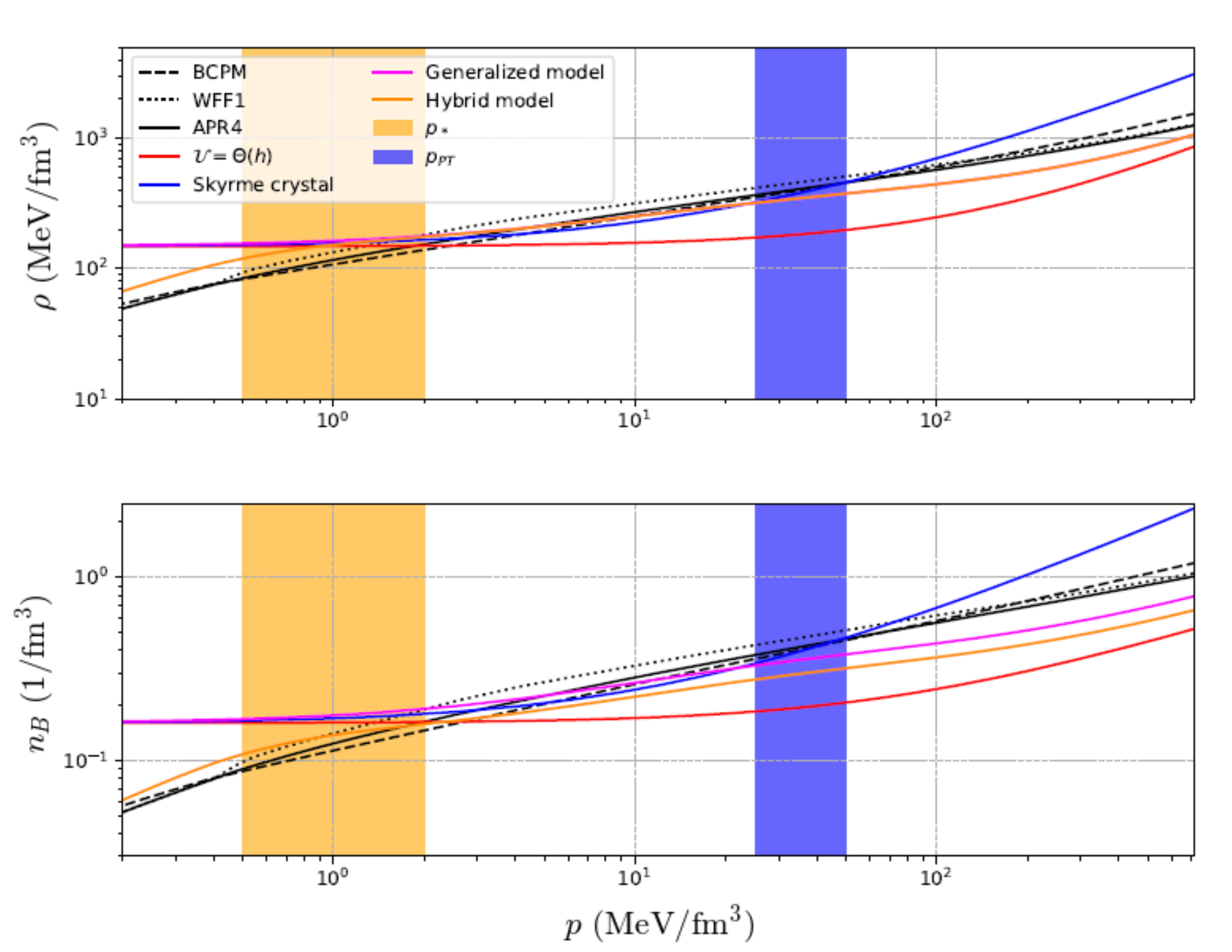}
    \caption{Energy and baryon density of the standard Skyrme crystal, the BPS model, the Generalized model with $p_{PT} = 50$ MeV/fm$^3$ and the Hybrid model, with the same $p_{PT}$ and $p_* = 2$ MeV/fm$^3$. We also show some usually considered EOS: APR4, WWF1, BCPM and we fill in orange and blue the regions of allowed values for $p_{PT}$ and $p_*$. This figure was originally published in \cite{Adam:2020yfv}}
    \label{Figure.EoS}
\end{figure}

Although we have constructed a general EOS which is able to describe matter inside NS it is still not complete. We have seen that the low density model (the Skyrme crystal) is only valid above nuclear saturation due to the presence of a minimum. For densities below saturation, finite-size and electromagnetic effects become important so we need to take them into account. The modelling of very large $B$ nuclei or the coupling to electromagnetic interactions is possible within the Skyrme model, however, how nuclear matter behaves below saturation is well known using standard methods like many-body calculations. Then for our purpose we may use one of these models to describe the low density regimes inside NS. Indeed, we will use the low-density model to describe nuclear matter until a certain pressure, denoted by $p_*$ and then we glue the generalized model \cref{generalized}, hence we are adding a crust to our generalized EOS which will mainly affect NS with low masses. We take the BCPM equation of state \cite{Vinhas_2015} as the model for low densities and perform a similar interpolation as before between this model and \cref{generalized}. For the same $\alpha(p)$ function, taking $\beta = 2$ and replacing $p_{PT}$ by $p_*$, we called the combination of the two models the Hybrid EOS,
\begin{equation}
    \rho_{\rm Hyb} = (1 - \alpha(p,p_*,2))\rho_{\rm BCPM} + \alpha(p,p_*,2)\rho_{\rm Gen}.
    \label{Interpolation}
\end{equation}
The resulting EOS, as well as some standard nuclear physics EOS, are shown in \cref{Figure.EoS}.

\begin{figure}[h!]
    \centering
    \includegraphics[scale=0.5]{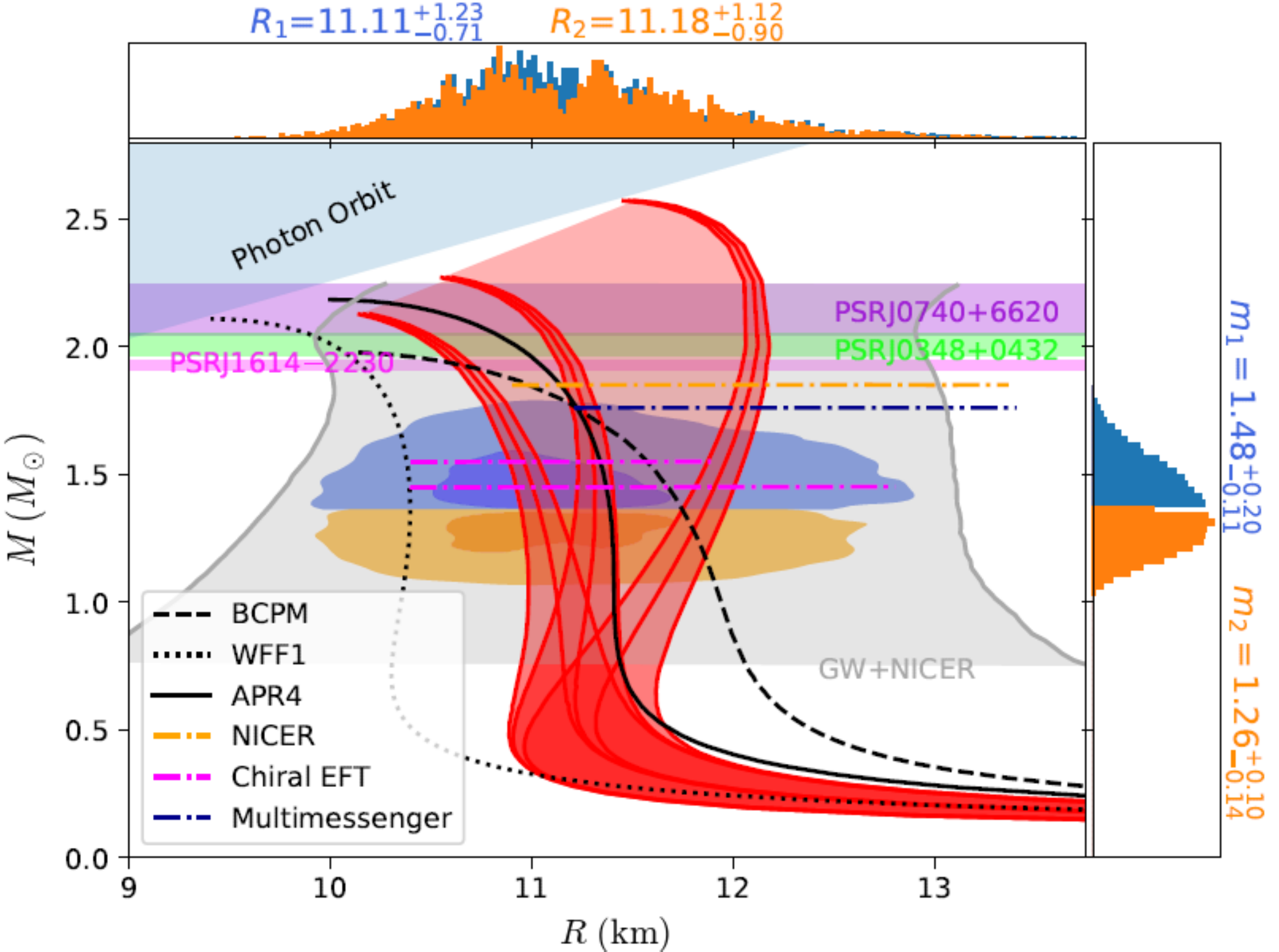}
    \caption{Mass-Radius curves of the Hybrid model with the values $p_{PT} = \{25, 40, 50\}$, $p_* = \{0.5, 1, 2\}$ MeV/fm$^3$. We compare our results with those of other EOS and constraints obtained from different analysis. This figure was originally published in \cite{Adam:2020yfv}}
    \label{Figure.MR}
\end{figure}

We sample the space of parameters $(p_{PT}, p_*)$ and restrict their values solving the TOV equations to compare the mass and radius results with some pulsar and GW measurements. We plot in \cref{Figure.MR} some representative results of the MR curves to show the accurate agreement between the generalized Skyrme EOS and observations. We use the constraints of maximum mass $M/M_{\odot}=2.16^{+0.17}_{-0.15}$ from the combined analysis of the GW170817 event and quasi-universal relations from NS in \cite{Rezzolla:2017aly}, which constraints the medium mass ($\sim 1.4M_{\odot}$) region, also the heavier pulsar observations PSR J1614 - 2230 ($1.928\pm 0.017 M_{\odot}$) \cite{Fonseca:2016tux}, PSR J0348 + 0432 ($2.01\pm 0.04 M_{\odot}$) \cite{Antoniadis:2013pzd}, PSR J0740 + 6620 ($2.14^{+0.10}_{-0.09} M_{\odot}$) \cite{NANOGrav:2019jur}, and some additional constraints from NICER, chiral EFT and multimessenger observations \cite{PhysRevD.101.123007}. As expected, it is found that $p_{PT}$ affects the maximum mass value, while $p_*$ mostly determines the radius of the medium mass range. The allowed ranges for the two transition pressures in order to satisfy the constraints are $p_{PT} \in [25, 50]$ MeV/fm$^3$ and $p_* \in [0.5, 2]$ MeV/fm$^3$. We even allowed for maximal masses $\sim 2.7M_{\odot}$ much larger than the observations, mainly motivated by the \cite{Abbott:2020uma} event in which a black hole and a secondary compact object falls in the \text{mass gap} ($2.5-2.7 M_{\odot}$) \cite{LIGOScientific:2020zkf}, with the possibility of being the heaviest stable NS ever observed. Besides we want to remark that we may easily accommodate these very large values of NS masses keeping reasonable values of the radii in the medium mass region, which is not an easy task in general for other nuclear models.

\begin{figure}[h!]
    \centering
    \includegraphics[scale=0.4]{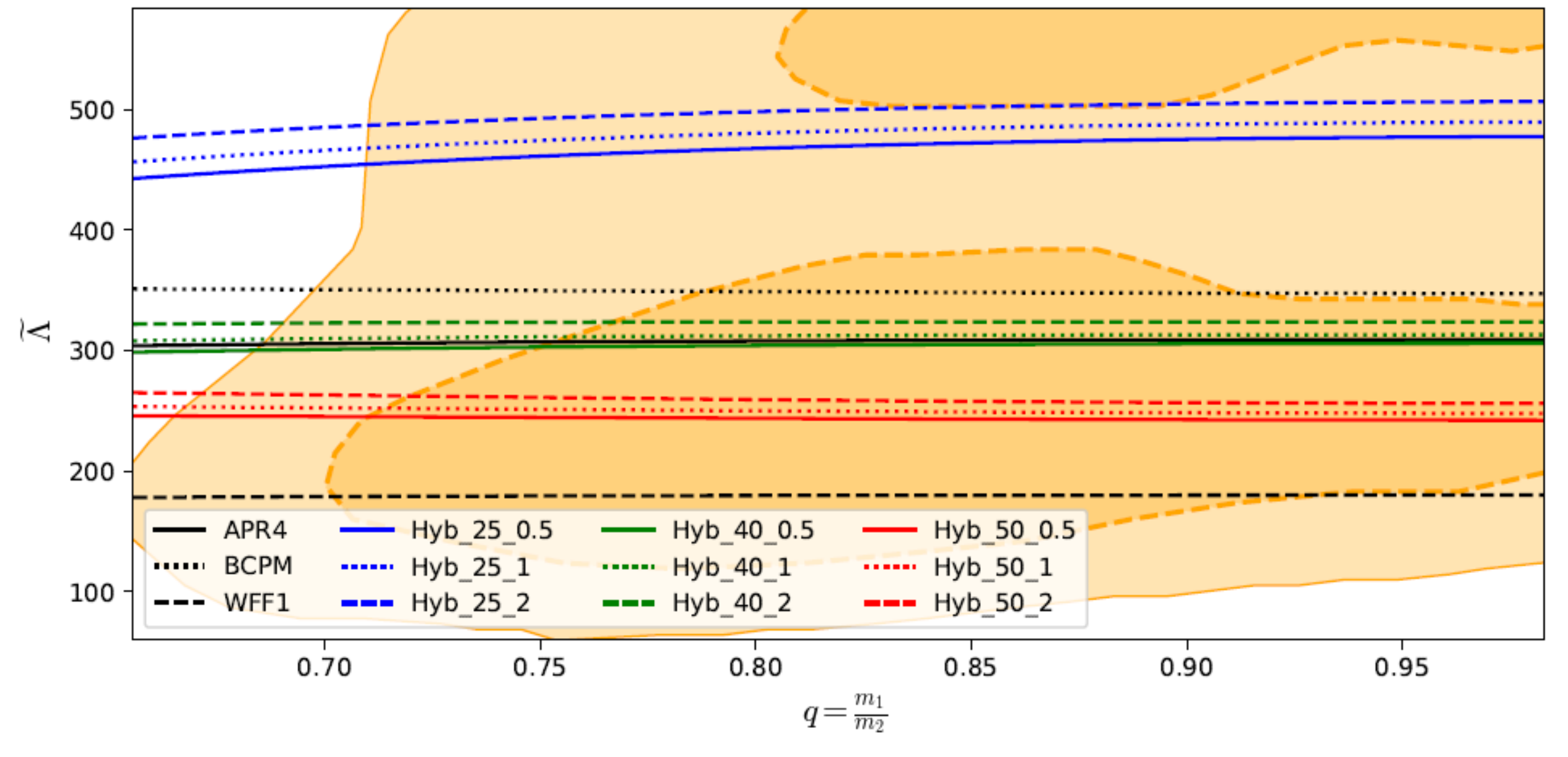}
    \caption{The effective tidal deformability $\widetilde{\Lambda}$ of the Hybrid model against the mass ratio. The orange regions show the PDFs obtained from \cite{LIGOScientific:2018hze}, and we adopt the notation Hyb\_$p_{PT}$\_$p_*$ for the curves. This figure was originally published in \cite{Adam:2020yfv}.}
    \label{Figure.LambdaTilde}
\end{figure}

It is also possible to extend the calculations and solve an additional equation to obtain the tidal deformability for a non-rotating NS \cite{Adam:2020yfv}. The importance of this magnitude is that it can be extracted from the waveform of the early phase inspiralling coalescence of two NS. Indeed what can be measured is a mass weighted averaged of the two individual deformabilities which is called effective tidal deformability $\widetilde{\Lambda}$. Similarly, the individual masses are not directly measured as well, but a magnitude called the \textit{chirp mass}, $M_c = m_1q^{3/5}/(1+q)^{1/5}$, with $q = m_1/m_2$ the ratio between the two masses, has been constrained at the $90\%$ confidence level in the GW170817 event \cite{LIGOScientific:2017vwq}. In \cref{Figure.LambdaTilde} we show the effective tidal deformability obtained from our Hybrid model with the estimations extracted from \cite{LIGOScientific:2018hze} of the effective tidal deformability for the measured values $M_c = 1.188^{+0.004}_{-0.002}$ and $q = 0.7 - 1$.
The estimated value of $\widetilde{\Lambda}$ from the GW170817 event is rather low (smaller than 800) and it is a new source to eliminate some EOS which may satisfy the mass-radius constraints but do not have a correct value of the deformability.

\subsection{Skyrme crystal EOS}
The results from the last section provide a strong motivation to use the full crystal solutions within the generalized Skyrme model, studied in the first section, to compute NS observables. Besides being solutions of the generalized Skyrme model, we have seen how the crystals allow to include isospin effects which furthermore led to the addition of leptons, leading to a much more realistic description of nuclear matter at high densities. However, these solutions and their properties depend on the values of the parameters that appear in the model, which we have set to some standard values. Then, in order to be acceptable candidates to describe realistic nuclear matter they must reproduce observables from nuclear physics and NS. In this section we will study how to fit the parameters to reproduce different nuclear observables using Skyrme crystals and, finally, we will extract the most important NS observables and compare them with the most fiducial measurements.

\subsubsection{Scan of the parameters}
A remarkable property of the standard Skyrme model without potential term (namely, the quadratic and quartic terms only) is that, with a suitable choice of units, one can factor out all dimensionful constants from the energy functional, so that the constants remaining inside it are just dimensionless numbers. This can be seen, for example, in \cref{Fit}.
As a consequence, one can just forget about the numerical values of the coupling constants and numerically find the different crystalline solutions at different unit cell lengths. Once the relation between the (adimensional) energy and length $E(L)$ is found, the values of the coupling constants can be adjusted \emph{a posteriori} in order to fit whatever observable we are interested in.
Unfortunately, the addition of the sextic and mass potential terms to the energy functional spoils this property, as there is no choice of units that allows to factor out all coupling constants in the energy functional. This means that, in order to be able to obtain solutions, one needs first to give specific values to (some of) the coupling constants appearing in the problem. However, this is problematic if one wants to fit the values of energy and density to some physical values. For example we want to identify the energy and the density at the minimum of the crystal ($L_0$) with the nuclear saturation point, however, one cannot know the value of $L_0$ without performing the numerical simulations, but in order to do so you need to fix the values of the parameters.

Hence, the fitting of the skyrmion crystal parameters in the generalized model to values at nuclear saturation is, in principle, a very difficult problem that needs to be solved iteratively until a self consistent solution is achieved. Given the computational cost of simulating a single unit cell at a given length, following this naive approach would make it almost impossible to realize a significant scan of parameters in a reasonable time.

\begin{figure}
    \centering
    \hspace*{-0.7cm}
    \includegraphics[scale=0.35]{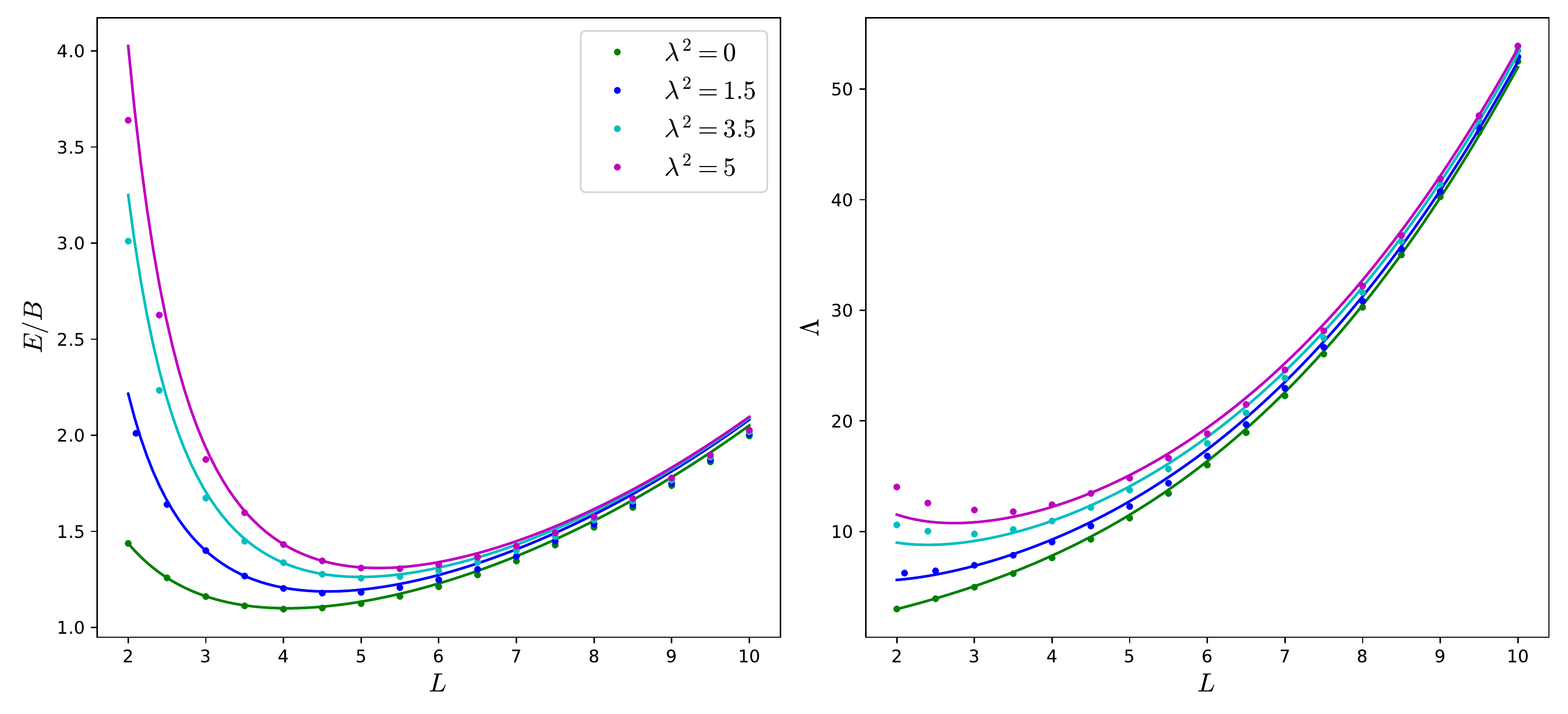}
    \caption{Energy and isospin moment of inertia of a skyrmion crystal as obtained via the full numerical minimization (dots) and the perfect scaling fit (solid curves), for the models with and without sextic term.}
    \label{fig:perfect scaling}
\end{figure}

However, from the results of our numerical simulations we have observed that the FCC$_+$ crystal displays an almost \emph{perfect scaling} property \cite{Adam:2021gbm} with the unit cell length. This property is stronger than the fact that each $E(L)$ curve may be fitted by \cref{Fit}, it means that each of the terms in the energy functional scales with $L$ independently as $E_i \propto L^{-i+3}$, where $i$ is again the number of spatial derivatives appearing in that particular term. The values of the perfect scaling constants, which we label by $K_i$, are then universal in the sense that they will not change for different values of the parameters. 
This perfect scaling property is a characteristic of the field configuration, and not only of its energy. For example, in 
\cref{fig:perfect scaling} it is observed that also the isospin moment of inertia of the crystal unit cell displays a sufficiently well perfect scaling. This is important in our analysis since we also want to fit the values of the symmetry energy and its derivatives.
Although the scaling is not perfect, in general the biggest deviations from the full numerical values of energy start far from the minimum, at which the perfect scaling fit is most precise.

Thus, we take advantage of this property in order to fit the magnitudes obtained from generalized Skyrme crystals to their physical 
values (up to a certain error). We do so following an iterative process based on five main steps:
\begin{enumerate}
    \item Due to the perfect scaling property, the (adimensional) energy and isospin moment of inertia approximately satisfy the following expressions,
    \begin{align}
        E(L, c_6, c_0) &= K_2 L + \frac{K_4}{L} + c_6\frac{K_6}{L^3} + c_0K_0 L^3,
        \label{e_PS}\\[2mm]
        \Lambda(L, c_6, c_0) &= \Lambda_2 L^3 + \Lambda_4 L + c_6\frac{\Lambda_6}{L},
    \end{align}
    and the values of  $K_i$ and $\Lambda_i$ are ''universal'', i.e., they do not depend on the parameters or on $L$. Hence, these values are obtained from the contribution of each term independently, for only one choice of the parameters and for a single value of the unit cell length $L$. For simplicity, we set $c_0$ and $c_6$ equal to one. The values of the universal scaling constants are shown in \cref{tab:scaling constants}.
    \begin{table*}[h!]
        \centering
        \begin{tabular}{|c|c|c|c|c|}
        \hline
             n & 0 & 2 & 4 & 6  \\
             \hline
             $K_{\rm n}$& 0.034 & 0.466 & 9.617 & 4.329\\ \hline
            $\Lambda_{\rm n}$& -- & 0.038 & 1.393 & 0.883\\ \hline
            \end{tabular}
        \caption{Perfect scaling parametrization constants}
        \label{tab:scaling constants}
    \end{table*}
    \item We fix the energy scale $E_s$ to a desired value in MeV. This is equivalent to fixing one of the three free parameters of the model, for instance, $f_{\pi}$.
    \item Then, we calculate $L_0$ by minimizing \eqref{e_PS} and the values of $E_0, n_0, S_0$ and $L_{\rm sym}$ for different pairs of values $(e,\lambda^2)$.
    \item When we find a set of parameters $(f_{\pi}, e, \lambda^2)$ that fits the nuclear magnitudes within their respective errors of at most 15\% then we calculate the corresponding EOS and solve the TOV system to obtain the mass-radius curve.
    \item Finally, we accept the sets of values that satisfy the constraints, $M_{\rm max} \geq 2 M_{\odot}$ and $R_{1.4M_{\odot}} \leq 12.5$ km. These constraints are motivated from pulsar measurements \cite{Antoniadis:2013pzd,Riley:2021pdl,Demorest:2010bx}. We find that there is more than one set of parameters, so there is a residual freedom in the choice of these values that satisfy the nuclear physics magnitudes at saturation and NS observables.
\end{enumerate}

\begin{figure}
    \centering
    \includegraphics[scale=0.5]{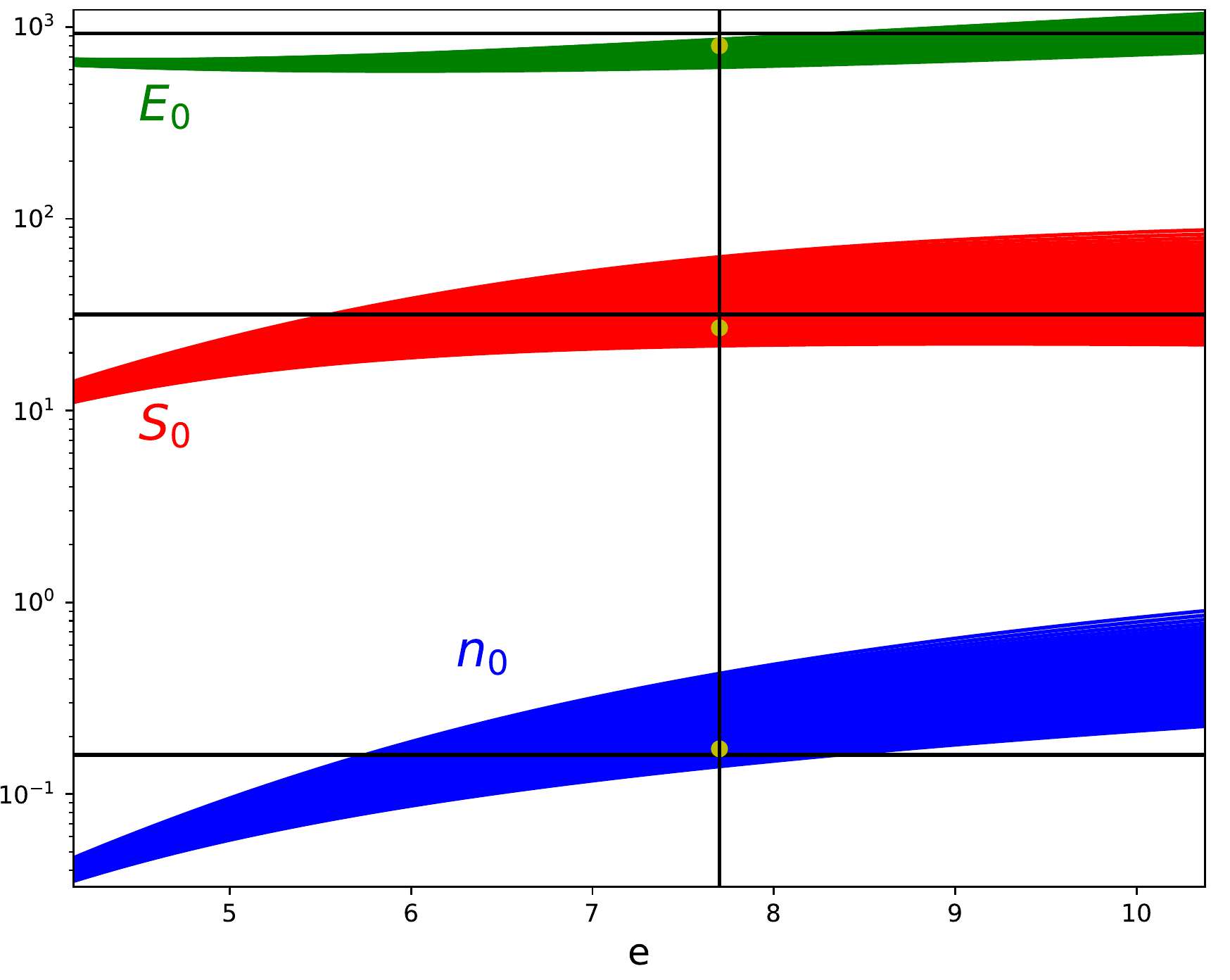}
    \caption{Scan of the parameters $(e, \lambda^2)$ for a fixed value of $E_s$. Each coloured line corresponds to a fixed value of $\lambda^2$ and the black horizontal lines are the experimental values of $E_0, n_0$ and $S_0$. The yellow dots are the optimal values and they lie in the black vertical line which is the corresponding value of $e$.}
    \label{fig:my_label}
\end{figure}

We plot in \cref{fig:my_label} the result of a scan for a fixed value of the energy scale, for other values of $E_s$ new sets of parameters were found.
\begin{figure*}
    \centering
    \hspace{-0.7cm}
    \includegraphics[scale=0.5]{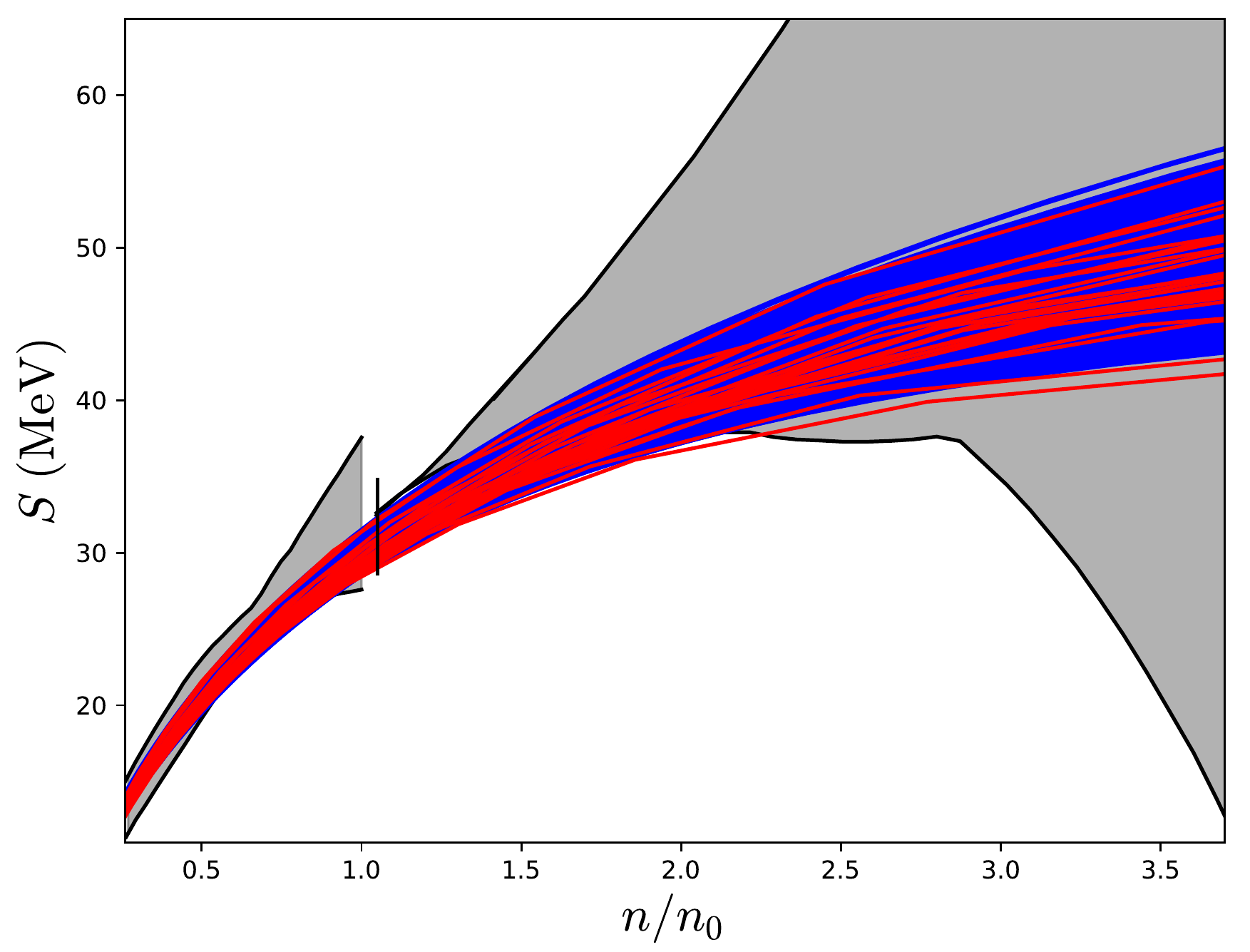}
    \caption{Symmetry energy of Skyrme crystals as a function of the density. We show 23 different curves from the scanned values. The shaded regions constrains the symmetry energy at sub-saturation \cite{Danielewicz:2013upa} and supra-saturation \cite{Li:2021thg} densities.}
    \label{fig:Symmetry_E}
\end{figure*}

The scan of parameters underscores the motivation for the generalized Skyrme model. The nuclear physics magnitudes are better fitted for very low or even null values of $\lambda^2$ since the sextic term reduces the value of $S_0$. However, those sets of parameters are not accepted since they do not satisfy the maximum mass requirement. This reflects the importance of the sextic term in the extension of the Skyrme model to very high densities as inside NS.

In \cref{fig:Symmetry_E} we plot 560 symmetry energy curves obtained from a first quick scan in blue, and in red we plot 23 representative cases from the larger set which have been fully minimized. We also represent at densities larger than $n_0$ some restrictions obtained from the most recent constraints of the analysis of neutron star observations, and at densities smaller than the saturation point which are more restrictive.

\begin{figure*}
    \centering
    \hspace{-1cm}
    \includegraphics[scale=0.4]{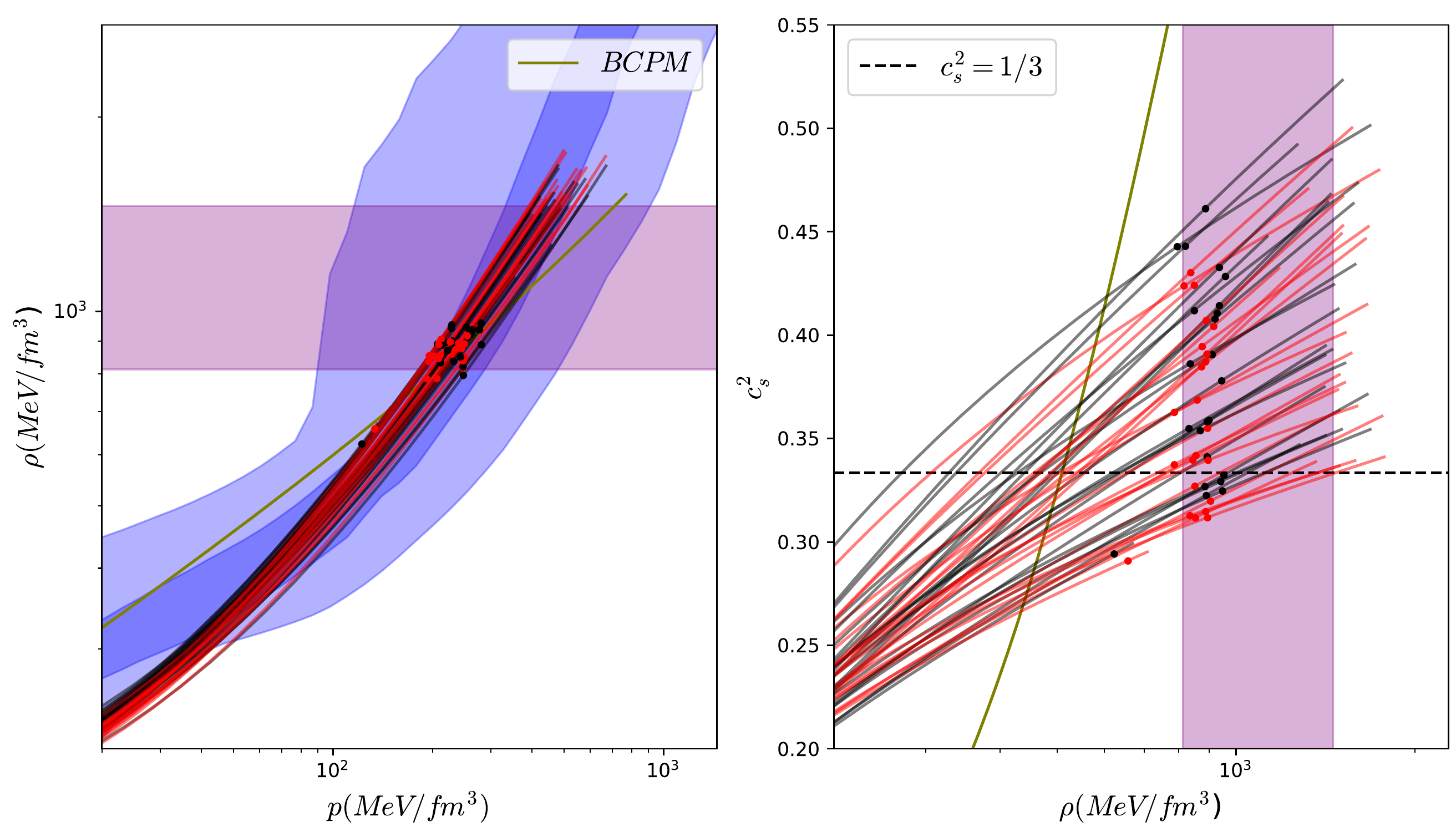}
    \caption{The EOS for the same 23 values shown before. In black we plot the resulting EOS without isospin effects, whilst in red we consider $npe\mu$ matter. We find a good agreement between our EOS and the shaded regions obtained from the analysis in \cite{Altiparmak:2022bke} at high densities. The purple region is an estimation for the range of the maximum central density inside NS, and the dots represent the maximum central energy densities in our models.}
    \label{Fig.EoS_Constraints}
\end{figure*}

We have obtained the EOS from these 23 sets of parameters and compare them with some constraints obtained from a recent analysis \cite{Altiparmak:2022bke}. In that work they build a huge number of physically well motivated EOS and compare the resulting NS with pulsars and GW observations. They conclude that the conformal limit in the speed of sound ($c^2_s = 1/3$) is expected to be surpassed inside NS. In \cref{Fig.EoS_Constraints} we show the EOS obtained from our analysis and a good agreement is found with their results. The majority of our EOS exceeds the conformal bound too, and all of them lie inside the constrained region in the $(\rho, p)$ diagram. We have cut the low density region in \cref{Fig.EoS_Constraints} due to the absence of a crust in our EOS. However it is remarkable how the Skyrme model correctly describes the high density regime, which corresponds to the core of the NS and hence the main responsible for the mass of the star. Also the BCPM \cite{Vinhas_2015} EOS is represented as an accepted candidate to compare with in the diagram.

In a more extensive analysis of the parameters, we found $\sim 10,000$ accepted sets of parameters. In order to obtain these values we made the scan with the steps: $\Delta E_s = 5$ MeV, $\Delta e = 0.01$, $\Delta \lambda^2 = 0.01$ MeV fm$^3$. As briefly mentioned before, the constraints on the symmetry energy yield rather stringent upper bounds on the sextic term coupling constant, we find that $\lambda^2 \lesssim 3.4$ MeV fm$^3$. Nevertheless, we remark that a lower bound for this constant can also be obtained from the maximum mass requirement of neutron star EOS \cite{Adam:2020aza}. In this analysis we have found a lower bound of $\lambda^2 \gtrsim 0.29$ MeV fm$^3$.

\begin{figure}
    \centering
    \includegraphics[scale=0.5]{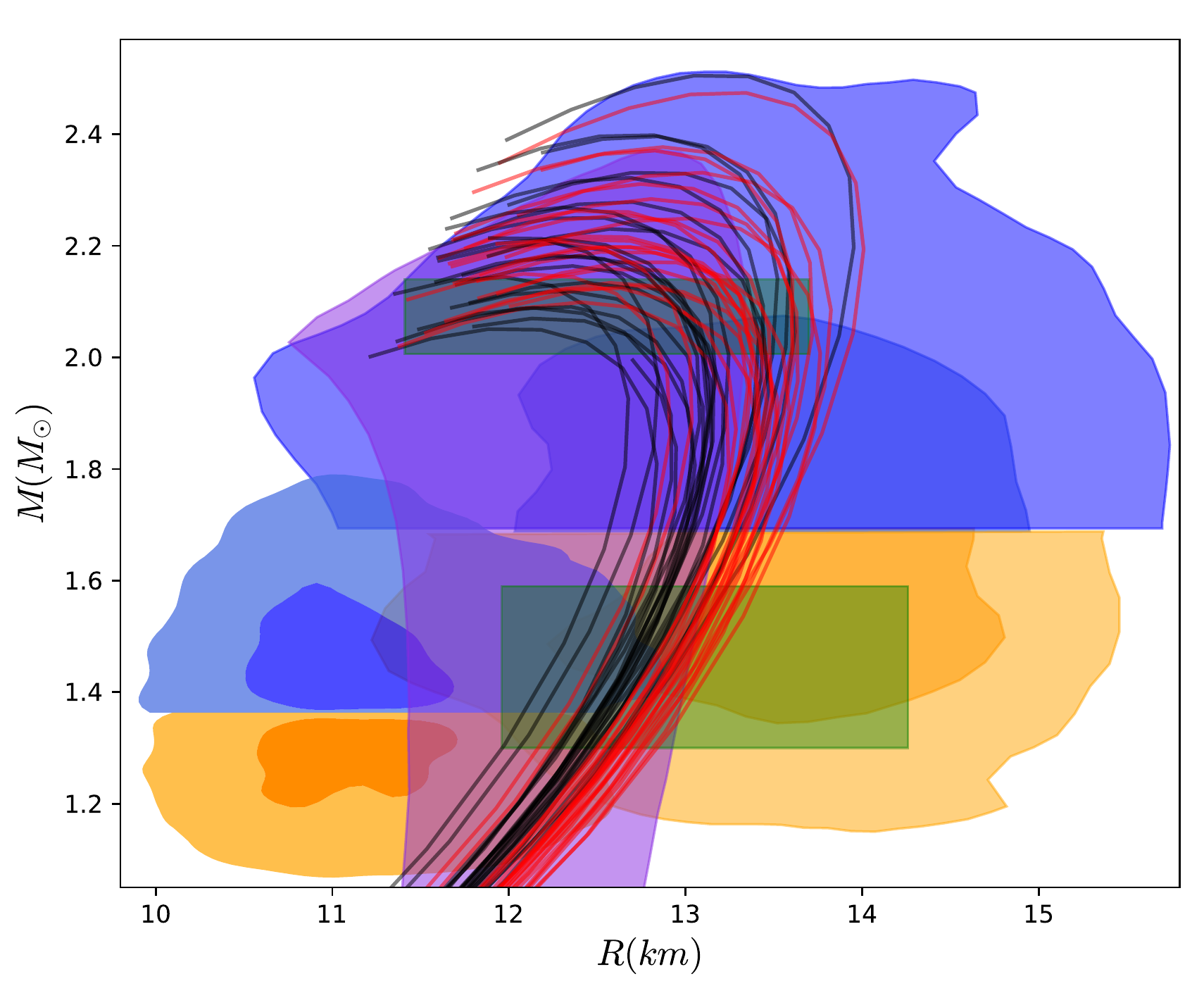}
    \caption{MR curves for the 23 representative sets of parameters considered. The colours of the lines represent the same as in \cref{Fig.EoS_Constraints}. The shaded regions correspond to GW (blue and orange) and pulsar (green) constraints.}
    \label{fig:MRcurves}
\end{figure}

We solved the system for the same 23 cases and in \cref{fig:MRcurves} we plot the MR curves, again with and without isospin effects for each case. The main conclusion is that the isospin always increases the radii of the stars. On the other hand, the isospin increases the masses of the stars with $M \lesssim 2.3M_{\odot}$, for larger values the masses are reduced. This effect is also visible in \cref{Fig.EoS_Constraints}, where the red curves lie below the black lines at lower densities, hence the reds are stiffer, whilst for very high densities the situation is slightly the opposite. Another keypoint is that all the sets of parameters obtained in this analysis allow to have a wide range of maximum masses, $M_{\max} \sim 2-2.5 M_{\odot}$. This was an important feature in the NS obtained in the last section, and it is still possible using the fully minimized Skyrme crystals. This is of great important for the Skyrme model since it would be able to describe possible high-mass measurements like \cite{Abbott:2020uma}. In addition, despite the difference in the maximum masses, the radii of the stars do not change as much when choosing some parameters or others, $R_{1.4M_{\odot}} \sim 12-13$ km. However a final comment about the radii of the NS requires the presence of a crust, since it will affect the radii of low mass NS.

As can be seen all the black lines satisfy (in good approximation) our $M_{\rm max}$ and $R_{1.4M_{\odot}}$ restrictions although they were imposed on the mass-radius curves resulting from the PS approximation. We have checked that the mass-radius curves obtained via the PS approximation are indeed quite similar to those of the fully minimized results for these 23 cases, so it confirms the PS approximation as a powerful and accurate tool for skyrmion crystals.

We also plot in \cref{fig:MRcurves} the most likely
mass-radius relations for the NS corresponding to GW170817 \cite{Abbott_2017} and GW190425 \cite{Abbott:2020uma} events. The green regions represent the estimations for the mass and radius values of J0030+0451 (bottom) \cite{Miller:2019cac}, and a more recent analysis of the PSR J0740+6620 mass and radius from NICER (top) \cite{Riley:2021pdl}. The purple region constraints the mass-radius curves from the statistical analysis done in \cite{Altiparmak:2022bke}, besides they also give an estimation for the maximum central energy density that a NS may support. We also do the comparison in \cref{Fig.EoS_Constraints} between the range of values that they obtain (purple region) and our values (dots).

The greatest difference between the NS obtained using the interpolation between the submodels in the last section and those obtained from the minimization of the generalized Skyrme model using crystal solutions is found in the radii. Although we still have freedom to reach very high masses $\sim 2.5M_{\odot}$, the radius increases when we consider the full model. The low-mass region ($\sim 1.4M_{\odot}$) in the MR curves of \cref{Figure.MR} and \cref{fig:MRcurves} differ due to the presence of a crust in the first case. We do not include a crust in the EOS shown in this section since it would be more interesting to consider an EOS which already has a crust entirely obtained from the Skyrme model. This is still an open problem due to the behaviour of the Skyrme crystals at low densities, but the study of the new lattices presented in the first section may lead to the correct description of the full EOS within the Skyrme model. Nonetheless we have seen that the inclusion of a crust via the simple construction \cref{Interpolation} increases the radius of the NS around 1 km.

On the other hand, the $2.5M_{\odot}$ NS radius in \cref{Figure.MR} is around 11.5 km, while the radius for the same mass in \cref{fig:MRcurves} reaches 13 km. These high-mass NS are hardly affected by the presence of a crust, so the numerical simulation of the Skyrme crystal leads to a stiffer EOS. Nevertheless, the high-mass region of the last plot may be sharply improved with the inclusion of strangeness degrees of freedom in the system. Taking into account this effect provides even more realistic EOS, besides it is known that it will decrease the maximal mass as well as the radius, leading to a softer EOS at high densities.
\subsection{Including kaons}

\begin{figure}
    \centering
    \includegraphics[scale=0.32]{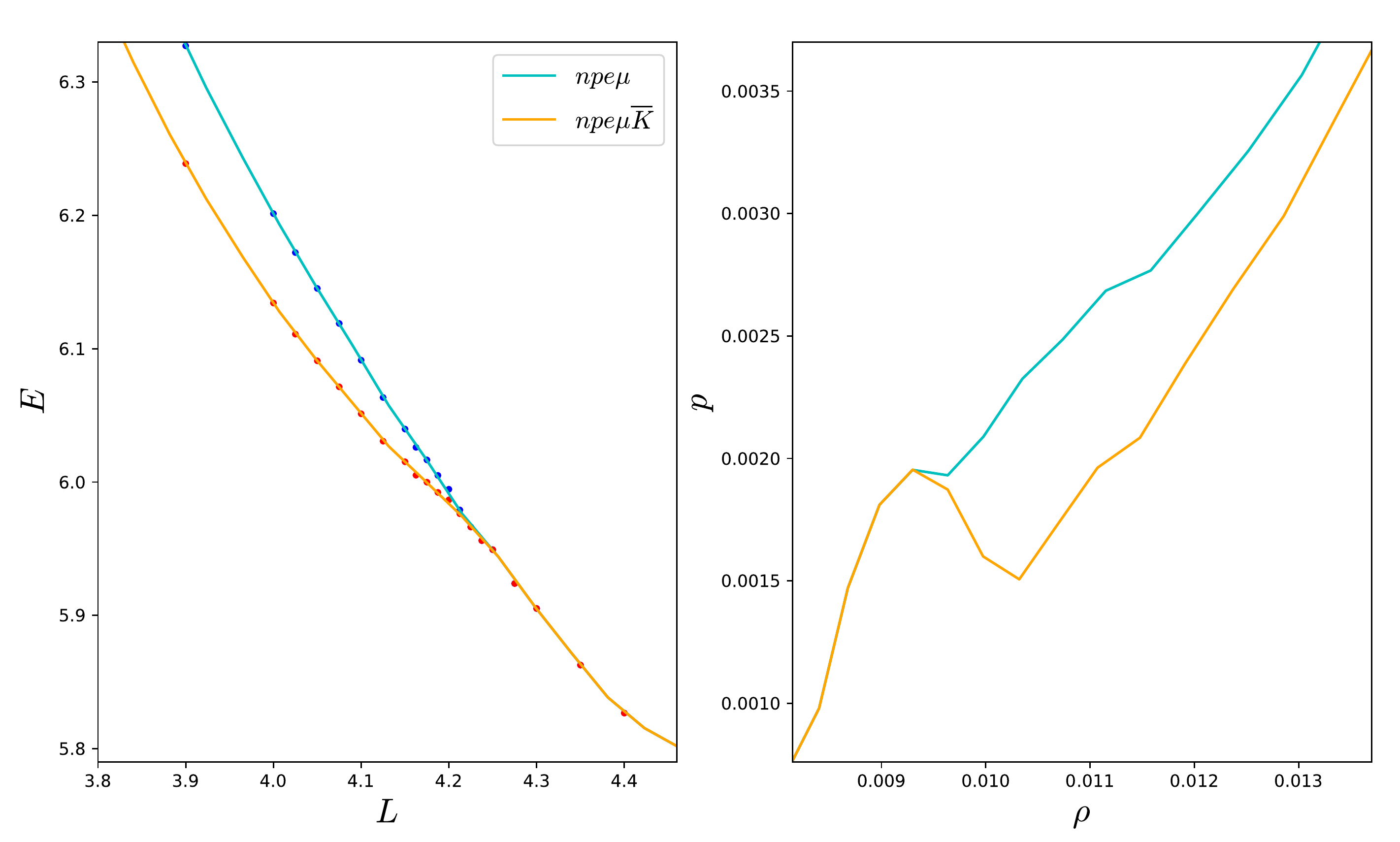}
    \caption{Left plot: energy against the side length of the crystal, calculated with more points near the condensation values for both branches and their interpolations. Right plot: pressure against the energy density from which we conclude that there is a first order phase transition. All magnitudes are shown in Skyrme units.}
    \label{Fig.FOT}
\end{figure}

The main ingredient to obtain the EOS for the Skyrme crystal is the energy dependence on the unit cell size. We have shown the results in the $npe\mu$ matter case above. However, once we include kaons, the change in the energy curve leads to a first order phase transition, see \cref{Fig.FOT}. This can be clearly seen in the right plot of \cref{Fig.FOT}, where we show the resulting EOS. There exists a region of diminishing pressure which implies a first order phase transition which, in principle,  can be treated by a Maxwell construction or a Gibbs construction.

The Maxwell construction (MC) is the standard procedure to obtain a physical EOS when a first order transition is present. Indeed, the MC has already been used for Skyrme crystals to describe the transition between crystals with different symmetries \cite{Adam:2021gbm}. The MC is based on a mixed phase of constant pressure which connects the two solutions. However, the MC is only correct if there exists a single conserved charge (in this case, the baryon number) for which the associated chemical potential is enforced to be common for both phases in the mixed phase \cite{Glendenning:1992vb}. If, instead, an additional charge is conserved, like the electric charge in the case of $npe\mu$ matter, the Gibbs conditions for the phase equilibrium,
\begin{equation}
    p^{\rm I} = p^{\rm II}, \hspace{2mm} \mu^{\rm I}_i = \mu^{\rm II}_i, \hspace{1.5mm} i = B, q
    \label{Gibbs_conds}
\end{equation}
cannot be both satisfied in a standard MC. In the last expression $\mu_B$ and $\mu_q$ represent the chemical potentials associated to the conserved baryon and electric charges, respectively. Instead, one should perform a Gibbs construction (GC) \cite{Glendenning:1992vb,Glendenning:1997ak}. Indeed, the GC has also been proven useful in the context of a hadron-to-quark phase transition inside NS \cite{Bhattacharyya:2009fg}.

We may write the chemical potential of each particle species as a linear combination of the chemical potentials associated to the conserved charges of our system:
\begin{equation}
    \mu_i = B_i\mu_B + q_i \mu_q,
\end{equation}
where $B_i$ and $q_i$ are the baryon number and electric charge of the particle species $i$.
Then we might identify the baryon and electric charge chemical potentials with the neutron and electron chemical potentials respectively.
The main difference between MC and GC is that, in the mixed phase, the first one imposes charge neutrality independently for both phases, whereas in the GC it is only imposed globally in the mixed phase. Considering a volume fraction $\chi$ of the kaon condensed phase, charge neutrality is imposed in the GC as:
\begin{equation}
    n^{MP}_q = (1-\chi)n^{\rm I}_q + \chi n^{\rm II}_q = 0.
    \label{Charge_neutral_MP}
\end{equation}
The mixed phase in the GC is calculated by identifying first the contributions to the pressure and charge densities in each phase separately. Then we have to solve the system of equations composed by \cref{Gibbs_conds,Charge_neutral_MP,cond2}. 


\begin{figure}
    \centering
    \includegraphics[scale=0.35]{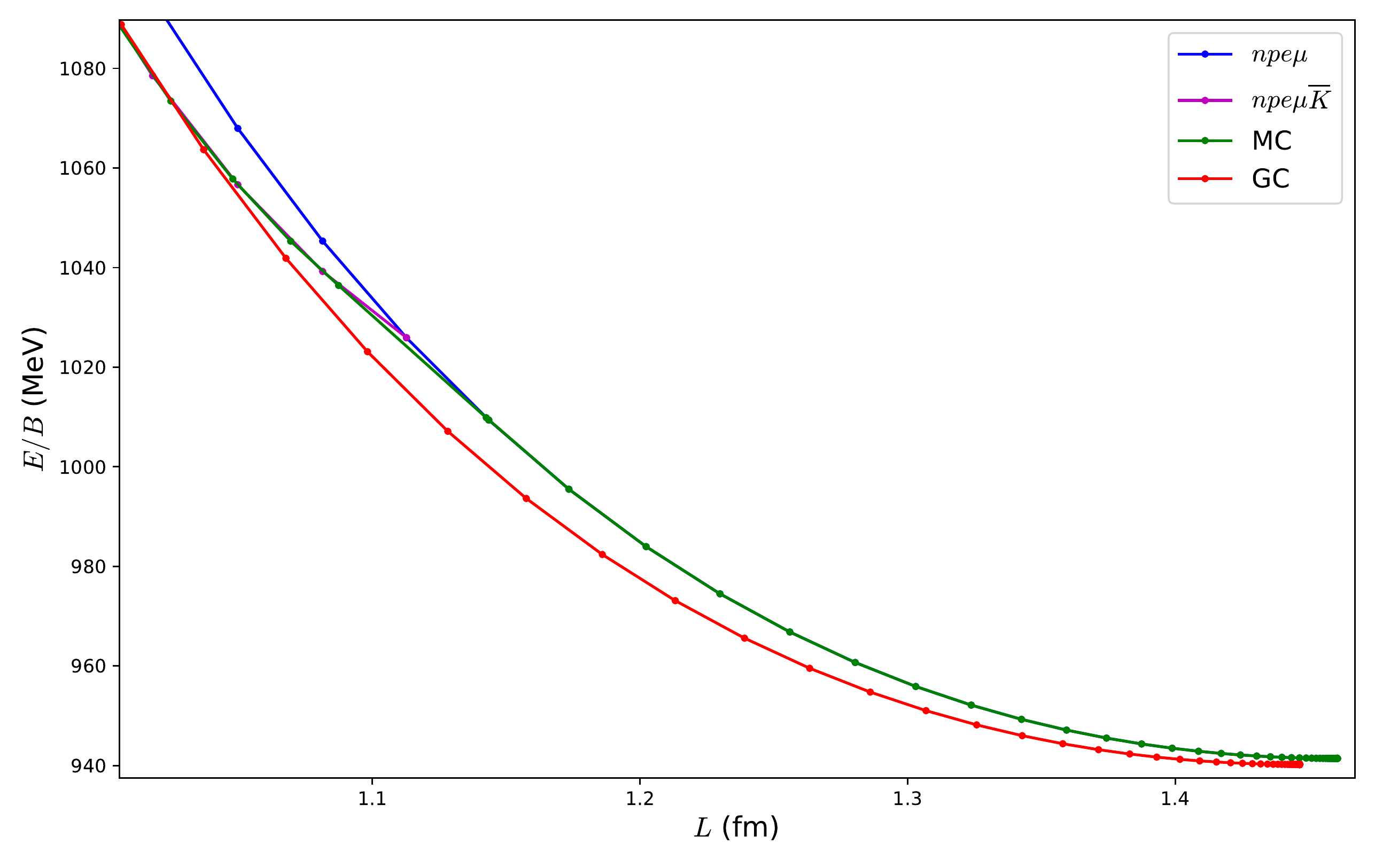}
    \caption{$E(L)$ curves for the two phases. The different slopes at the point of phase separation indicate a first-order phase transition. We also show the curves resulting from a Maxwell construction (MC) and a Gibbs construction (GC).}
    \label{Comparison_MCGC}
\end{figure}

\begin{figure}[htb!]
    \centering
    \includegraphics[scale=0.37]{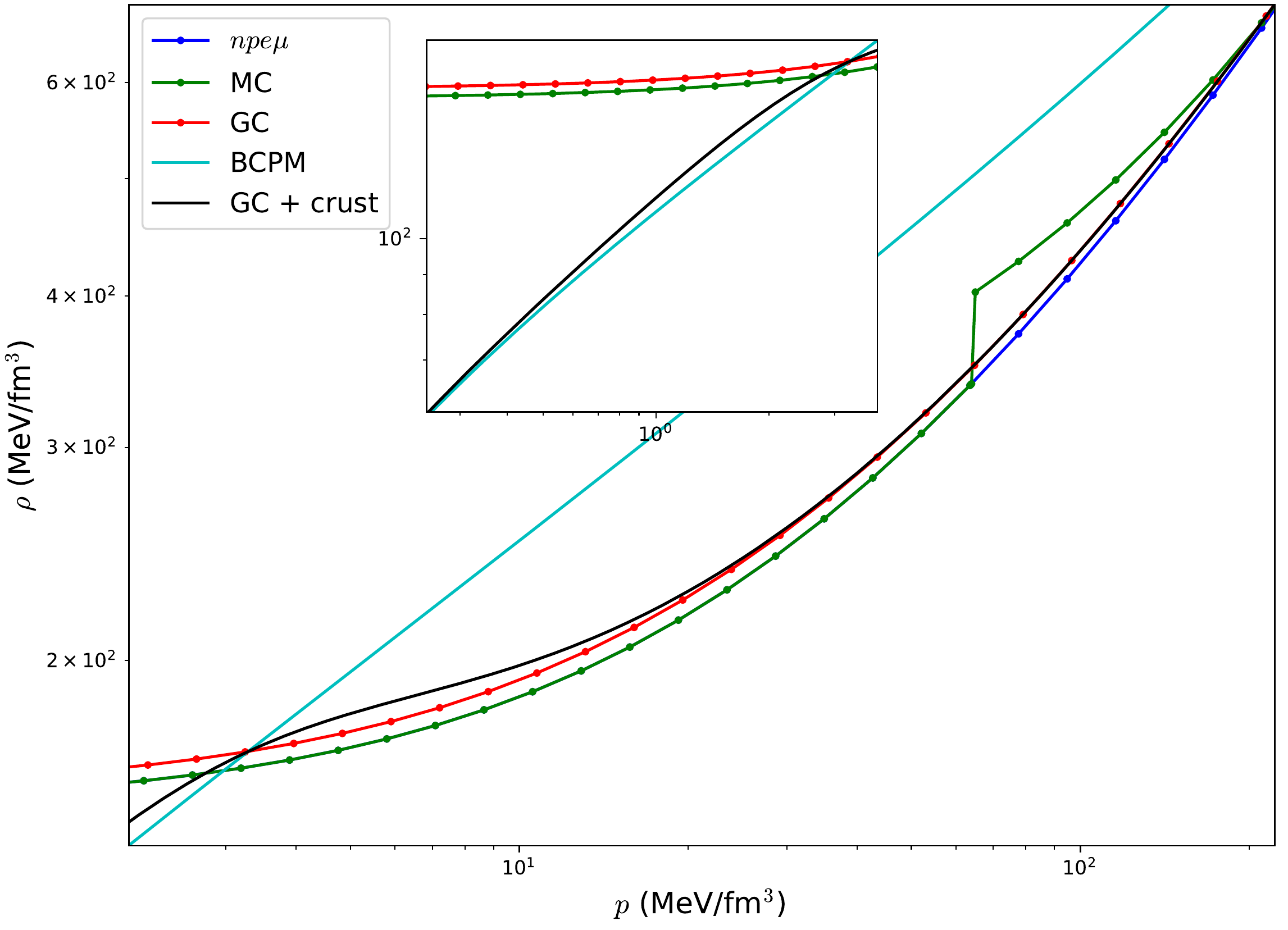}
    \caption{EOS for the three different cases that we have constructed. The jump in the MC due to the first order transition and the different behaviour of the GC at low densities are clearly visible. We also show the standard nuclear physics EOS of \cite{Vinhas_2015} (BCPM) and a hybrid EoS obtained by joining the BCPM EOS at low pressure with the GC EOS at high pressure. This figure was originally published in \cite{Adam:2022cbs}.}
    \label{EOS}
\end{figure}

We show our results in \cref{Comparison_MCGC} and in \cref{EOS}, using in all figures the parameter set 1 from \cref{tab:Condensation}.
We find that the mixed phase of the GC and, therefore, the density at which the kaon condensate sets in, is smaller density than the value obtained in \cref{tab:Condensation}. This is also found in \cite{Glendenning:1997ak}, for which the GC mixed phase extends to a larger region than the one obtained from the MC, because the mixed phase in the GC no longer is for constant pressure. In our case,  even the minimum of $E(L)$ is shifted to slightly lower values, see \cref{Comparison_MCGC} and, hence, the use of the GC affects the low density regime of the EOS, as can be seen in \cref{EOS}. 

\begin{figure}[htb!]
    \centering
    \includegraphics[scale=0.45]{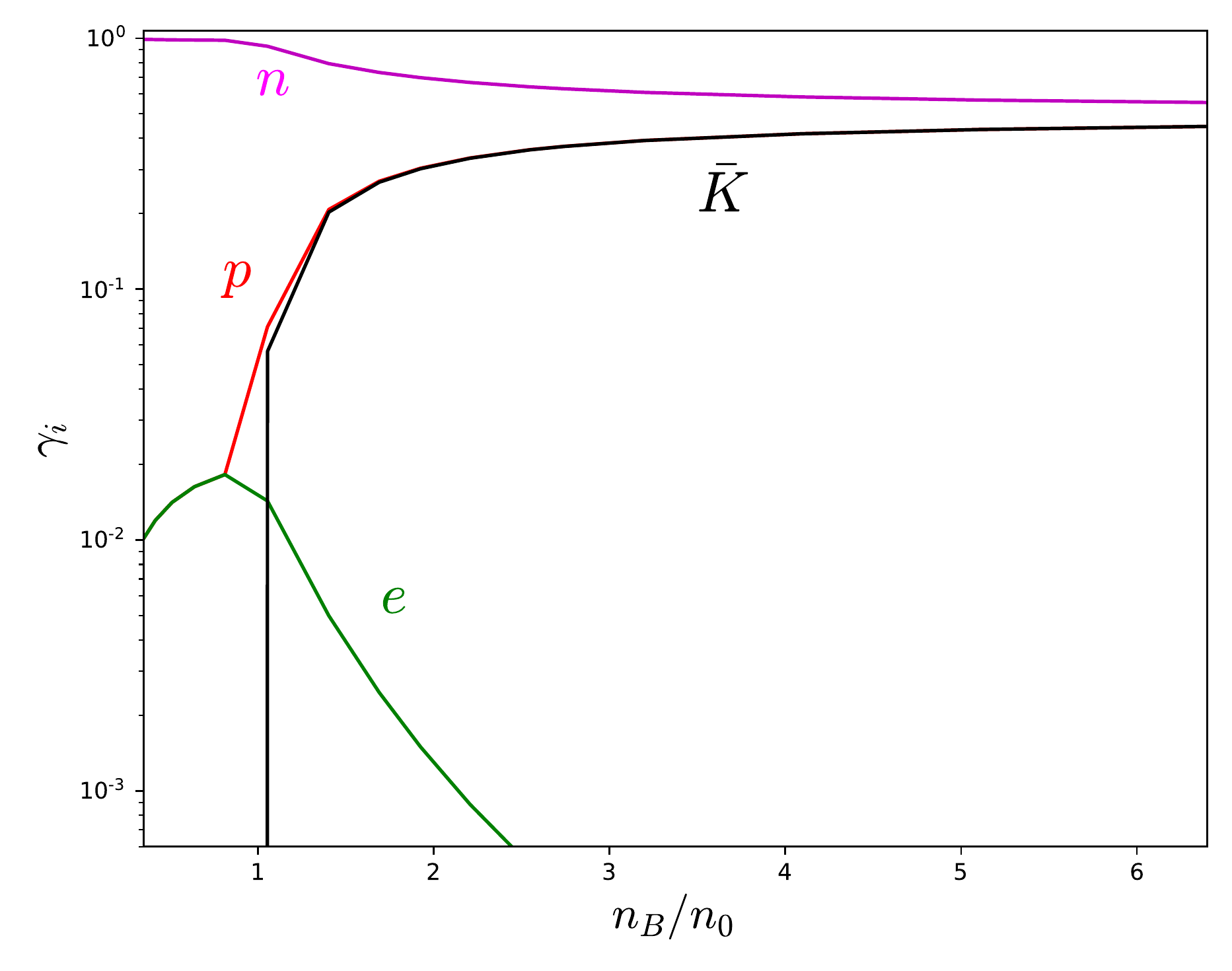}
    \caption{Particle fractions for the Gibbs Construction of the kaon condensation phase transition. This figure was originally published in \cite{Adam:2022cbs}.}
    \label{FractionsGC}
\end{figure}

We may also calculate the particle fractions in the mixed phase of the GC using an expression equivalent to \cref{Charge_neutral_MP} for each particle. We show the new particle fractions in \cref{FractionsGC}. Besides, during the mixed phase, we find that there are more protons in the second phase than in the first one. However, the presence of more kaons than protons in the second phase results in a partial negative charge density. That negative charge is compensated by the overall positive charge density of the first phase. In both phases the number of electrons is much less than that of protons and kaons.

\begin{figure*}[h]
   \centering
    \includegraphics[scale=0.5]{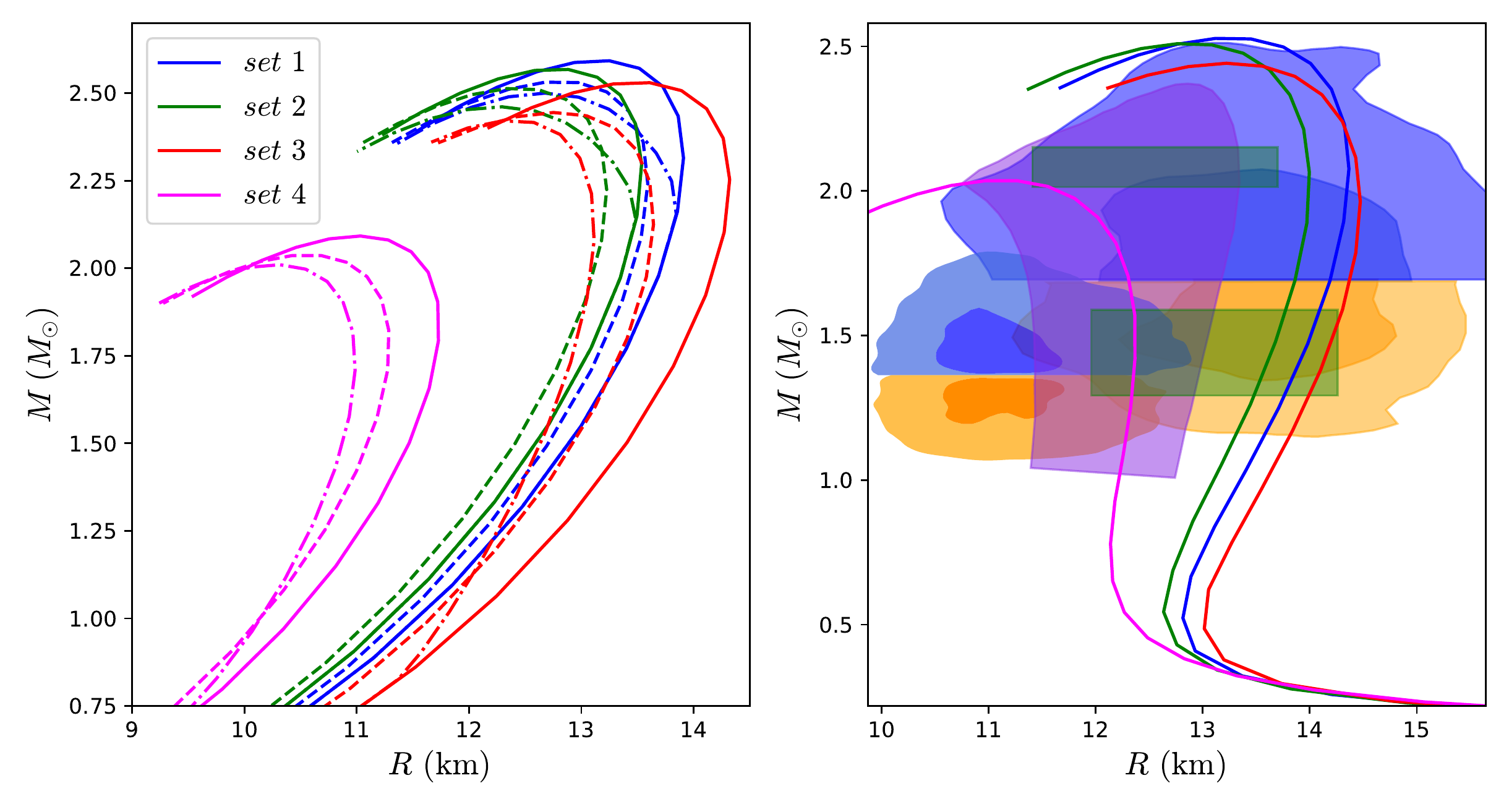}
    \caption{Mass-Radius curves of NS with a kaon condensed core. The different sets of parameters that we consider are shown with different colors. {\em Left panel:}  Solid lines represent $npe\mu$ matter, dashed-dotted lines are obtained with a MC and the dashed with the GC. {\em Right panel:} The effect of adding a standard nuclear physics crust to the Skyrme crystal EoS with kaon condensate obtained from the Gibbs construction (GC). This figure was originally published in \cite{Adam:2022cbs}.}
    \label{MRcurves}
\end{figure*}

Once we have obtained the EOS by performing a Gibbs construction, the corresponding NS solutions can be obtained. We show in the left panel of \cref{MRcurves} the Mass-radius curves for the 4 different sets of parameters given in \cref{tab:Condensation}.

In the right panel of  \cref{MRcurves}, we show the same curves but after the addition of a crust to the NS. We do so by joining the GC equations of state of the Skyrme crystal with 
the BCPM EOS \cite{Vinhas_2015} for low densities, exactly as in Section 5.2.
The two EOS join at the pressure $p=p_*$ where they coincide, i.e., where $\rho_{\rm BCPM} (p_*)= \rho_{\rm, crystal}(p_*)$.
In terms of the baryon density, the joining occurs at $n_{B,*} \sim 1.1 n_0$ for the parameter sets 1-3, and for $n_{B,*} \sim 1.2 n_0$ for the set 4. Again, as in 
\cite{Adam:2020yfv}, we assume a smooth joining between the two EOS (concretely, described by a quadratic interpolation) in order to avoid an artificial phase transition at $p_*$.
We also plot in \cref{MRcurves} the most likely
mass-radius relations for the NS corresponding to GW170817 \cite{LIGOScientific:2017vwq} and GW190425 \cite{Abbott:2020uma} events (orange and blue regions). The green regions represent the estimations for the mass and radius values of PSR J0740+6620 (top) \cite{Miller:2019cac} and J0030+0451 (bottom) \cite{Riley:2021pdl}. The purple region constraints the mass-radius curves from the statistical analysis done in \cite{Altiparmak:2022bke}. We find that the NS resulting from the addition of a crust to the Skyrme crystal EOS with a nonzero kaon condensate agree very well with these recent constraints. Further, the softening of the EOS due to the presence of kaons and the resulting smaller NS radii are important for this agreement.

\section{Conclusions and Outlook}
It was the main purpose of the present paper to report and review recent progress on the modeling of nuclear matter in terms of Skyrme crystals and its application to neutron stars. To put our results in perspective, in the first instant let us underline that, despite its conceptual strengths and some partial successes, the Skyrme model in its current state of exploration is not yet quantitatively competitive with more standard methods of nuclear physics in the description of the large number of available data on nuclei and nuclear matter at low densities. We believe, nevertheless, that the investigation of the Skyrme model as a possible model to describe nuclear matter is a worthy enterprise, for the following reasons. 

Firstly, once the assumption of periodicity of skyrmionic matter is accepted, the Skyrme model actually simplifies at higher densities and shows a more universal behavior which does not depend on many details (e.g., potential terms) which would be very relevant at low densities. The results reported in the present review are a clear demonstration of this fact. The Skyrme model, therefore, provides a rather simple alternative for the description of nuclear matter at high densities which is based on the qualitative assumptions that {\em i)} in the high-pressure region repulsive nuclear forces are more important than, e.g., degeneracy pressure, {\em ii)} the extended character of nucleons -  which is a built-in property of the model -  is important, and {\em iii)} the deconfinement transition is irrelevant inside NS. The inclusion of further hadronic degrees of freedom, on the other hand, is in principle straightforward, and we considered the condensation of kaons as a relevant example.

Secondly, the methods and calculations presented in this article can be applied with only minor modifications to more extended versions of the Skyrme model, where additional terms and further degrees of freedom (e.g., vector mesons) are included.  In other words, once promising candidates among these extended Skyrme models have been identified, their application to the study of nuclear matter at high densities and the resulting NS is relatively straightforward. The difficult part in finding these promising models is related to the calculation of skyrmion solutions, particularly for higher baryon number $B$, and to the identification and calculation of the most relevant quantum corrections. It is our hope that powerful up-to-date computing resources together with modern machine learning techniques like neural networks or other artificial-intelligence based methods will allow to find such promising models via a systematic exploration of the parameter space of the extended Skyrme models. In any case, the progress achieved in recent years is encouraging. 

In our investigation, we placed special emphasis on the importance of the sextic term of the Skyrme model for a viable description of strongly interacting matter at high densities. Further, the impact of this term on the resulting EOS and properties of NS was studied in detail. We also included the important effects of isospin quantization, relevant for the modeling of non-symmetric nuclear matter, and of kaon condensation on the skyrmionic matter properties. The overall behavior of the resulting models of nuclear matter at high density and the resulting NS is already in good agreement with observational data for certain choices of the coupling constants of the underlying Skyrme crystal. Still, there remain some minor differences with the EOS and NS that are most favored by recent observations. In particular, there are indications that the Skyrme crystal EOS is slightly too stiff in the intermediate density range (close to saturation), resulting in mass-radius curves with slightly bigger radii than the most likely ones according to a statistical analysis of observational data. On the other hand, these differences typically do not exceed one kilometer, demonstrating that the Skyrme crystal framework in its current state of development already provides a very reasonable description. There are several possible ways to overcome the remaining discrepancies.

There are, in fact, still some open questions remaining in the Skyrme crystal framework, and many interesting ways in which the results reviewed here can be extended. For instance, the ground state of the Skyrme model with periodic boundary conditions for values of the lattice parameter $L$ slightly above the value at the minimum is not known in the large baryon number limit. 
 That is to say, skyrmion matter in this region is not correctly described by the Skyrme crystal and will most likely be more inhomogeneous, see, e.g., \cref{Figure.Lattice}, which would result in a softening of the EOS there. 
This implies that some observables, such as the compression modulus, which measures the curvature of the energy vs lattice length, will not be well reproduced in the Skyrme crystal at the saturation point. Further investigations on the low density limit of the model are also required in order to be able to describe the physics of neutron star crusts, without recurring to a matching with other nuclear EOS.
In particular, a more realistic description of these intermediate and low density regions will require the addition of further terms to the lagrangian and  
 the inclusion of further degrees of freedom (DOF) such as hyperons or additional mesons. 
 Additional DOF, in general, tend to soften the EOS, and some of these DOF may have a significant impact on the EOS. This question should be further investigated.

Further, a more meticulous treatment of the quantum corrections to the crystal EOS should include, in addition to the isospin quantum effects, also the effect of quantum fluctuations associated to non-zero modes like, e.g., small fluctuation of the pion fields on top of the crystal background. The excitation of such modes could play a relevant role in the finite temperature case, but they might also be important in the zero temperature limit, in which case they contribute to the one loop correction to the energy of the crystal, or the Casimir energy. However, the computation of Casimir energies for solitons in three dimensions without spherical symmetry is in general extremely difficult, especially in non-renormalizable theories such as the Skyrme model, and in the case of skyrmion crystals some sort of mean field approximation will probably be necessary.
A further natural extension of the EOS presented here is the inclusion of the effects of finite temperature and/or a large magnetic field, as well as the study of transport properties of skyrmion crystals, which are crucial to describe nonequilibrium processes of nuclear matter.

Finally, we want to comment on the relevance of our findings for the generalized nuclear effective field theory (GnEFT) based on the hidden local symmetry \cite{Ma-Yang-2023} which we already mentioned in the introduction. Owing to their different field contents, many results are difficult to compare directly between the two theories. There are, however, some results in the GnEFT which are based on the assumption of a skyrmion-to-half-skyrmion phase transition somewhere above two times the nuclear saturation density \cite{MaRho2020}. All that we can say about this issue is that in all our investigations we did not find a sign of this phase transition. For zero pion mass, we formally do find a skyrmion-to-half-skyrmion phase transition (from the FCC to the FCC$_+$ crystal), but this transition is located on the thermodynamically unstable branch, i.e., at a baryon density below the density $n_0$ where the energy takes its minimum value, see \cref{Figure.E_L}. Because of this instability, the Skyrme crystal and its resulting EOS are physically irrelevant in this region and, further, we know that there exist other, more inhomogeneous Skyrme matter solutions with lower energy, see \cref{Figure.Lattice}. 
It was natural for our purposes to identify $n_0$ with the nuclear saturation density, but the present argument, in fact, does not depend on this identification. In other words, our results for the generalized Skyrme model agree with the standard Skyrme model results in  \cite{kugler1989skyrmion}, which the authors of that paper summarized as
"Thus the phase transition between a crystal of half-Skyrmions to a crystal of Skyrmions that was investigated in Refs. 5 -- 11 and 14 is not accessible, it appears on a thermodynamically unstable branch of the phase diagram" (the reference numbers are the references of that paper). Of course, even in the IR limit the effective coupling constants in the Skyrme-type model appearing in the GnEFT are different from ours, and at higher densities the impact of additional fields is difficult to gauge. But taking into account that a skyrmion-to-half-skyrmion  phase transition in a physically relevant branch of the Skyrme crystal EOS has not been found in the full numerical study of any Skyrme model, this transition is probably rather unlikely to happen.

\section*{Acknowledgements}

The 
authors acknowledge financial support from the Ministry of Education, Culture, and Sports, Spain (Grant No. PID2020-119632GB-I00), the Xunta de Galicia (Grant No. INCITE09.296.035PR and Centro singular de investigación de Galicia accreditation 2019-2022), the Spanish Consolider-Ingenio 2010 Programme CPAN (CSD2007-00042), and the European Union ERDF.
 AGMC is grateful to the Spanish Ministry of Science, Innovation and Universities, and the European Social Fund for the funding of his predoctoral research activity (\emph{Ayuda para contratos predoctorales para la formaci\'on de doctores} 2019). MHG thanks the Xunta de Galicia (Consellería de Cultura, Educación y Universidad) for the funding of his predoctoral activity through \emph{Programa de ayudas a la etapa predoctoral} 2021. AW was supported by the Polish National Science Centre 
(NCN 2020/39/B/ST2/01553).

\end{document}